\documentclass[11pt]{article}

\PassOptionsToPackage{nosort}{cite}

\usepackage{amssymb}
\usepackage{dsfont}
\usepackage{mathrsfs}
\usepackage{enumitem}
\usepackage[T1]{fontenc}
\usepackage{chet}
\usepackage[font=small,format=hang,labelfont={sf,bf}]{caption}
\usepackage{tikz}
\usepackage{float}
\usepackage{tensor}
\usepackage{mathtools}

\usetikzlibrary{intersections,arrows.meta,patterns,external}
\newcommand{\lsp}{\hspace{1pt}}
\newcommand{\llsp}{\hspace{0.5pt}}
\newcommand{\lnsp}{\hspace{-1pt}}
\newcommand{\llnsp}{\hspace{-0.5pt}}
\newcommand{\veps}{\varepsilon}

\renewcommand{\geq}{\geqslant}
\renewcommand{\leq}{\leqslant}
\newcommand{\nc}[3]{\lsp\tensor[^{#2}]{#1}{_{#3}}}

\allowdisplaybreaks

\definecolor{darkblue}{rgb}{0.1,0.1,0.7}
\hypersetup{colorlinks,
           linkcolor={darkblue},
           citecolor={darkblue},
           urlcolor={darkblue},
           pdftitle={Gradient Flows and the Curvature of Theory Space},
           pdfdisplaydoctitle
}

\tikzset{cross/.style={path picture={
      \draw[black]
            (path picture bounding box.south east) --
            (path picture bounding box.north west)
            (path picture bounding box.south west) --
            (path picture bounding box.north east);}}}

\title{Gradient Flows and the Curvature of Theory Space}

\author{William H.\ Pannell and Andreas Stergiou\emails{(\href{mailto:william.pannell@kcl.ac.uk}{william.pannell}, \href{mailto:andreas.stergiou@kcl.ac.uk}{andreas.stergiou})@kcl.ac.uk}}

\affiliation{Department of Mathematics, King's College London, Strand, London WC2R 2LS, United Kingdom}

\abstract{The metric and potential associated with the gradient property of renormalisation group flow in multiscalar models in $d=4-\varepsilon$ dimensions are studied. The metric is identified with the Zamolodchikov metric of nearly marginal operators on the sphere. An explicit form for the associated Ricci scalar in $d=4-\varepsilon$ is derived, which shows that the space of multiscalar field theories is curved. The potential is identified with a quantity $\widetilde{F}$ that was previously  proposed as a weakly monotonic function interpolating between the $a$-theorem in four dimensions and the $F$-theorem in three dimensions. This implies that the $\widetilde{F}$-theorem can be extended perturbatively to a theorem about gradient flow in $d=4-\varepsilon$.}

\date{February 2025}

\begin{document}

\maketitle

\toc

\section{Introduction}
With the advent of Wilson's renormalisation group (RG) techniques\cite{Wilson:1971bg}, it is understood that perturbative quantum field theories (QFTs) do not exist in isolation, but rather are to be seen as living along flows in the space of interaction couplings. If the couplings giving the strength of interactions between the various fields in the theory are denoted $\lambda^I$, where $I$ is to be understood as a generalised index, then this renormalisation flow is governed by a dynamical system
\begin{equation}
    \frac{d\lambda^I}{d\ln\mu}=\beta^I\,,
\end{equation}
where $\mu$ is an energy scale and $\beta^I$, the beta function, is a vector field which may be calculated via the renormalisation of divergent diagrams.

Since Zamolodchikov's celebrated $c$-theorem for two-dimensional QFTs\cite{Zamolodchikov:1986gt}, there has been considerable interest in studying the irreversibility of RG flows in QFTs in a variety of spacetime dimensions, with the loss of information as the flow progresses reflected in the existence of a function decreasing along the flow from the ultraviolet (UV) to the infrared (IR). There has been some success in establishing analogous theorems in three dimensions\cite{Myers:2010xs,Jafferis:2011zi,Klebanov:2011gs,Casini:2012ei}, four dimensions\cite{Cardy:1988cwa, Jack:1990eb, Osborn:1991gm, Jack:2013sha, Komargodski:2011vj,Hartman:2023qdn}, and for defects placed inside of bulk QFTs\cite{Casini:2016fgb,Cuomo:2021rkm,Jensen:2015swa}, using both QFT techniques as well as information theoretic properties. These theorems come in three flavours, depending upon the strength of the statement they are able to make:
\begin{enumerate}
    \item \textit{Weak}, there exists $A$ such that $A_{\text{UV}} > A_{\text{IR}}$,
    \item \textit{Strong}, there exists $A$ that monotonically decreases along any RG flow from UV to IR,
    \item \textit{Strongest}, there exist $A$ and Riemannian $G_{IJ}$ such that $\partial_I A=G_{IJ}\beta^{J}$.
\end{enumerate}
Here, $\partial_I$ refers to a derivative with respect to the interaction couplings $\lambda^I$. It should be noted that each successive case is strictly stronger than the previous, e.g.\ weak monotonicity is implied by strong monotonicity, so that the holy grail of these theorems is a non-perturbative proof of the strongest version, i.e.\ gradient flow. As already mentioned, the tensor $G_{IJ}$ which appears cannot be arbitrary, but must be positive definite and symmetric. Both of these conditions are essential for gradient flow, as if we relax our requirements and drop our demand of symmetry one simply arrives back at strong RG-monotonicity. In fact it was in this form, $\partial_I A=T_{IJ}\beta^{J}$ for non-symmetric $T_{IJ}$, that Jack and Osborn originally derived a perturbative proof of the strong $a$-theorem in four dimensions\cite{Jack:1990eb}. Though there does not exist an all-loops proof that it is possible to extend this $T_{IJ}$ to a Riemannian $G_{IJ}$, the possibility of gradient flow has been investigated perturbatively for multiscalar theories in three\cite{Jack:2015tka}, four\cite{Wallace:1974dx,Wallace:1974dy,Jack:2018oec,Pannell:2024sia} and six\cite{Gracey:2015fia} dimensions, and for more general gauge-fermion-scalar theories in four dimensions\cite{Poole:2019kcm}. Gradient flow has also been discussed in two dimensions \cite{Das:1988vd}. Recent work by the authors \cite{Pannell:2024sia} has demonstrated that multiscalar theories are gradient through six loops in both $d=4$ and $d=4-\varepsilon$, as long as the beta function is replaced by a suitable modification, $B$, which removes ambiguities with field redefinitions. The weight of perturbative evidence thus suggests that the strongest level of monotonicity theorem can be expected to hold in a wide class of QFTs.

As a symmetric, positive definite rank-two tensor, $G_{IJ}$ possesses all of the properties necessary to interpret it as a Riemannian metric on the space of couplings. However, as the monotonic $A$-function is usually the object of interest, this identification with a metric has never been pursued beyond a naming convention, and it is entirely unknown what sort of geometry this metric corresponds to. In this paper we intend to take this analogy seriously and ask how demanding gradient flow is reflected in the geometric properties of the space of couplings associated with the metric one is led to write down. One may hope that its explicit relationship with the beta function $\beta^I$ will permit $G_{IJ}$ to encode information about renormalisation within the fabric of the space of couplings itself.

Here, we will restrict our attention to the simple case of multiscalar models. Due to their relative simplicity, and also their applicability to a number of physically interesting systems \cite{Pelissetto:2000ek}, the renormalisation of multiscalar models has been a well-studied topic, especially within the context of the $\varepsilon$ expansion. For our model we will take $N$ massless scalar fields $\phi_i$, $i=1,\ldots,N$, interacting via a generic quartic interaction term, with an action
\begin{equation}
    S=\int d^{\lsp 4-\varepsilon}x\,\big(\tfrac{1}{2}\partial_\mu\phi_i\partial^\mu\phi_i+\tfrac{1}{4!}\lambda_{ijkl}\phi_i\phi_j\phi_k\phi_l\big)\,.
\end{equation}
The gradient properties of this model have been investigated since the 1970's, and it was recently shown via explicit calculation to undergo gradient flow at least through six-loops\cite{Wallace:1974dx, Wallace:1974dy, Jack:2018oec, Pannell:2024sia}.

In this paper we study the simplest geometric question one may ask of a manifold: is it flat? After reviewing the details of gradient flow for multiscalar models in section \ref{sec:review}, in section \ref{sec:ricci} we address this question by explicitly computing the Ricci scalar of the metric $G_{IJ}$ at next-to-leading order in the interaction coupling $\lambda_{ijkl}$. To place the purely formal solution to gradient flow on physical grounds, in section \ref{sec:Ftilde} we demonstrate that it is possible to choose the remaining  unfixed freedom in the solution for $A$ such that it matches the $\widetilde{F}$ proposed in \cite{Giombi:2014xxa} as an RG-monotonic function in $d=4-\varepsilon$. Finally, in section \ref{sec:metric} we show that by fixing the remaining freedom one can also match $G_{IJ}$ with the Zamolodchikov norm $\langle O_I O_J\rangle$ of quartic operators, leading to a situation in $d=4-\varepsilon$ quite analogous to the $c$-theorem in two dimensions. We then show that for $A=\widetilde{F}$ and $G_{IJ}=\langle O_I O_J\rangle$, the Ricci scalar must be non-zero, so that the space of theories is most naturally considered to be curved, rather than flat. We conclude in section \ref{sec:conclusion} and a few appendices contain technical details and results.

\section{Review of gradient flow}\label{sec:review}
Before turning to the main body of this work, for the convenience of the reader let us briefly review gradient flow in multiscalar field theories and its perturbative solution. We begin in $d=4-\varepsilon$, rather than a fixed integer dimension as is normally the case, so that it is important to remember that all of the arbitrary coefficients which will appear are not simply numbers, but will generically be polynomials in $\varepsilon$. We will not attempt to attack the problem non-perturbatively, and will instead write down the functions $A$ and $G_{IJ}$ as series in the interaction tensor $\lambda^{I}$
\begin{equation}
    A=\sum_{n=2}^\infty\sum_m\nc{a}{n\llnsp}{m} \nc{A}{n\lnsp}{m}\,, \qquad G_{IJ}=\nc{g}{0\lnsp}{}\delta_{IJ}+\sum_{n=1}^\infty\sum_m\nc{g}{n\lnsp}{m}\left(\nc{G}{n}{m}\right)_{IJ}\,,
\end{equation}
with the goal of finding a solution for the unfixed coefficients $\nc{a}{n\llnsp}{m}$ and $\nc{g}{n\lnsp}{m}$ satisfying
\begin{equation}
    \partial_I A=G_{IJ}\beta^J
    \label{eq:gradflow}
\end{equation}
order by order in $\lambda$. In these series, the index $n$ indexes a sum over orders in $\lambda$ while the index $m$ denotes a sum over the distinct tensor structures which appear at that order, so that, for instance, $\nc{A}{n\lnsp}{m}$ is the $m$-th $\text{O}(\lambda^n)$ vacuum bubble. Symmetry of the metric is ensured by taking the constituent tensor structures $\left(\nc{G}{n}{m}\right)_{IJ}$ to be symmetric from the start, while positive definiteness is expected to hold perturbatively from the presence of the leading $\delta_{IJ}$ term. The tensor structures themselves which appear in these series rapidly become unwieldy as the number of $\lambda$'s increases, and for clarity we will express them only diagrammatically. As an example of this diagrammatic representation, the terms contributing to (\ref{eq:gradflow}) through $\text{O}(\lambda^3)$ are
\begin{equation}
\begin{split}
A=&\nc{a}{2}{}\,\begin{tikzpicture}[scale=0.5,baseline=(vert_cent.base)]
        \node (vert_cent) at (0,0) {$\phantom{\cdot}$};
        \draw (0,0) circle (1cm);
        \filldraw (-1,0) circle (2pt);
        \filldraw (1,0) circle (2pt);
        \draw (1,0) to[out=135, in=45] (-1,0);
        \draw (1,0) to[out=225, in=315] (-1,0);
    \end{tikzpicture}+\nc{a}{3\lnsp}{}\,\begin{tikzpicture}[scale=0.5,baseline=(vert_cent.base)]
        \node (vert_cent) at (0,0) {$\phantom{\cdot}$};
        \node (center) at (0,0) {};
        \def\radius{1cm};
        \node[inner sep=0pt] (top) at (90:\radius) {};
        \node[inner sep=0pt] (left) at (210:\radius) {};
        \node[inner sep=0pt] (right) at (330:\radius) {};
        \draw (center) circle[radius=\radius];
        \filldraw (top) circle[radius=2pt];
        \filldraw (left) circle[radius=2pt];
        \filldraw (right) circle[radius=2pt];
        \draw (top) to[out=270, in=30] (left);
        \draw (top) to[out=270, in=150] (right);
        \draw (left) to[out=30, in=150] (right);
    \end{tikzpicture}+\nc{a}{4\lnsp}{1}\,
    \begin{tikzpicture}[scale=0.5,baseline=(vert_cent.base)]
      \node (vert_cent) at (0,0) {$\phantom{\cdot}$};
      \node (center) at (0,0) {};
      \def\radius{1cm};
      \node[inner sep=0pt] (n1) at (45:\radius) {};
      \node[inner sep=0pt] (n2) at (135:\radius) {};
      \node[inner sep=0pt] (n3) at (225:\radius) {};
      \node[inner sep=0pt] (n4) at (315:\radius) {};
      \draw (center) circle[radius=\radius];
      \filldraw (n1) circle[radius=2pt];
      \filldraw (n2) circle[radius=2pt];
      \filldraw (n3) circle[radius=2pt];
      \filldraw (n4) circle[radius=2pt];
      \draw (n1) to[out=225, in=135] (n4);
      \draw (n1) -- (n4);
      \draw (n2) to[out=315, in=45] (n3);
      \draw (n2) -- (n3);
    \end{tikzpicture}
  + \nc{a}{4\lnsp}{2}\,
  \begin{tikzpicture}[scale=0.5,baseline=(vert_cent.base)]
      \node (vert_cent) at (0,0) {$\phantom{\cdot}$};
      \node (center) at (0,0) {};
      \def\radius{1cm};
      \node[inner sep=0pt] (n1) at (0:\radius) {};
      \node[inner sep=0pt] (n2) at (90:\radius) {};
      \node[inner sep=0pt] (n3) at (180:\radius) {};
      \node[inner sep=0pt] (n4) at (270:\radius) {};
      \draw (center) circle[radius=\radius];
      \filldraw (n1) circle[radius=2pt];
      \filldraw (n2) circle[radius=2pt];
      \filldraw (n3) circle[radius=2pt];
      \filldraw (n4) circle[radius=2pt];
      \draw (n1) to[out=180, in=270] (n2);
      \draw (n2) to[out=270, in=0] (n3);
      \draw (n3) to[out=0, in=90] (n4);
      \draw (n4) to[out=90, in=180] (n1);
    \end{tikzpicture}
  + \nc{a}{4\lnsp}{3}\,
  \begin{tikzpicture}[scale=0.5,baseline=(vert_cent.base)]
      \node (vert_cent) at (0,0) {$\phantom{\cdot}$};
      \node (center) at (0,0) {};
      \def\radius{1cm};
      \node[inner sep=0pt] (n1) at (0:\radius) {};
      \node[inner sep=0pt] (n2) at (90:\radius) {};
      \node[inner sep=0pt] (n3) at (180:\radius) {};
      \node[inner sep=0pt] (n4) at (270:\radius) {};
      \draw (center) circle[radius=\radius];
      \filldraw (n1) circle[radius=2pt];
      \filldraw (n2) circle[radius=2pt];
      \filldraw (n3) circle[radius=2pt];
      \filldraw (n4) circle[radius=2pt];
      \draw (n1) to[out=150, in=30] (n3);
      \draw (n1) to[out=210, in=-30] (n3);
      \draw (n2) to[out=300, in=60] (n4);
      \draw (n2) to[out=240, in=120] (n4);
    \end{tikzpicture}+\text{O}(\lambda^5)\,,
    \label{eq:A2loops}
\end{split}
\end{equation}

\begin{equation}
\begin{split}
G_{IJ}=&\nc{g}{0\lnsp}{}\lsp\begin{tikzpicture}[scale=0.65,
    vertex/.style={draw,circle,fill=black,minimum size=2pt,inner sep=0pt},
    arc/.style={thick},baseline=(vert_cent.base)]
    \node (vert_cent) at (0,-0.75) {$\phantom{\cdot}$};
    \foreach [count=\i] \coord in {
(1.00,0), (-1,0),(-1,-0.5),(1,-0.5)}{
        \node[] (p\i) at \coord {};
    }
    \foreach [count=\i] \coord in {
(-1,-1), (1,-1),(-1,-1.5),(1,-1.5)}{
        \node[] (d\i) at \coord {};
    }
    \draw (d1) edge (d2);
    \draw (d3) edge (d4);
    \draw (p1) edge (p2);
    \draw (p3) edge (p4);
\end{tikzpicture}+\nc{g}{1\lnsp}{}\lsp\begin{tikzpicture}[scale=0.65,
    vertex/.style={draw,circle,fill=black,minimum size=2pt,inner sep=0pt},
    arc/.style={thick},baseline=(vert_cent.base)]
    \node (vert_cent) at (0,-0.75) {$\phantom{\cdot}$};
    \node[vertex] (c) at (0,-0.25) {};
    \foreach [count=\i] \coord in {
(1.00,0), (-1,0),(-1,-0.5),(1,-0.5)}{
        \node[] (p\i) at \coord {};
    }
    \foreach [count=\i] \coord in {
(-1,-1), (1,-1),(-1,-1.5),(1,-1.5)}{
        \node[] (d\i) at \coord {};
    }
    \draw (c) edge (p1)
                   edge (p2)
                   edge (p3)
                   edge (p4);
    \draw (d1) edge (d2);
    \draw (d3) edge (d4);
\end{tikzpicture}+\nc{g}{2\lnsp}{1}\lsp
    \begin{tikzpicture}[scale=0.65,
    vertex/.style={draw,circle,fill=black,minimum size=2pt,inner sep=0pt},
    arc/.style={thick},baseline=(vert_cent.base)]
    \node (vert_cent) at (0,-0.75) {$\phantom{\cdot}$};
    \node[vertex] (c1) at (0,-0.25) {};
    \node[vertex] (c2) at (0,-1.25) {};
    \foreach [count=\i] \coord in {
(1.00,0), (-1,0),(-1,-0.5),(1,-0.5)}{
        \node[] (p\i) at \coord {};
    }
    \foreach [count=\i] \coord in {
(-1,-1), (1,-1),(-1,-1.5),(1,-1.5)}{
        \node[] (d\i) at \coord {};
    }
    \draw (c1) edge (p1)
                   edge (p2)
                   edge (p3)
                   edge (d1);
    \draw (c2) edge (p4)
                   edge (d2)
                   edge (d3)
                   edge (d4);
\end{tikzpicture}+\nc{g}{2\lnsp}{2}\lsp
    \begin{tikzpicture}[scale=0.65,
    vertex/.style={draw,circle,fill=black,minimum size=2pt,inner sep=0pt},
    arc/.style={thick},baseline=(vert_cent.base)]
    \node (vert_cent) at (0,-0.75) {$\phantom{\cdot}$};
    \node[vertex] (c1) at (0,-0.25) {};
    \node[vertex] (c2) at (0,-1.25) {};
    \foreach [count=\i] \coord in {
(1.00,0), (-1,0),(-1,-0.5),(1,-0.5)}{
        \node[] (p\i) at \coord {};
    }
    \foreach [count=\i] \coord in {
(-1,-1), (1,-1),(-1,-1.5),(1,-1.5)}{
        \node[] (d\i) at \coord {};
    }
    \draw (c1) edge (p1)
                   edge (p2)
                   edge (p3)
                   edge (p4);
    \draw (c2) edge (d1)
                   edge (d2)
                   edge (d3)
                   edge (d4);
\end{tikzpicture}+\nc{g}{2\lnsp}{3}\lsp
    \begin{tikzpicture}[scale=0.65,
    vertex/.style={draw,circle,fill=black,minimum size=2pt,inner sep=0pt},
    arc/.style={thick},baseline=(vert_cent.base)]
    \node (vert_cent) at (0,-0.75) {$\phantom{\cdot}$};
    \node[vertex] (c1) at (-0.25,-0.5) {};
    \node[vertex] (c2) at (0.25,-0.5) {};
    \foreach [count=\i] \coord in {(-1,0),(-1,-0.5),(-1,-1)}{
        \node[] (p\i) at \coord {};
    }
    \foreach [count=\i] \coord in {(1,0),(1,-0.5),(1,-1)}{
        \node[] (d\i) at \coord {};
    }
        \node[] (d4) at (1,-1.5) {};
    \node[] (p4) at (-1,-1.5) {};
    \draw (c1) edge (p1)
                   edge (p2)
                   edge (p3)
                   edge (c2);
    \draw (c2) edge (d1)
                   edge (d2)
                   edge (d3);
    \draw (p4) edge (d4);
\end{tikzpicture}\\
    &+\nc{g}{2\lnsp}{4}\lsp\begin{tikzpicture}[scale=0.65,
    vertex/.style={draw,circle,fill=black,minimum size=2pt,inner sep=0pt},
    arc/.style={thick},baseline=(vert_cent.base)]
    \node (vert_cent) at (0,-0.75) {$\phantom{\cdot}$};
    \node[vertex] (c1) at (0,0) {};
    \node[vertex] (c2) at (0,-1) {};
    \foreach [count=\i] \coord in {(-1,0),(-1,-0.5),(-1,-1)}{
        \node[] (p\i) at \coord {};
    }
    \foreach [count=\i] \coord in {(1,0),(1,-0.5),(1,-1)}{
        \node[] (d\i) at \coord {};
    }
    \node[] (d4) at (1,-1.5) {};
    \node[] (p4) at (-1,-1.5) {};
    \draw (c1) edge (p1)
                   edge (p2)
                   edge (d1)
                   edge (c2);
    \draw (c2) edge (p3)
                   edge (d2)
                   edge (d3);
    \draw (p4) edge (d4);
\end{tikzpicture}+\nc{g}{2\lnsp}{5}\lsp
    \begin{tikzpicture}[scale=0.65,
    vertex/.style={draw,circle,fill=black,minimum size=2pt,inner sep=0pt},
    arc/.style={thick},baseline=(vert_cent.base)]
    \node (vert_cent) at (0,-0.75) {$\phantom{\cdot}$};
    \node[vertex] (c1) at (-0.25,-0.25) {};
    \node[vertex] (c2) at (0.25,-0.25) {};
    \foreach [count=\i] \coord in {(-1,0),(-1,-0.5),(-1,-1)}{
        \node[] (p\i) at \coord {};
    }
    \foreach [count=\i] \coord in {(1,0),(1,-0.5),(1,-1)}{
        \node[] (d\i) at \coord {};
    }
    \node[] (d4) at (1,-1.5) {};
    \node[] (p4) at (-1,-1.5) {};
    \draw (c1) edge (p1)
                   edge (p2)
                   edge[bend right=60] (c2)
                   edge[bend left=60] (c2);
    \draw (c2) edge (d1)
                   edge (d2);
    \draw (p3) edge (d3);
    \draw (p4) edge (d4);
\end{tikzpicture}+\nc{g}{2\lnsp}{6}\lsp
    \begin{tikzpicture}[scale=0.65,
    vertex/.style={draw,circle,fill=black,minimum size=2pt,inner sep=0pt},
    arc/.style={thick},baseline=(vert_cent.base)]
    \node (vert_cent) at (0,-0.75) {$\phantom{\cdot}$};
    \node[vertex] (c1) at (0,0) {};
    \node[vertex] (c2) at (0,-0.5) {};
    \foreach [count=\i] \coord in {(-1,0),(-1,-0.5),(-1,-1)}{
        \node[] (p\i) at \coord {};
    }
    \foreach [count=\i] \coord in {(1,0),(1,-0.5),(1,-1)}{
        \node[] (d\i) at \coord {};
    }
    \node[] (d4) at (1,-1.5) {};
    \node[] (p4) at (-1,-1.5) {};
    \draw (c1) edge (p1)
                   edge (d1)
                   edge[bend right=60] (c2)
                   edge[bend left=60] (c2);
    \draw (c2) edge (p2)
                   edge (d2);
    \draw (p3) edge (d3);
    \draw (p4) edge (d4);
\end{tikzpicture}+\nc{g}{2\lnsp}{7}\lsp
    \begin{tikzpicture}[scale=0.65,
    vertex/.style={draw,circle,fill=black,minimum size=2pt,inner sep=0pt},
    arc/.style={thick},baseline=(vert_cent.base)]
    \node (vert_cent) at (0,-0.75) {$\phantom{\cdot}$};
    \node[vertex] (c1) at (-0.5,0) {};
    \node[vertex] (c2) at (0.5,0) {};
    \foreach [count=\i] \coord in {(-1,0),(-1,-0.5),(-1,-1)}{
        \node[] (p\i) at \coord {};
    }
    \foreach [count=\i] \coord in {(1,0),(1,-0.5),(1,-1)}{
        \node[] (d\i) at \coord {};
    }
    \node[] (d4) at (1,-1.5) {};
    \node[] (p4) at (-1,-1.5) {};
    \draw (c1) edge (p1)
                   edge (c2)
                   edge[bend right=60] (c2)
                   edge[bend left=60] (c2);
    \draw (c2) edge (d1);
    \draw (p2) edge (d2);
    \draw (p3) edge (d3);
    \draw (p4) edge (d4);
\end{tikzpicture}+\text{O}(\lambda^3)\,.
\end{split}
\end{equation}
We should note that in the diagrammatic representation of the metric we implicitly symmetrise over the labelling of the indices on the two sides of each diagram. The beta function itself will also have an expansion in $\lambda$ like
\begin{equation}
    \beta^I=\sum_{n=1}^\infty\sum_m (\nc{\mathfrak{b}}{n}{m}+\varepsilon\nc{\mathfrak{e}}{n}{m}) (\nc{\beta}{n}{m})^I\,,
\end{equation}
where the coefficients $\{\nc{\mathfrak{b}}{n}{m},\nc{\mathfrak{e}}{n}{m}\}$ depend upon the specific choice of renormalisation scheme. In the $\overline{\text{MS}}$ scheme, which will be used for explicit computations, the terms in the beta function through two loops are
\begin{equation}
    \beta=-\varepsilon\,\begin{tikzpicture}[scale=0.5,baseline=(vert_cent.base)]
        \node (vert_cent) at (0,0) {$\phantom{\cdot}$};
        \filldraw (0,0) circle (2pt);
        \draw (-1,1)--(0,0);
        \draw (-1,-1)--(0,0);
        \draw (1,1)--(0,0);
        \draw (1,-1)--(0,0);
    \end{tikzpicture}+\begin{tikzpicture}[scale=0.5,baseline=(vert_cent.base)]
        \node (vert_cent) at (0,0) {$\phantom{\cdot}$};
        \draw (0,0) circle (1cm);
        \filldraw (-1,0) circle (2pt);
        \filldraw (1,0) circle (2pt);
        \draw (-1.7,0.7)--(-1,0);
        \draw (-1.7,-0.7)--(-1,0);
        \draw (1.7,0.7)--(1,0);
        \draw (1.7,-0.7)--(1,0);
    \end{tikzpicture}-\begin{tikzpicture}[scale=0.5,baseline=(vert_cent.base)]
      \node (vert_cent) at (0,0) {$\phantom{\cdot}$};
      \draw (-1,-1)--(1,1);
      \draw (-1,1)--(1,-1);
      \filldraw (0,0) circle[radius=2pt];
      \filldraw (1,1) circle[radius=2pt];
      \filldraw (1,-1) circle[radius=2pt];
      \draw (1,1)--(2,1);
      \draw (1,-1)--(2,-1);
      \draw (1,1) to[out=240, in=120] (1,-1);
      \draw (1,1) to[out=300, in=60] (1,-1);
    \end{tikzpicture}
    + \frac{1}{12}\,
    \begin{tikzpicture}[scale=0.5,baseline=(vert_cent.base)]
      \node (vert_cent) at (0,0) {$\phantom{\cdot}$};
      \draw (-1,0)--(1,0);
      \draw (-0.8,1)--(0,0);
      \draw (-0.8,-1)--(0,0);
      \filldraw (0,0) circle[radius=2pt];
      \filldraw (0.7,0) circle[radius=2pt];
      \filldraw (2.3,0) circle[radius=2pt];
      \draw (1,0)--(3,0);
      \draw (0.7,0) to[out=60, in=120] (2.3,0);
      \draw (0.7,0) to[out=-60, in=-120] (2.3,0);
    \end{tikzpicture}+\text{O}(\lambda^4)\,,
\end{equation}
where for the beta function there is an implicit sum over inequivalent permutations of external indices. The existence of a solution does not depend upon the particular renormalisation scheme one chooses, so in general we will leave the coefficients arbitrary.

Plugging in these series expansions into the gradient flow equation (\ref{eq:gradflow}), one finds that it becomes
\begin{equation}
\begin{split}
    \sum_{n=2}^\infty\sum_m\nc{a}{n\llnsp}{m}\partial_I\nc{A}{n\lnsp}{m}=&\sum_{n=1}^\infty\bigg(\nc{g}{0\lnsp}{}\sum_m\nc{\mathfrak{b}}{n}{m}(\nc{\beta}{n}{m})_I+\varepsilon\nc{g}{0\lnsp}{}\sum_m\nc{\mathfrak{e}}{n}{m}(\nc{\beta}{n}{m})_I\\&+\sum_{n'=1}^{n-1}\sum_{m,m'}\nc{\mathfrak{b}}{(n-n')}{m}\nc{g}{n'}{m'}(\nc{G}{n'}{m'})_{IJ}(\nc{\beta}{(n-n')}{m})^J\\&+\varepsilon\sum_{n'=1}^{n-1}\sum_{m,m'}\nc{\mathfrak{e}}{(n-n')}{m}\nc{g}{n'}{m'}(\nc{G}{n'}{m'})_{IJ}(\nc{\beta}{(n-n')}{m})^J\bigg)\,.
    \label{eq:gradflowtosolve}
\end{split}
\end{equation}
Working order by order in $\lambda$, one solves this equation by matching the coefficients of the distinct tensor structures which appear on either side. The resulting equations are solved for the unknowns $\nc{a}{n\llnsp}{m}$ and $\nc{g}{n\lnsp}{m}$. The existence of a solution to these equations was first verified through three loops in the 1970's\cite{Wallace:1974dx,Wallace:1974dy}, and more recently through four loops\cite{Jack:2018oec} and now six loops\cite{Pannell:2024sia}. The number of equations grows much more rapidly than the number of unknowns one has to solve them with, indicating that this system of equations must be under-constrained. Indeed, while through three loops one finds that there generically exists a solution, beginning at four loops there begin to appear constraints on the coefficients $\{\nc{\mathfrak{b}}{n}{m},\nc{\mathfrak{e}}{n}{m}\}$ of the beta function themselves \cite{Jack:2018oec}. These constraints are scheme-independent and satisfied by the $\overline{\text{MS}}$ values, though at six loops they require a modification of the beta function to address certain ambiguities\cite{Jack:2018oec,Pannell:2024sia}. The details of these constraints will not be of interest to us in this paper, as we will largely restrict ourselves to lower-loop computations. At low loop orders it is straightforward to write down the solution to (\ref{eq:gradflowtosolve}) explicitly, and in Appendix \ref{app:gradsol} we do so through $\text{O}(\lambda^4)$.

\section{Computation of the Ricci scalar}\label{sec:ricci}
If we are to interpret the rank-two tensor, $G_{IJ}$, which appears in the gradient flow equation as a metric on the space of couplings, it is natural to ask what type of geometry gradient flow imposes on the space of theories. As physical statements must be scheme-invariant, and as scheme changes can be realised by diffeomorphisms on the space of couplings\cite{Jack:2016tpp}, any meaningful statement we are to make must be written in terms of diffeomorphism-invariant geometric quantities. The simplest, and most natural, question to ask is if the space of theories must be curved, that is to say whether or not there exists a consistent, theory-independent choice of freedom in the gradient flow solution such that the Ricci scalar $R$ associated to the metric $G_{IJ}$ is zero. To evaluate this question, we will begin by writing the Ricci scalar in a $\lambda$-expansion as
\begin{equation}
    R=\sum_{n=0}^\infty\sum_m \nc{R}{n\lnsp}{m}\lsp\nc{T}{n}{m}\,,
\end{equation}
where $\nc{T}{n}{m}$ is the $m$-th vacuum bubble at $\text{O}(\lambda^n)$. The coefficients $\nc{R}{n\lnsp}{m}$ will be expressed in terms of the coefficients $\nc{g}{n\lnsp}{m}$ of the metric using
\begin{equation}
    R=G^{IJ}\left(\Gamma^{K}_{IK}\Gamma^{L}_{JL}-\Gamma^{K}_{LI}\Gamma^{L}_{KJ}+\partial_K\Gamma^{K}_{IJ}-\partial_I\Gamma^{K}_{JK}\right),
    \label{eq:riccidef}
\end{equation}
whose applicability does not require that we apply the solution to gradient flow or specify if we are working in $d=4$ or $d=4-\varepsilon$.

We will evaluate the Ricci scalar perturbatively in the coupling $\lambda$, which requires taking the perturbative expansion for the metric $G_{IJ}$ and identifying which terms will contribute to (\ref{eq:riccidef}) at a given order when the correct derivatives are taken. At leading order, $\text{O}(\lambda^0)$, we will require at most $\text{O}(\lambda^2)$ terms in the metric, as there are at most two derivatives acting on a single metric diagram. After taking derivatives, we will be left with a string of Kronecker deltas $\delta_{ij}$ which will contract to give a polynomial in $N$. However, as the indices in (\ref{eq:riccidef}) are generalised indices $I=(ijkl)$, each contracted index in $R$ hides 24 index permutations, so that each term in the metric will contribute $\sim \lnsp 10^9$ terms to $R$, which makes it a daunting computational task to completely evaluate even $\nc{R}{0\lnsp}{}$. However, we find at leading order that it is possible to determine $R$ by using direct, brute-force methods to explicitly evaluate all of the tensor contractions. The tensor structures are initially constructed in \textit{Mathematica} using the package \href{http://www.xact.es}{\texttt{xAct}}~\cite{xact}, by taking a single `seed' structure and then acting with proper index permutations. The list of different permutations is then fed into a program written in \href{https://www.nikhef.nl/~form/}{\texttt{FORM}}, which is able to efficiently evaluate and reduce the products of tensors and sum the result. Much more efficient, though difficult to reproduce on a computer, is the evaluation of these contractions by representing the tensor structures diagrammatically. This diagrammatic method, which we use to explicitly compute $\nc{R}{0\lnsp}{}$ in Appendix \ref{sec:diagcalc}, reduces the problem to its combinatorial core, and bypasses the difficulty with the total number of terms by performing the generalised index contractions one at a time without the need to write out all of the other indices explicitly. However, its inability to be reproduced on a computer limits its scope, as the number of terms in the metric rises quickly as the order in $\lambda$ increases. The resulting expression one obtains for $\nc{R}{0\lnsp}{}$ is rather long, and can be found explicitly for the interested reader in (\ref{eq:Ricci0}) of Appendix \ref{app:explicit}.

So far we have not applied the fact that the metric $G_{IJ}$ has arisen as a result of demanding that multiscalar systems undergo gradient flow, and have simply written down the most general expansion for $\nc{R}{0\lnsp}{}$ which is compatible with the diagrammatic expansion the metric must have. Though it appears as if there is a great deal of freedom in $\nc{R}{0\lnsp}{}$, many of the coefficients $\nc{g}{2\lnsp}{m}$ will be fixed by imposing the solution for gradient flow. In $d=4-\varepsilon$, these coefficients will not simply be numbers, but instead will generally be functions of $\varepsilon$, so that $\nc{R}{0\lnsp}{}$ will have its own $\varepsilon$ expansion. Plugging in the solution outlined in Appendix \ref{app:gradsol}, one finds that
\begin{align}
    \nc{R}{0\lnsp}{}&=\frac{N (N-1)(N+2)(N+4) }{6912}\bigg(6 \nc{a}{5\lnsp}{3} \left(3 N^2+18 N+31\right)-(\nc{g}{1\lnsp}{})^2 \left(3 N^2+18 N+31\right)\nonumber\\& +6 \nc{g}{1\lnsp}{} \left(3 N^2+16 N+25\right)+216 \nc{g}{2\lnsp}{1} N+648\nc{g}{2\lnsp}{1}+10 N^3+7 N^2-126 N-63\bigg)\nonumber\\&+\frac{ N (N-1)(N+2)(N+4) \varepsilon}{82944}\bigg(\nc{g}{1\lnsp}{} \left(288 \nc{{a\smash{\mathrlap{'}}}}{2\llnsp}{}\lsp \big(3 N^2+16N+25\big)-20 N^3-14 N^2+288 N+234\right)\nonumber\\&+48 \nc{{a\smash{\mathrlap{'}}}}{2\llnsp}{}\lsp \big(6 \nc{a}{5\lnsp}{3} \left(3 N^2+18 N+31\right)+216 \nc{g}{2\lnsp}{1} (N+3)+10 N^3+7 N^2-126 N-63\big)\nonumber\\&+9 \big(-2 \nc{a}{5\lnsp}{3} \left(69 N^2+412 N+707\right)+56 N^3+527 N^2+1806 N+72 \left(N^2+6 N+5\right) \zeta(3)\nonumber\\&
    +2751\big)-72 \nc{{a\smash{\mathrlap{'}}}}{2\llnsp}{}\lsp (\nc{g}{1\lnsp}{})^2 \left(3 N^2+18 N+31\right)+288 \nc{g}{3\lnsp}{2}\left(3 N^2+18 N+31\right)\nonumber\\&-288 \nc{g}{3\lnsp}{3} \left(3 N^2+18 N+31\right)\bigg)+\text{O}(\varepsilon ^2)\,,
\end{align}
where have used the $\overline{\text{MS}}$ values for the beta function coefficients, found e.g.\ in \cite{Kompaniets:2017yct}, and we omit higher order in $\varepsilon$ terms as they contain a great number of coefficients which will only be fixed by demanding gradient flow at higher loop orders. Though there still remain unfixed degrees of freedom, it is important to note that the leading contribution to $\nc{R}{0\lnsp}{}$ at large $N$ is already entirely fixed by gradient flow,
\begin{equation}
    \nc{R}{0\lnsp}{}=\frac{5}{3456}N^7+\text{O}(N^6,\,\varepsilon N^7)\,.
\end{equation}
It is thus not possible to choose the remaining coefficients in a theory-independent, i.e.\ $N$-independent, manner to set $\nc{R}{0\lnsp}{}=0$, so that we are led to conclude that gradient flow naturally induces a curvature on the space of couplings. It is also interesting to note that though there appear to be three unfixed degrees of freedom, at each order in $N$ the coefficients appear only in two distinct combinations, so that there are really only two degrees of freedom that must be specified to determine the $\varepsilon$-independent part of $\nc{R}{0\lnsp}{}$.

\subsection{Scheme independence}
A non-trivial check on the veracity of (\ref{eq:Ricci0}) is its independence from the renormalisation scheme used to compute it. Recast in the language of differential geometry, a generic scheme change is realised by a diffeomorphism on the couplings $\lambda^I$ of the form
\begin{equation}\label{eq:lamchange}
    \lambda^{I}\rightarrow \lambda\smash{\mathrlap{'}}^{\,\,I}=\lambda^{I}+\sum_{n=2}^\infty\sum_m\nc{r}{n}{m} \tensor[^n]{T}{_{m}}^I(\lambda)\,,
\end{equation}
where the scheme change parameters $\nc{r}{n}{m}$ are arbitrary, and $m$ indexes a sum over the tensor structures at $\text{O}(\lambda^n)$ with one free generalised index. The constant term in the Ricci scalar cannot change under this diffeomorphism, so that it must hold that (\ref{eq:Ricci0}) is unchanged if we change all of the metric coefficients to their primed equivalents. As a rank-two tensor, the metric will change in the usual way under diffeomorphisms,
\begin{equation}\label{eq:metchange}
    G_{IJ}\rightarrow g\smash{\mathrlap{'}}_{IJ}=\frac{\partial\lambda^K}{\partial\lambda\smash{\mathrlap{'}}^{\,\,I}}\frac{\partial\lambda^L}{\partial\lambda\smash{\mathrlap{'}}^{\,\,J}}G_{KL}\,,
\end{equation}
where $g\smash{\mathrlap{'}}_{IJ}$ will also have an expansion in terms of $\lambda\smash{\mathrlap{'}}$,
\begin{equation}
g\smash{\mathrlap{'}}_{IJ}=\nc{g}{0\lnsp}{} \delta_{IJ}+\sum_{n=1}^\infty \sum_m \nc{{g\smash{\mathrlap{'}}}}{\lnsp n}{m}(\nc{G}{n}{m})_{IJ}(\lambda')\,.
\end{equation}
To solve for the coefficients $\nc{{g\smash{\mathrlap{'}}}}{n\lnsp}{m}$ in terms of the original coefficients $\nc{g}{n\lnsp}{m}$ it is easiest to invert (\ref{eq:metchange}) to
\begin{equation}
    G_{IJ}=\frac{\partial\lambda\smash{\mathrlap{'}}^{
    \,\,K}}{\partial\lambda^I}\frac{\partial\lambda\smash{\mathrlap{'}}^{\,\,L}}{\partial\lambda^J}\frac{}{}g\smash{\mathrlap{'}}_{KL}\,,
\end{equation}
and then use (\ref{eq:lamchange}) to expand the right-hand side in terms of $\lambda$. After matching the various tensor structures, we find that
\begin{equation}
\begin{split}
    \nc{{g}}{1\lnsp}{}'&=\nc{g}{1\lnsp}{}-4 \nc{g}{0\lnsp}{} \lsp\nc{r}{2}{}\,,\qquad\nc{g\smash{\mathrlap{'}}}{2\lnsp}{1}= \nc{g}{2\lnsp}{1}-2 \nc{g}{0\lnsp}{} \lsp\nc{r}{3}{1}\,,\!\!\qquad\nc{{g\smash{\mathrlap{'}}}}{2\lnsp}{2}= \frac{1}{3} \left(6 \nc{g}{0\lnsp}{} \big(\nc{r}{2}{}\big)^2-6 \nc{g}{0\lnsp}{} \lsp\nc{r}{3}{2}-2 \nc{g}{1\lnsp}{} \lsp\nc{r}{2}{}+3 \nc{g}{2\lnsp}{2}\right),\\
    \nc{{g\smash{\mathrlap{'}}}}{2\lnsp}{3}&= \nc{g}{2\lnsp}{3}-2 \nc{g}{0\lnsp}{} \lsp\nc{r}{3}{1}\,,\qquad\nc{{g\smash{\mathrlap{'}}}}{2\lnsp}{4}= \frac{1}{3} \left(24\nc{g}{0\lnsp}{} \big(\nc{r}{2}{}\big)^2-12 \nc{g}{0\lnsp}{} \lsp\nc{r}{3}{3}-8 \nc{g}{1\lnsp}{} \lsp\nc{r}{2}{}+3 \nc{g}{2\lnsp}{4}\right),\\
    \nc{{g\smash{\mathrlap{'}}}}{2\lnsp}{5}&= \frac{1}{3} \left(10 \nc{g}{0\lnsp}{} \big(\nc{r}{2}{}\big)^2-12 \nc{g}{0\lnsp}{} \lsp \nc{r}{3}{2}-3 \nc{g}{1\lnsp}{} \lsp\nc{r}{2}{}+3 \nc{g}{2\lnsp}{5}\right),\\
    \nc{{g\smash{\mathrlap{'}}}}{2\lnsp}{6}&= \frac{1}{3} \left(8 \nc{g}{0\lnsp}{} \big(\nc{r}{2}{}\big)^2-6 \nc{g}{0\lnsp}{} \lsp \nc{r}{3}{3}-2\nc{g}{1\lnsp}{} \lsp \nc{r}{2}{}+3 \nc{g}{2\lnsp}{6}\right),\qquad\nc{{g\smash{\mathrlap{'}}}}{2\lnsp}{7}= \nc{g}{2\lnsp}{7}-2 \nc{g}{0\lnsp}{} \lsp \nc{r}{3}{1}\,.
    \label{eq:scheme}
\end{split}
\end{equation}
Plugging in this into (\ref{eq:Ricci0}), we see that all of the scheme change parameters cancel, so that $\nc{R}{0\lnsp}{}$ is indeed scheme-independent.

The requirement of scheme invariance also allows us to `bootstrap' the full expression for $\nc{R}{0\lnsp}{}$ using only the $\nc{g}{2\lnsp}{m}$ metric contributions by noticing that it is always possible to choose $\nc{r}{2}{}$ such that $\nc{{g\smash{\mathrlap{'}}}}{1\lnsp}{}=0$. That is to say, it is possible to treat $\nc{g}{1\lnsp}{}$ in (\ref{eq:Ricci0}) as though it were pure scheme, and thus fix the $\nc{g}{1\lnsp}{}$ dependence to cancel any scheme dependence. Surveying the form of (\ref{eq:riccidef}) we see that if $\nc{g}{1\lnsp}{}$ is to enter into $\nc{R}{0\lnsp}{}$, it must enter in the combination $\big(\nc{g}{1\lnsp}{}\big)^2$. Consequently, the purely $\nc{g}{1\lnsp}{}$ part of $\nc{R}{0\lnsp}{}$ can be written as
\begin{equation}
    \big(\nc{g}{1\lnsp}{}\big)^2\left(a_1\lsp N+a_2\lsp N^2+a_3\lsp N^3+a_4\lsp N^4+a_5\lsp N^5+a_6\lsp N^6+a_7\lsp N^7\right)
\end{equation}
for some coefficients $a_i$ which must be fixed. As we can treat $\nc{g}{1\lnsp}{}$ as pure scheme, we can replace this with
\begin{equation}
    \big(-4\nc{g}{0\lnsp}{}\lsp \nc{r}{2}{}\big)^2\left(a_1\lsp N+a_2\lsp N^2+a_3\lsp N^3+a_4\lsp N^4+a_5\lsp N^5+a_6\lsp N^6+a_7\lsp N^7\right)\,,
\end{equation}
which must absorb any scheme variance in the purely $\nc{g}{2\lnsp}{m}$ terms. Using (\ref{eq:scheme}), and setting $\nc{g}{1\lnsp}{}=0$, we find that the purely $\nc{g}{2\lnsp}{m}$ part of $\nc{R}{0\lnsp}{}$ will change under a scheme change by
\begin{equation}
    -\frac{31 N (\nc{r}{2}{})^2}{54\nc{g}{0\lnsp}{}}-\frac{41 N^2 (\nc{r}{2}{})^2}{216\nc{g}{0\lnsp}{}}+\frac{167 N^3 (\nc{r}{2}{})^2}{432\nc{g}{0\lnsp}{}}+\frac{11 N^5 (\nc{r}{2}{})^2}{144\nc{g}{0\lnsp}{}}+\frac{127 N^4 (\nc{r}{2}{})^2}{432\nc{g}{0\lnsp}{}}+\frac{N^6 (\nc{r}{2}{})^2}{144\nc{g}{0\lnsp}{}}\,,
\end{equation}
which uniquely fixes the coefficients $a_i$ to be
\begin{equation}
\begin{split}
    a_1&=\frac{31}{864(\nc{g}{0\lnsp}{})^3}\,,\qquad a_2=\frac{41}{3456(\nc{g}{0\lnsp}{})^3}\,,\qquad a_3=-\frac{167}{6912(\nc{g}{0\lnsp}{})^3}\,,\\
    a_4&=-\frac{127}{6912(\nc{g}{0\lnsp}{})^3}\,,\qquad a_5=-\frac{11}{2304(\nc{g}{0\lnsp}{})^3}\,,\qquad a_6=-\frac{1}{2304(\nc{g}{0\lnsp}{})^3}\,,\qquad a_7=0\,.
\end{split}
\end{equation}
Examining (\ref{eq:Ricci0}), we see that these are precisely the coefficients of the $(\nc{g}{1\lnsp}{})^2$ terms present in $\nc{R}{0\lnsp}{}$.

\subsection{Next-to-leading order}
To compute the next order term in the Ricci scalar, $\nc{R}{1\lnsp}{}$, we will need to include terms in the metric up through $\text{O}(\lambda^3)$, of which there are 18. As there is only a single scalar one can produce using one $\lambda_{ijkl}$ tensor, namely $\lambda_{iijj}$, each term in $\nc{R}{1\lnsp}{}$ will be this scalar times an $N$-dependent factor. As in $\nc{R}{0\lnsp}{}$ there will be two types of terms: terms containing two derivatives acting on one term from $\nc{G}{3}{}$, and products of derivatives of $\nc{G}{1}{}$ and derivatives of $\nc{G}{2}{}$. The latter terms present a problem computationally, as the product means that evaluating them in terms of explicit tensors involves summing many more terms than before. Happily, we will be able to reconstruct the full form of $\nc{R}{1\lnsp}{}$ using just the simpler terms coming purely from $\nc{G}{3}{}$ and the bootstrap method outlined at the end of the last section. As there are only two general classes of terms, this bootstrap method will still be able to completely fix all of the $\nc{g}{1\lnsp}{} \nc{g}{2\lnsp}{m}$ and $(\nc{g}{1\lnsp}{})^3$ coefficients from the requirement of scheme invariance.

For the terms coming from $\nc{G}{3}{}$, we again rely on \href{https://www.nikhef.nl/~form/}{\texttt{FORM}} to evaluate the tensor contractions, after having produced the derivatives in $\textit{Mathematica}$. After finding the partial result, we add to it a polynomial in $N$, $(\nc{g}{1\lnsp}{})^3$ and $\nc{g}{1\lnsp}{}\nc{g}{2\lnsp}{m}$, and fix the coefficients by demanding scheme invariance. Once again the length of the explicit expression forces us in the interest of space to relegate it to (\ref{eq:Ricci1}) of Appendix \ref{app:explicit}.

Despite using this bootstrapping method to derive the form of $\nc{R}{1\lnsp}{}$ from the $\nc{G}{3}{}$ contributions and scheme invariance, the scheme independence still provides a non-trivial check on the validity of this result. The scheme change of the $\nc{g}{3\lnsp}{m}$ coefficients contains coefficients $\nc{r}{4}{m}$ which do not appear in the transformation of the lower order $\nc{g}{2\lnsp}{m}$ and $\nc{g}{1\lnsp}{}$ coefficients. Consequently, these terms cannot be absorbed by the scheme change of any $(\nc{g}{1\lnsp}{})^3$ and $\nc{g}{1\lnsp}{}\nc{g}{2\lnsp}{m}$ contributions, and thus must cancel on their own. Indeed, the $\nc{g}{3\lnsp}{m}$ terms in (\ref{eq:Ricci1}) have the correct relative coefficients such that all $\nc{r}{4}{m}$ terms cancel. As an explicit example, notice that $\nc{R}{1\lnsp}{}$ does not contain $\nc{g}{3\lnsp}{10}$ or $\nc{g}{3\lnsp}{12}$. This can be explained by scheme invariance, for those coefficients are the only $\text{O}(\lambda^3)$ metric coefficients whose scheme change contains $\nc{r}{4}{9}$ and $\nc{r}{4}{10}$ respectively. Thus, those contributions must vanish.

Once again, we can fix some of the coefficients appearing in $\nc{R}{1\lnsp}{}$ by applying the solution to gradient flow, this time through four loops. One finds that at leading order
\begin{equation}
\begin{split}
\nc{R}{1\lnsp}{}&=-\frac{(N-1)(N+4)}{20736}\Big(6 \nc{a}{5\lnsp}{3} \big(3 \nc{g}{1\lnsp}{} \left(2 N^3+47 N^2+251 N+426\right)-2 \left(2 N^3+94 N^2+504 N+801\big)\right)\\&\qquad\quad-2 (\nc{g}{1\lnsp}{})^3 \big(2 N^3+47 N^2+251N+426\big)+2 (\nc{g}{1\lnsp}{})^2 \big(8 N^3+235 N^2+1239 N+2016\big)\\&\qquad\quad+3 \nc{g}{1\lnsp}{} \big(432 \nc{g}{2\lnsp}{1} (2 N+7)+22 N^3+329 N^2+1713 N+3060\big)\\&\qquad\quad-72 \nc{g}{3\lnsp}{3} \left(2 N^3+47 N^2+251 N+426\right)-1296 \nc{g}{3\lnsp}{1} (2 N+7)\\&\qquad\quad+108 \big(4 N^3-11 N^2-219 N-355+8 \left(N^3+13 N^2+76 N+138\right) \zeta(3)\big)\Big)\lambda_{iijj}+\text{O}(\varepsilon)\,,
\label{eq:Ricci1grad}
\end{split}
\end{equation}
where as with $\nc{R}{0\lnsp}{}$ higher order terms in $\varepsilon$ contain many more coefficients which have yet to be fixed at four loops.

Armed with the expressions for $\nc{R}{0\lnsp}{}$ and $\nc{R}{1\lnsp}{}$, it is now possible to compute the curvature of theory space through $\text{O}(\varepsilon)$ at any fixed point, $R_{\text{CFT}}$. To see that this quantity is capable of distinguishing distinct CFTs, we note that there exists no $N$-independent choice of the unfixed coefficients in (\ref{eq:Ricci1grad}) such that $\nc{R}{1\lnsp}{}=0$, so that as long as $\lambda^1_{iijj}\neq\lambda^2_{iijj}$ for two fixed points, $\lambda^1_{ijkl}$ and $\lambda^2_{ijkl}$, the difference
\begin{equation}
    \Delta R=R_1-R_2=\nc{R}{1\lnsp}{}(\lambda^1_{iijj}-\lambda^2_{iijj})+\text{O}(\lambda^2)
\end{equation}
will be non-zero. The value of $\lambda_{iijj}$ at various fixed points can be found to $\text{O}(\varepsilon)$ in \cite{Osborn:2020cnf}, where it was called $a_0$. As can be seen in the tables of that paper, all inequivalent fixed points differ in their value of $a_0$, so that $\Delta R$ will be non-zero between all multiscalar CFTs. Often of particular interest in the perturbative study of multiscalar CFTs is the $N$-dependence of the behaviour of RG flows between the $O(N)$ and hypercubic fixed points. One finds that the order-$\veps$ difference in curvature between these two CFTs is
\begin{equation}
\begin{split}
    R_{O(N)}-R_{\text{Cubic}}=&-\frac{(N-1)(N+4)(N-4)^2\varepsilon}{62208(N+8)}\bigg(\nc{g}{1\lnsp}{} \big(18\nc{a}{5\lnsp}{3} (2N^3+47N^2+251N+426)\\&\qquad\qquad+1296 \nc{g}{2\lnsp}{1} (2 N+7)+66 N^3+987 N^2+5139 N+9180\big)\\&\qquad\qquad-12\nc{a}{5\lnsp}{3} (2N^3+94N^2+504N+801)\\&\qquad\qquad-2(\nc{g}{1\lnsp}{})^3 (2N^3+47N^2+251N+426)\\&\qquad\qquad+2(\nc{g}{1\lnsp}{})^2 (8N^3+235N^2+1239N+4032)\\&\qquad\qquad-72 \nc{g}{3\lnsp}{3} \left(2 N^3+47N^2+251 N+426\right)-1296 \nc{g}{3\lnsp}{1} (2 N+7)\\&\qquad\qquad+108 \left(4 N^3-11 N^2-219 N-355+8 (N+3) (N^2+10N+46) \zeta(3)\right)\bigg)\,.
\end{split}
\end{equation}
This expression does not appear to change sign at $N=4$, as the value of $A_{O(N)}-A_{\text{Cubic}}$ does at leading order in $\veps$. Therefore, the Ricci scalar does not seem useful in discerning which CFT flows to which.

\section{Connection of potential with \texorpdfstring{$\boldsymbol{\widetilde{F}}$}{F-tilde}}\label{sec:Ftilde}
Having determined the general form of the Ricci scalar, let us briefly detour in an attempt to provide physical motivation for these mathematical constructs. RG monotonicity theorems, like the $F$-theorem in three dimensions and the $a$-theorem in four dimensions, are usually derived within the context of a fixed, integer spacetime dimension. Within the context of dimensional regularisation, on the other hand, perturbative RG flows are only accessible for an unfixed, and non-integer dimension. As it is expected that the fixed points of these non-integer dimensional flows become the usual CFTs in the limit of integer dimension, it is natural to ask whether or not there exists a monotonicity theorem for these flows, interpolating between the theorems in integer dimension. In \cite{Giombi:2014xxa,Fei:2015oha} it was conjectured that a quantity derived from the free energy of the QFT evaluated on a sphere,
\begin{equation}
    \widetilde{F}=\sin\left(\frac{\pi d}{2}\right)\log Z_{S^d}\,,
\end{equation}
is weakly monotonic for all $3\leq d\leq4$, and smoothly interpolates between $F$ for $d=3$ and $a$ for $d=4$. In those papers the monotonicity properties of this function were shown explicitly in a perturbative computation of $Z_{S^d}$ at various fixed points through three loop order. It is natural, then, to ask whether or not this conjecture is reflected in the monotonic function, $A$, which one finds as a result of gradient flow.

The identification of $\widetilde{F}$ with $A$ is to take place only at the fixed points themselves, so that we can interpret $A$ as a function which extends the weak monotonicity properties of $\widetilde{F}$ to strong monotonicity throughout the entire space of theories. This identification is not to come without caveats, however, as we would like a single choice of the function $A$ to capture $\widetilde{F}$ for not just one value of $N$, but all numbers of scalar fields. That is to say, we must choose the unfixed coefficients $\nc{a}{n\llnsp}{m}$ in an $N$-independent manner. In \cite{Fei:2015oha} the value of $\widetilde{F}$ at the $O(N)$ fixed point is computed to be
\begin{equation}
\begin{split}
    \widetilde{F}_{O(N)}-N\widetilde{F}_s=&-\frac{\pi N(N+2)}{576(N+8)^2}\varepsilon^3-\frac{\pi N (N + 2)(13N^2 + 370N + 1588)}{6912(N + 8)^4}\varepsilon^4\\&+\frac{\pi N(N+2)}{414720(N + 8)^6}\bigg(10368(N + 8)(5N + 22)\zeta(3) - 647N^4 - 32152N^3 - 606576N^2 \\&\qquad\qquad\qquad\qquad- 3939520N- 8451008+30\pi^2(N + 8)^4\bigg)\varepsilon^5+\text{O}(\varepsilon^6)\,,
    \label{eq:Ftilde}
\end{split}
\end{equation}
where $\widetilde{F}_s$ is the value of $\widetilde{F}$ for a single free scalar field. To compare this with our solution for $A$, we must first go to the $O(N)$ fixed point,
\begin{equation}
    \lambda_{ijkl}=\lambda(\delta_{ij}\delta_{kl}+\delta_{ik}\delta_{jl}+\delta_{il}\delta_{jk})\,,
    \label{eq:toON1}
\end{equation}
where $\lambda$ can be found to be\cite{Fei:2015oha,Kleinert:2001ax}
\begin{equation}
\begin{split}
    \lambda=&\frac{1}{N+8}\varepsilon+\frac{3(3N+14)}{(N+8)^3}\varepsilon^2\\&-\frac{1}{(N+8)^5}\left(\frac{1}{8} \left(33 N^3-110 N^2-1760 N-4544\right)+12 (N+8) (5 N+22) \zeta(3)\right)\varepsilon^3\\
    &-\frac{1}{3 (N+8)^7}\bigg(\frac{1}{16} \left(5 N^5+2670 N^4+5584 N^3-52784 N^2-309312 N-529792\right)\\&\qquad+\frac{1}{5} \pi ^4 (N+8)^3(5 N+22)-6 (N+8)\left(63 N^3-82 N^2-3796 N-9064\right) \zeta(3)\\&\qquad-120 (N+8)^2\left(2
   N^2+55 N+186\right) \zeta(5)\bigg)\varepsilon^4+\text{O}(\varepsilon^5)\,.
    \label{eq:toON2}
\end{split}
\end{equation}
To match with $\widetilde{F}$ through $\text{O}(\varepsilon^5)$, we will only need to consider terms through $\text{O}(\lambda^5)$ in $A$, which will require only the solution to gradient flow through three loops exhibited in Appendix \ref{app:gradsol}. As the value of $\widetilde{F}$ we are matching with has been calculated specifically in the $\overline{\text{MS}}$ renormalisation scheme, we must use the $\overline{\text{MS}}$ values of the beta function coefficients $\nc{\mathfrak{b}}{n}{m}$ for consistency. Plugging in (\ref{eq:toON1}) and (\ref{eq:toON2}), one finds that the vacuum bubbles in $A$ evaluate to
\begin{equation}
\begin{split}
A_{O(N)}=&-\frac{N (N+2)}{2 (N+8)^2}\varepsilon ^3+\frac{ N (N+2) \left(12 \nc{{a}}{2\llnsp}{}' (N+8)^2-\nc{g}{1\lnsp}{} (N+8)^2-27 (3 N+14)\right)}{12 (N+8)^4}\varepsilon ^4\\&+\frac{1}{160 (N+8)^6}\bigg(N (N+2)
   \big(720 \nc{a}{2\llnsp}{}' (N+8)^2 (3 N+14)-10 \nc{a}{5\lnsp}{3} (N+8)^4-10 \nc{g}{1\lnsp}{} N^4\\&\qquad\qquad-500 \nc{g}{1\lnsp}{} N^3-7560 \nc{g}{1\lnsp}{} N^2-45440 \nc{g}{1\lnsp}{} N-94720 \nc{g}{1\lnsp}{}+27 N^4+762 N^3-5244 N^2\\&\qquad\qquad-72960 N-184512+1152(N+8)(5N+22) \zeta (3)\big)\bigg)\varepsilon ^5+\text{O}(\varepsilon ^6)\,.
   \label{eq:AON}
\end{split}
\end{equation}
In order to match with $(\ref{eq:Ftilde})$, one finds that we must rescale the entire solution for $A$ by a factor of $\pi/288$. This overall rescaling will not alter the solution to (\ref{eq:gradflow}) as long as we introduce the same rescaling of the metric $G_{IJ}$. Once this rescaling has been performed, we find that $A$ matches with $\widetilde{F}$ as long as we fix
\begin{equation}
    \nc{a}{2\llnsp}{}'=-\tfrac{13}{24}+\tfrac{1}{12}\nc{g}{1\lnsp}{}\,,\qquad \nc{a}{5\lnsp}{3}=-\tfrac{1}{9} \left(3 \pi ^2-89\right)-\nc{g}{1\lnsp}{}\,.
    \label{eq:Fmatchingcond}
\end{equation}
Though these conditions have been computed in $\overline{\text{MS}}$, as the value of $\widetilde{F}$ is a physical quantity, we also expect the value of $A$ to be physical, and thus scheme invariant, at fixed points. This may be checked explicitly using the scheme variance of the various coefficients $\nc{a}{n\llnsp}{m}$, $\nc{g}{n\lnsp}{m}$ and $\nc{\mathfrak{b}}{n}{m}$, and one indeed finds that the value of $A$ is scheme-independent at the $O(N)$ fixed point.\footnote{While it may seem like the conditions in (\ref{eq:Fmatchingcond}) are explicitly scheme-variant, as $\nc{g}{1\lnsp}{}$ depends on the scheme while $\nc{{a}}{2\llnsp}{}'$ does not, the constants which appear in fact come from beta function coefficients $\nc{\mathfrak{b}}{n}{m}$, which will cancel the apparent scheme-variance.}

It is not sufficient that $A$ can be chosen to match with $\widetilde{F}$ at only a single, fixed point: one choice of the unfixed coefficients must cause $A$ and $\widetilde{F}$ to match at every fixed point. To show that this is indeed the case, consider the value of $\widetilde{A}$ with coefficients given by (\ref{eq:Fmatchingcond}) at the cubic fixed point, where the interaction tensor is given by
\begin{equation}
    \lambda_{ijkl}=\lambda(\delta_{ij}\delta_{kl}+\delta_{ik}\delta_{jl}+\delta_{il}\delta_{jk})+g\lsp\delta_{ijkl}\,.
\end{equation}
The values of $\lambda$ and $g$ can be computed to be\cite{Kleinert:2001ax}
\begin{equation}
\begin{split}
    \lambda=&\frac{1}{3N}\varepsilon-\frac{19N^2-125N+106}{81N^3}\varepsilon^2\\&-\frac{1}{17496N^5}\big(1955N^4+41971N^3-229974N^2+360640N-179776\\&-2592N(2N^3-6N^2-7N+14)\zeta_3\big)\varepsilon^3\,,
\end{split}
\end{equation}
\begin{equation}
\begin{split}
    g=&\frac{N-4}{3N}\varepsilon+\frac{(N-1)(17N^2+110N-424)}{81N^3}\varepsilon^2\\&+\frac{1}{17496N^5}\big(709N^5+11713N^4+180562N^3-989656N^2+1500224N-719104\\& -2592N(N+2)(N^3+6N^2-32N+28)\zeta_3\big)\varepsilon^3\,.
\end{split}
\end{equation}
Plugging these expressions into the expansion for $A$, and rescaling by a factor of $\pi/288$, one finds that its value at the cubic fixed point is
\begin{equation}
\begin{split}
    \frac{\pi}{288}A_{\text{Cubic}}=&-\frac{\pi(N^2+N-2)}{15552N}\varepsilon^3-\frac{\pi(N-1)(73N^3+36N^2+432N-424)}{559872N^3}\varepsilon^4\\&-\frac{\pi}{302330880N^5}\big(53621N^6-14259N^5-224610N^4+2161280N^3\\&\qquad\qquad-5542944N^2+5364672N-1797760-810\pi^2N^4(N^2+N-2)\\&\qquad\qquad-10368N(N-1)(N^4-4N^3+16N^2+6N-28)\zeta(3)\big)\varepsilon^5+\text{O}(\varepsilon^6)\,,
\end{split}
\end{equation}
which matches the value of $\widetilde{F}_{\text{Cubic}}-N\widetilde{F}_s$ given in \cite{Fei:2015oha}. The value of $\widetilde{F}$ at a generic fixed point will only differ from $\widetilde{F}_{O(N)}$ or $\widetilde{F}_{\text{Cubic}}$ in that different expressions are used for the $\lambda_{ijkl}$ which appear multiplying the vacuum diagrams. As the value of $A$ will differ between different fixed points in exactly the same way, fixing the coefficients of the tensor structures in $A$ such that $A_{O(N)}\propto \widetilde{F}_{O(N)}$ must fix $A\propto \widetilde{F}$ at all fixed points. Thus, we can safely claim that the choice (\ref{eq:Fmatchingcond}) guarantees that
\begin{equation}
    \widetilde{F}-N\widetilde{F}_s=\frac{\pi}{288}A\,,
\end{equation}
at fixed points, so that $A$ provides a natural, RG-monotonic extension of $\widetilde{F}$. It thus appears that our $A$ turns the conjectured weak $\widetilde{F}$-theorem into a theorem which interpolates between gradient flow in $d=4$ and $d=3$.

Much as we were able to use scheme-invariance to bootstrap the expression for $R$, it is possible to partially derive the $N$-dependence of $A$ using the condition that any choice of the free parameters must be $N$-independent. Examining the $\text{O}(\varepsilon^4)$ term in (\ref{eq:AON}), one immediately sees that any choice we make of the unfixed coefficients $\nc{{a}}{2\llnsp}{}'$ and $\nc{g}{1\lnsp}{}$ will produce a term which scales with $N$ like $N(N+2)/(N+8)^2$. The $N$-dependence of $\widetilde{F}_{O(N)}$ thus must go like
\begin{equation}
    \widetilde{F}_{O(N)}\supset \frac{ N (N+2) \left(a(N+8)^2-27 (3 N+14)\right)\pi}{3456 (N+8)^4}\varepsilon ^4\,,
\end{equation}
for some $N$-independent value of $a$, and it is the matching with explicit computations that forces the choice $a=-13/2$ giving the first equation in (\ref{eq:Fmatchingcond}). A similar analysis will work for the $\varepsilon^6$ piece, which will involve coefficients $\nc{a}{6}{m}$ which appear first in the gradient flow equations at four-loops, so that
\begin{alignat}{2}
    \widetilde{F}_{O(N)}\supset
    \bigg(& a_1\frac{\pi N(N+2) (N^3+38N^2+244N+608) }{23040 (N+8)^5}\nonumber\\&-a_2\frac{\pi N(N+2) (3N^3+284N^2+2232N+5824)}{15360 (N+8)^5}\nonumber \\&-a_3\frac{\pi N (N+2)^2}{1536 (N+8)^4}+a_4\frac{\pi N(N+2)(N^3+68N^2+1224N+3648)}{1920 (N+8)^5}\nonumber \\&+\frac{\pi ^3 N (N+2) \big(5 (N^3-112N^2+1424N+5248)+72 \pi ^2 (5 N+22)\big)}{345600 (N+8)^5}\nonumber \\&-\frac{5 \pi  N (N+2) \left(2 N^2+55 N+186\right) \zeta(5)}{72 (N+8)^6}\nonumber \\&+\frac{\pi  N (N+2) \left(9 N^5+676 N^4+22584
   N^3+416576 N^2+2610496 N+4867328\right) \zeta(3)}{23040 (N+8)^7}\nonumber \\&+\frac{\splitfrac{\pi  N (139 N^7+265876 N^6+5902788
   N^5+31827232 N^4-108375758438400
   N^3}{-1443607089315840 N^2-4275842879324160
   N-3924070713262080)}}{829440 (N+8)^8}\bigg)\varepsilon^6\,,
\end{alignat}
for some $N$-independent choice of the coefficients $a_1$, $a_2$, $a_3$ and $a_4$.

\section{Connection of metric with \texorpdfstring{$\boldsymbol{\langle\phi^4_I\phi^4_J\rangle}$}{<ϕ4 ϕ4>}}\label{sec:metric}
Having found a physical reflection of $A$ in $\widetilde{F}$, one can ask what physical interpretation remains for the metric. As any physical counterpart must have the correct number of indices, one is limited in what one can conjecture, and it is natural to, following Zamolodchikov, ask whether or not the metric can be identified with $\langle\phi^4_I(x)\phi^4_J(y)\rangle$, the two-point function of perturbing operators. As with $\widetilde{F}$, this identification would occur only at the fixed points themselves, with $G_{IJ}$ serving as a sort of continuation of $\langle\phi^4_I(x)\phi^4_J(y)\rangle$ onto the whole space of couplings. Specifically, at a given fixed point this two-point function will take the form
\begin{equation}
    \langle\phi^4_I(x)\phi^4_J(y)\rangle=\frac{C_{IJ}}{|x-y|^{2\Delta_I}}\,,
    \label{eq:twoptgen}
\end{equation}
where $\Delta_I$ is the dimension of the quartic operator $\phi_I^4$. As we have already identified $A$ with $\widetilde{F}$, which is computed on $S^d$ rather than $\mathbb{R}^d$, we will also perform the computation of the two-point function on the sphere, so that the length $|x-y|$ is replaced by $s(\hat{x},\hat{y})$, the chordal distance. Here we have defined dimensionless coordinates $\hat{x}=x/R$.

In order to fix the first two coefficients in the metric, $G_{IJ}$, we wish to consider all contributions to the two point function $\langle\phi^4_I(\hat{x})\phi^4_J(\hat{y})\rangle$ up to $\text{O}(\lambda)$. The scalar propagator used in this computation will be
\begin{equation}
    \begin{tikzpicture}[scale=0.5,baseline=(vert_cent.base)]
        \node (vert_cent) at (0,0) {$\phantom{\cdot}$};
        \node[inner sep=0pt] (l) at (-1.5,0) {};
        \node[inner sep=0pt] (r) at (1.5,0) {};
        \node[xshift=-8.5pt] at (-1.25,0) {$x$};
        \node[xshift=8.5pt] at (1.25,-0.1) {$y$};
        \draw (l) to (r);
    \end{tikzpicture}=\frac{C_\phi}{(s(\hat{x},\hat{y}))^{d-2}}\,,\qquad C_\phi=\frac{\Gamma\left(\frac{d-2}{2}\right)}{4\pi^{d/2}}\,,
\end{equation}
where we work in stereographic coordinates on the sphere. For simplicity, we choose the second insertion to lie at the north pole, which in stereographic coordinates corresponds to taking the limit $\hat{y}\rightarrow \infty$. There is a single $\text{O}(\lambda^0)$ diagram given by
\begin{equation}
    \begin{tikzpicture}[scale=0.5,baseline=(vert_cent.base)]
        \node (vert_cent) at (0,0) {$\phantom{\cdot}$};
        \draw (0,0) circle (1.25cm);
        \node[draw,circle,thick,black,fill=white,inner sep=2pt] (l) at (-1,0) {$I$};
        \node[draw,circle,thick,black,fill=white,inner sep=2pt] (r) at (1,0) {$J$};
        \draw (l) to[out=300, in=240] (r);
        \draw (l) to[out=60, in=120] (r);
    \end{tikzpicture}=\mu^{4\varepsilon}\frac{24}{(s(\hat{x},\infty))^{4d-8}}\delta_{IJ}\,,
\end{equation}
where the factor of $24$ in the numerator arises from the $4!$ ways of Wick contracting the individual fields. At the next order there is again a single diagram,
\begin{equation}
    \begin{tikzpicture}[scale=0.5,baseline=(vert_cent.base)]
        \node (vert_cent) at (0,0) {$\phantom{\cdot}$};
        \draw (0,0) circle (1.25cm);
        \node[draw,circle,thick,black,fill=white,inner sep=2pt] (l) at (-1,0) {$I$};
        \node[draw,circle,thick,black,fill=white,inner sep=2pt] (r) at (1,0) {$J$};
        \node[inner sep=0pt] (t) at (0,1.25) {};
        \filldraw (t) circle[radius=2pt];
        \draw (l) to[out=300, in=240] (r);
        \draw (l) to[out=30, in=270] (t);
        \draw (r) to[out=150, in=270] (t);
    \end{tikzpicture}=-\mu^{5\varepsilon}(\lambda_{ijmn}\delta_{ko}\delta_{lp}+\text{perms.})\int d^{\lsp d}\hat{y}\sqrt{g(\hat{y})}\,\frac{C_\phi^6}{(s(\hat{x},\infty)s(\hat{x},\hat{y})s(\hat{y},\infty))^{2d-4}}\,,
    \label{eq:lambdacontribution}
\end{equation}
where the sum is over the 72 distinct structures arising from permutations of $i,j,k,l$ and of $m,n,o,p$. We can evaluate the integral in (\ref{eq:lambdacontribution}) using two distinct methods. First, following \cite{Drummond:1975yc}, it is possible to write the chordal distance using spherical harmonics generalised to the $n$-sphere as
\begin{equation}
    s(\hat{x},\hat{y})^{2\Delta}=\sum_{\ell,m}\frac{(2R)^{2\Delta+d}\pi^\frac{d}{2}\Gamma\left(\Delta+\frac{d}{2}\right)\Gamma\left(\ell-\Delta\right)}{\Gamma\left(\ell+d+\Delta\right)\Gamma\left(-\Delta\right)}Y^*_{\ell,m}(\hat{x})Y_{\ell,m}(\hat{y})\,,
\end{equation}
where $R$ is the radius of the sphere (not to be confused with the Ricci scalar above), with the spherical harmonics obeying the usual orthonormality relations
\begin{equation}
    \int d^{\lsp d}\hat{x}\sqrt{g(\hat{x})}\, Y^*_{\ell,m}(\hat{x})Y_{\ell',m'}(\hat{x})=\delta_{\ell,\ell'}\delta_{m,m'}\,.
\end{equation}
One then finds that (\ref{eq:lambdacontribution}) becomes
\begin{equation}
    -(\lambda_{ijmn}\delta_{ko}\delta_{lp}+\text{perms.})\frac{\mu^{5\varepsilon}C_\phi^6}{(s(x,\infty))^{2d-4}}\sum_{\ell,m}\left(\frac{(2R)^{4-d}\pi^\frac{d}{2}\Gamma\left(2-\frac{d}{2}\right)\Gamma\left(\ell+d-2\right)}{\Gamma\left(\ell+2\right)\Gamma\left(d-2\right)}\right)^2Y^*_{\ell,m}(\hat{x})Y_{\ell,m}(\infty)\,.
    \label{eq:partialworkharmonics}
\end{equation}
Using the fact that $d=4-\varepsilon$, one finds that up to $\text{O}(\varepsilon^2)$ terms
\begin{equation}
    \begin{aligned}
        \left(\frac{(2R)^{4-d}\pi^\frac{d}{2}\Gamma\left(2-\frac{d}{2}\right)\Gamma\left(\ell+d-2\right)}{\Gamma\left(\ell+2\right)\Gamma\left(d-2\right)}\right)^2\to\left(\frac{\pi^\frac{d}{2}\Gamma\left(2-\frac{d}{2}\right)^2\Gamma\left(4-\frac{d}{2}\right)}{\Gamma(4-d)}\right)\\
        &\hspace{-2.3cm}\times\left(\frac{(2R)^{8-2d}\pi^\frac{d}{2}\Gamma(4-d)\Gamma\left(\frac{3d}{2}-4+\ell\right)}{\Gamma\left(\frac{3d}{2}-4\right)\Gamma\left(4-\frac{d}{2}+\ell\right)}\right),
    \end{aligned}
\end{equation}
so that in the computation of (\ref{eq:lambdacontribution}) up to $\text{O}(\varepsilon)$ we can safely replace the sum in (\ref{eq:partialworkharmonics}) by $(s(x,\infty))^{8-3d}$. We thus find the final expression
\begin{equation}
\begin{split}
    \begin{tikzpicture}[scale=0.5,baseline=(vert_cent.base)]
        \node (vert_cent) at (0,0) {$\phantom{\cdot}$};
        \draw (0,0) circle (1.25cm);
        \node[draw,circle,thick,black,fill=white,inner sep=2pt] (l) at (-1,0) {$I$};
        \node[draw,circle,thick,black,fill=white,inner sep=2pt] (r) at (1,0) {$J$};
        \node[inner sep=0pt] (t) at (0,1.25) {};
        \filldraw (t) circle[radius=2pt];
        \draw (l) to[out=300, in=240] (r);
        \draw (l) to[out=30, in=270] (t);
        \draw (r) to[out=150, in=270] (t);
    \end{tikzpicture}=&-(\lambda_{ijmn}\delta_{ko}\delta_{lp}+\text{perms.})\frac{\mu^{5\varepsilon}C_\phi^6}{(s(\hat{x},\infty))^{5d-12}}\frac{\pi^\frac{d}{2}\Gamma\left(2-\frac{d}{2}\right)^2\Gamma\left(4-\frac{d}{2}\right)}{\Gamma(4-d)}+O(\varepsilon^2)\\
    =&\left(-\frac{1}{64\pi^8\varepsilon}-\frac{1+5\gamma_E+5\log\pi+10\log(s(\hat{x},\infty)\mu)}{128\pi^8}\right)\frac{(\lambda_{ijmn}\delta_{ko}\delta_{lp}+\text{perms.})}{16\pi^2\,s(\hat{x},\infty)^8}+\text{O}(\varepsilon)\,.
\end{split}
\label{eq:lambdacontributionfinal}
\end{equation}

We can also try to compute the integral in (\ref{eq:lambdacontribution}) by using the following explicit expressions\cite{Fei:2015oha} for the chordal distance and for the measure $d^{\lsp d}\hat{y}\sqrt{g(\hat{y})}$:
\begin{equation}
    s(\hat{x},\hat{y})=\frac{2R|\hat{x}-\hat{y}|}{\sqrt{(1+\hat{x}^2)(1+\hat{y}^2)}}\,,\qquad\qquad d^{\lsp d}\hat{y}\sqrt{g(\hat{y})}=\frac{(2R)^d\lsp d^d\hat{y}}{(1+\hat{y}^2)^d}\,.
\end{equation}
The integral in (\ref{eq:lambdacontribution}) then becomes
\begin{equation}
    \int d^{\lsp d}\hat{y}\, (2R)^{12-5d}(1+\hat{x}^2)^{2d-4}\left((\hat{x}-\hat{y})^2\right)^{2-d}(1+\hat{y}^2)^{d-4}\,,
\end{equation}
which is a particular instance of a wider class of integral,
\begin{equation}
    I(\Delta)=\int d^{\lsp d}\hat{y}\, (2R)^{d-3\Delta}(1+\hat{x}^2)^{\Delta}\left((\hat{x}-\hat{y})^2\right)^{-\frac{\Delta}{2}}(1+\hat{y}^2)^{\Delta-d}\,.
\end{equation}
Introducing Feynman parameters, one finds after some algebra that
\begin{equation}
    I(\Delta)=(2R)^{d-3\Delta}(1+\hat{x}^2)^{\Delta}\frac{\pi^\frac{d}{2}\Gamma\left(\frac{d-\Delta}{2}\right)}{\Gamma\left(\frac{\Delta}{2}\right)\Gamma\left(d-\Delta\right)}\int_0^1 d\alpha\, \alpha^{\frac{\Delta}{2}-1}(1-\alpha)^{\frac{d-\Delta}{2}-1}(1+\hat{x}^2\alpha)^{\frac{\Delta-d}{2}}\,.
\end{equation}
As $x^2\geq0$ one can recognise this as an integral expression for the hypergeometric function ${_2F_1}$, with the final result that
\begin{equation}
    I(\Delta)=(2R)^{d-3\Delta}(1+\hat{x}^2)^{\Delta}\frac{\pi^\frac{d}{2}\Gamma\left(\frac{d-\Delta}{2}\right)^2}{\Gamma\left(\frac{d}{2}\right)\Gamma\left(d-\Delta\right)}{_2F_1}\left(\frac{d-\Delta}{2},\frac{\Delta}{2};\frac{d}{2},-\hat{x}^2\right).
\end{equation}
Setting $\Delta=2d-4$ we thus find the general expression for the $\text{O}(\lambda)$ contribution
\begin{equation}
    \begin{tikzpicture}[scale=0.5,baseline=(vert_cent.base)]
        \node (vert_cent) at (0,0) {$\phantom{\cdot}$};
        \draw (0,0) circle (1.25cm);
        \node[draw,circle,thick,black,fill=white,inner sep=2pt] (l) at (-1,0) {$I$};
        \node[draw,circle,thick,black,fill=white,inner sep=2pt] (r) at (1,0) {$J$};
        \node[inner sep=0pt] (t) at (0,1.25) {};
        \filldraw (t) circle[radius=2pt];
        \draw (l) to[out=300, in=240] (r);
        \draw (l) to[out=30, in=270] (t);
        \draw (r) to[out=150, in=270] (t);
    \end{tikzpicture}=-(\lambda_{ijmn}\delta_{ko}\delta_{lp}+\text{perms.})\frac{\mu^{5\varepsilon}C_\phi^6(2R)^{4-d}}{(s(\hat{x},\infty))^{4d-8}}\frac{\pi^\frac{d}{2}\Gamma\left(2-\frac{d}{2}\right)^2}{\Gamma\left(\frac{d}{2}\right)\Gamma\left(4-d\right)}{_2F_1}\left(2-\frac{d}{2},d-2;\frac{d}{2},-\hat{x}^2\right).
\end{equation}
When expanding this expression we use the following identities for the derivatives of ${_2F_1}$:
\begin{equation}
    {_2F_1}^{(0,1,0,0)}\left(0,2;2,-\hat{x}^2\right)= {_2F_1}^{(0,0,1,0)}\left(0,2;2,-\hat{x}^2\right)=0\,,\qquad {_2F_1}^{(1,0,0,0)}\left(0,2;2,-\hat{x}^2\right)=-\log(1+\hat{x}^2)\,,
\end{equation}
and find the same expression as in (\ref{eq:lambdacontributionfinal}).

The divergence in the $\text{O}(\lambda)$ diagram must be accounted for by renormalising the quartic operator $\phi^4_I$. We introduce the two counterterm graphs
\begin{equation}
    \begin{tikzpicture}[scale=0.5,baseline=(vert_cent.base)]
        \node (vert_cent) at (0,0) {$\phantom{\cdot}$};
        \draw (0,0) circle (1.25cm);
        \node[draw,circle,thick,black,cross,fill=white,inner sep=2pt] (l) at (-1,0) {$\phantom{\cdot}$};
        \node[xshift=-10pt] at (-1,0) {$I$};
        \node[draw,circle,thick,black,fill=white,inner sep=2pt] (r) at (1,0) {$J$};
        \draw (l) to[out=300, in=240] (r);
        \draw (l) to[out=60, in=120] (r);
    \end{tikzpicture}+\begin{tikzpicture}[scale=0.5,baseline=(vert_cent.base)]
        \node (vert_cent) at (0,0) {$\phantom{\cdot}$};
        \draw (0,0) circle (1.25cm);
        \node[draw,circle,thick,black,fill=white,inner sep=2pt] (l) at (-1,0) {$I$};
        \node[draw,circle,thick,black,cross,fill=white,inner sep=2pt] (r) at (1,0) {$\phantom{\cdot}$};
        \node[xshift=10pt] at (1,0) {$J$};
        \draw (l) to[out=300, in=240] (r);
        \draw (l) to[out=60, in=120] (r);
    \end{tikzpicture}=-\frac{48\lsp C_\phi^4}{(s(\hat{x},\infty))^{4d-8}}\frac{\delta Z^{\phi^4}_{IJ}}{\varepsilon}\,.
\end{equation}
Summing with the expression in (\ref{eq:lambdacontributionfinal}), one finds that the counterterm must be chosen to be
\begin{equation}
    \delta Z^{\phi^4}_{ijkl,mnop}=-\tfrac{1}{12}(\lambda_{ijmn}\delta_{ko}\delta_{lp}+\text{perms.})\,.
    \label{eq:counterterm}
\end{equation}
It is possible to remove factors of $\gamma_E+\log4\pi$ in these expressions by working in the $\overline{\text{MS}}$ scheme rather than the MS scheme, in which we shift $\mu^2\rightarrow\mu^2 e^{-\gamma_E}/4\pi$. As $\widetilde{F}$ was computed in the $\overline{\text{MS}}$ scheme in \cite{Fei:2015oha}, we must also use this scheme if we are to consistently match $A$ with $\widetilde{F}$ and $G_{IJ}$ with the two-point function simultaneously.

At a fixed point we may write the final two-point function in the form
\begin{equation}
    \langle[O]^\alpha(\hat{x})[O]^\beta(\infty)\rangle=\frac{C_0^\alpha+\varepsilon\lsp C_1^\alpha}{\big(\frac12 s(\hat{x},\infty)\big)^{4d-8+2\varepsilon\gamma_\alpha}}\delta_{\alpha\beta}\,,
    \label{eq:Twoptfunct}
\end{equation}
where the $[O]^\alpha(\hat{x})$ are renormalised quartic operators. The factor of $1/2$ is an overall normalisation to simplify the resulting form of $C_0^\alpha$ and $C_1^\alpha$. These operators can be related to the $\phi^4_I$ operators we began with via
\begin{equation}
    [O]^\alpha(\hat{x})=n^\alpha_I\phi^4_I(\hat{x})\,,
\end{equation}
where the tensor structure $n^\alpha_I$ is an eigenvector of the mixing matrix
\begin{equation}
    S_{IJ}n^\alpha_J=-\varepsilon(1-\gamma^\alpha) n^\alpha_I\,.
\end{equation}
Here, $C_0^\alpha$ and $C_1^\alpha$ may be determined from the $C_{IJ}$ of (\ref{eq:twoptgen}) by contracting with two factors of $n^\alpha_I$. The mixing matrix can be determined equivalently via the expression $S_{IJ}=\partial_I\beta_J$ or from the counterterm (\ref{eq:counterterm}). Both methods produce the one loop result
\begin{equation}
    S_{ijkl,mnop}=-\varepsilon\lsp\delta_{ijkl,mnop}+\tfrac{1}{12}(\lambda_{ijmn}\delta_{ko}\delta_{lp}+\text{perms.})\,,
\end{equation}
where $\delta_{ijkl,mnop}$ represents a symmetrised product of four Kronecker deltas. The two-point functions of renormalised quartic operators can then be written down as
\begin{equation}
\begin{split}
    \langle[O]^\alpha(\hat{x})[O]^\beta(\infty)\rangle=& \langle \phi^4_I(\hat{x})\phi^4_J(\infty)\rangle n^\alpha_I\lsp n^\beta_J\\
    =&\left(\begin{tikzpicture}[scale=0.5,baseline=(vert_cent.base)]
        \node (vert_cent) at (0,0) {$\phantom{\cdot}$};
        \draw (0,0) circle (1.25cm);
        \node[draw,circle,thick,black,fill=white,inner sep=2pt] (l) at (-1,0) {$I$};
        \node[draw,circle,thick,black,fill=white,inner sep=2pt] (r) at (1,0) {$J$};
        \draw (l) to[out=300, in=240] (r);
        \draw (l) to[out=60, in=120] (r);
    \end{tikzpicture}+\begin{tikzpicture}[scale=0.5,baseline=(vert_cent.base)]
        \node (vert_cent) at (0,0) {$\phantom{\cdot}$};
        \draw (0,0) circle (1.25cm);
        \node[draw,circle,thick,black,fill=white,inner sep=2pt] (l) at (-1,0) {$I$};
        \node[draw,circle,thick,black,fill=white,inner sep=2pt] (r) at (1,0) {$J$};
        \node[inner sep=0pt] (t) at (0,1.25) {};
        \filldraw (t) circle[radius=2pt];
        \draw (l) to[out=300, in=240] (r);
        \draw (l) to[out=30, in=270] (t);
        \draw (r) to[out=150, in=270] (t);
    \end{tikzpicture}+\begin{tikzpicture}[scale=0.5,baseline=(vert_cent.base)]
        \node (vert_cent) at (0,0) {$\phantom{\cdot}$};
        \draw (0,0) circle (1.25cm);
        \node[draw,circle,thick,black,cross,fill=white,inner sep=2pt] (l) at (-1,0) {$\phantom{\cdot}$};
        \node[xshift=-10pt] at (-1,0) {$I$};
        \node[draw,circle,thick,black,fill=white,inner sep=2pt] (r) at (1,0) {$J$};
        \draw (l) to[out=300, in=240] (r);
        \draw (l) to[out=60, in=120] (r);
    \end{tikzpicture}+\begin{tikzpicture}[scale=0.5,baseline=(vert_cent.base)]
        \node (vert_cent) at (0,0) {$\phantom{\cdot}$};
        \draw (0,0) circle (1.25cm);
        \node[draw,circle,thick,black,fill=white,inner sep=2pt] (l) at (-1,0) {$I$};
        \node[draw,circle,thick,black,cross,fill=white,inner sep=2pt] (r) at (1,0) {$\phantom{\cdot}$};
        \node[xshift=10pt] at (1,0) {$J$};
        \draw (l) to[out=300, in=240] (r);
        \draw (l) to[out=60, in=120] (r);
    \end{tikzpicture}\right) n^\alpha_I n^\beta_J+\text{O}(\lambda^2)\,.
\end{split}
\end{equation}
Using the definition of $\gamma^\alpha$, and the form of the diagrams given above, we find that $C_0$ and $C_1$ are given by
\begin{equation}
    C_0^\alpha=\frac{1}{(4\pi)^8}\,,\qquad C_1^\alpha=-\frac{\gamma^\alpha}{(4\pi)^8}\,.
\end{equation}

To match the two-point function with the metric $G_{IJ}$ at the fixed point, we must use the $n^\alpha_I$ to change coordinates from the original $\phi^4_I$ operators to the renormalised $[O]^\alpha$ operators, so that the metric we are interested in becomes
\begin{equation}
    G_{\alpha\beta}=n^\alpha_I\lsp G_{IJ}\lsp n^\beta_J\,.
\end{equation}
To $\text{O}(\lambda)$ the expansion for the metric will include only two terms
\begin{equation}
    G_{IJ}=\nc{g}{0\lnsp}{}\begin{tikzpicture}[xscale=0.6,yscale=0.5,
    vertex/.style={draw,circle,fill=black,minimum size=2pt,inner sep=0pt},
    arc/.style={},baseline=(vert_cent.base)]
    \node (vert_cent) at (0,-1.25) {$\phantom{\cdot}$};
    \foreach [count=\i] \coord in {
(1.00,0), (-1,0),(-1,-0.75),(1,-0.75)}{
        \node[] (p\i) at \coord {};
    }
    \foreach [count=\i] \coord in {
(-1,-1.5), (1,-1.5),(-1,-2.25),(1,-2.25)}{
        \node[] (d\i) at \coord {};
    }
    \draw (p1) edge (p2);
    \draw (p4) edge (p3);
    \draw (d1) edge (d2);
    \draw (d3) edge (d4);
\end{tikzpicture}
+\nc{g}{1\lnsp}{}\begin{tikzpicture}[xscale=0.6,yscale=0.5,
    vertex/.style={draw,circle,fill=black,minimum size=2pt,inner sep=0pt},
    arc/.style={},baseline=(vert_cent.base)]
    \node (vert_cent) at (0,-1.25) {$\phantom{\cdot}$};
    \node[vertex] (c) at (0,-0.375) {};
    \foreach [count=\i] \coord in {
(1.00,0), (-1,0),(-1,-0.75),(1,-0.75)}{
        \node[] (p\i) at \coord {};
    }
    \foreach [count=\i] \coord in {
(-1,-1.5), (1,-1.5),(-1,-2.25),(1,-2.25)}{
        \node[] (d\i) at \coord {};
    }
    \draw (c) edge (p1)
                   edge (p2)
                   edge (p3)
                   edge (p4);
    \draw (d1) edge (d2);
    \draw (d3) edge (d4);
\end{tikzpicture}\,,
\end{equation}
so that
\begin{equation}
    G_{\alpha \beta}=\big(\nc{g}{0\lnsp}{}+\tfrac16\varepsilon\gamma^\alpha\nc{g}{1\lnsp}{}\big)\delta_{\alpha \beta}\,.
\end{equation}
Demanding that the gradient flow equations are satisfied and matching with $\widetilde{F}$ requires that $\nc{g}{0\lnsp}{}=1+(\frac{13}{12}-\frac{1}{6}\nc{g}{1\lnsp}{})\veps$, so that
\begin{equation}
    G_{\alpha\beta}=\delta_{\alpha \beta}+\varepsilon\left(\frac{13-2\nc{g}{1\lnsp}{}}{12}+\frac16\gamma^\alpha\nc{g}{1\lnsp}{}\right)\delta_{\alpha\beta}\,.
\end{equation}
It is impossible to choose $\nc{g}{1\lnsp}{}$ such that $G_{\alpha\beta}=2^{2\Delta_\alpha}(C^\alpha_0+\varepsilon\lsp C^\alpha_1)\delta_{\alpha\beta}$ for all $\alpha$. However, if one relaxes the assumption of direct equality one finds that the choice
\begin{equation}
    \nc{g}{1\lnsp}{}=-6\,,
\end{equation}
solves the expression
\begin{equation}
    G_{\alpha\beta}=2^{2\Delta_\alpha}\left(1+\tfrac{25}{12}\varepsilon\right)(C^\alpha_0+\varepsilon\lsp C^\alpha_1)\delta_{\alpha\beta}\,.
\end{equation}
One can interpret this as demanding that the nearly-marginal operators appearing in the two-point functions are rescaled versions of the operators in (\ref{eq:Twoptfunct}),
\begin{equation}
    [O']^\alpha(\hat{x})=2^{\Delta_\alpha}\sqrt{1+\tfrac{25}{12}\varepsilon}\,[O]^\alpha(\hat{x})\,.
\end{equation}
Note that the term under the square root is positive for all meaningful values of $\varepsilon$. That we are forced to rescale the operators is a consequence of simultaneously demanding that the identification of $G_{\alpha\beta}$ with $C_{\alpha\beta}$ holds in the free theory while including an $\text{O}(\varepsilon)$ part in the coefficient of $\delta_{IJ}$. In order to simultaneously match both the Zamalodchikov norm and $\widetilde{F}$, recall that we must rescale both $A$ and $G_{IJ}$ by an overall factor of $\pi/288$, so that the real operators that we wish to work with are
\begin{equation}
    [O']^\alpha(\hat{x})=2^{\Delta_\alpha}\sqrt{\tfrac{\pi}{288}(1+\tfrac{25}{12}\varepsilon)}\,[O]^\alpha(\hat{x})\,.
\end{equation}
With this choice of quartic operators, we thus find that there exists a generalisation of $\widetilde{F}$ such that the expression
\begin{equation}
    \partial_I\widetilde{F}=C_{IJ}\beta^J
\end{equation}
holds even outside of fixed points, at least to leading order in $\varepsilon$.

The identification of $A$ with $\widetilde{F}$ and $G_{IJ}$ with $C_{IJ}$ has forced us to fix a number of the arbitrary coefficients appearing in the gradient flow solution. Consequently, the constant piece of the Ricci scalar can be evaluated for this choice of $A$ and metric to be
\begin{equation}
    \begin{split}
        \nc{R}{0\lnsp}{}=\frac{(N-1) N (N+2) (N+4)}{72\pi}\bigg(&648 \lsp\nc{g}{2\lnsp}{1} (N+3)+30N^3+231N^2+1098N\\&+2629-6 \pi ^2 (3 N (N+6)+31)\bigg)+\text{O}(\varepsilon)\,,
    \end{split}
\end{equation}
where the $\text{O}(\varepsilon)$ terms include a number of unfixed coefficients $\nc{g}{3\lnsp}{n}$ of the $\text{O}(\lambda^3)$ contributions to the metric and where we have included the overall rescaling by $\pi/288$. The coefficient $\nc{g}{2\lnsp}{1}$ which is left unfixed here may be fixed by demanding that $G_{IJ}$ matches $C_{IJ}$ to $\text{O}(\varepsilon^2)$. However, the $\text{O}(\varepsilon N^7)$ term is now entirely fixed to be
\begin{equation}
   \nc{R}{0\lnsp}{}=\tfrac{1}{12\pi}\big(5+\tfrac{31}{6}\lsp\varepsilon\big)N^7+\text{O}(N^6,\,\varepsilon^2 N^7)\,.
\end{equation}
It is interesting to note that the $\text{O}(\varepsilon)$ correction is always positive, so that theory space remains compact at large $N$.

\section{Conclusion}\label{sec:conclusion}
Recent work by the authors \cite{Pannell:2024sia} and also earlier work \cite{Wallace:1974dx, Wallace:1974dy, Jack:2018oec} have provided extensive evidence in favour of a gradient flow interpretation of RG flow, at least in a wide class of examples built with scalar fields in $d=4$ and $d=4-\varepsilon$ dimensions. Motivated by this observation, which hinges on highly non-trivial constraints satisfied by perturbative beta function coefficients, we have sought here a more physical recasting of the relevant potential and metric that feature in the gradient equation. To that end, we have managed to identify the potential with a certain quantity $\widetilde{F}$ that at fixed points is derived from the sphere free energy, and the metric with a quantity that at fixed points becomes the Zamolodchikov metric of two-point functions of classically-marginal operators on the sphere.

Although the potential and metric are renormalisation scheme-dependent, the gradient equation is scheme-independent and thus conveys physical information. In fact, the potential and metric are fully determined at fixed points by requiring their identification with physical quantities of choice, but in general they retain some arbitrariness even at fixed points so as to be able to be matched to potentially different physical quantities that characterise fixed points.

In an attempt to provide a novel set of such quantities, we have computed here the first two orders of the Ricci scalar associated with the gradient metric. This has shown that the space of quantum field theories with $N$ scalar fields is curved and in particular compact at large $N$. At leading order the Ricci scalar does not depend on the couplings and is simply $N$-dependent, while at second order it is proportional to the invariant $\lambda_{iijj}$, which was called $a_0$ in \cite{Osborn:2020cnf, Hogervorst:2020gtc} and was found there to be bounded above by $a_0\leq N(N+2)/(N+8)$ at fixed points. The $a_0$ bound is the value of $a_0$ attained at the $O(N)$ fixed point, so in some sense the $O(N)$ model sets a limit on the curvature of the space of theories at fixed points permitted for given $N$.

The scope of this paper is limited only to the study of gradient flow in purely scalar systems, and thus covers only a small corner of the full space of all perturbative QFTs. Gradient flow has been studied previously in a number of different sectors of this diverse space, including for more general gauge-fermion-scalar systems\cite{Poole:2019kcm,Davies:2021mnc}, and it should be possible to extend the ideas presented in this paper to define $R$ throughout all of theory space. Another region of theory space of interest in the modern study of QFTs is that of theories obtained by the insertion of a defect in an otherwise critical bulk. Though, depending on the dimension of the defect, theorems have been established about the weak monotonicity of RG flows\cite{Cuomo:2021rkm,Jensen:2015swa}, whether or not these theories undergo gradient flow has not been widely studied. In a semi-classical limit it has been shown that the beta functions for multiscalar line defects are indeed gradient with a flat metric\cite{Rodriguez-Gomez:2022gif,CarrenoBolla:2023sos,CarrenoBolla:2023vrv}, but gradient properties of the beta functions have only been shown at very low loop orders in line\cite{Pannell:2023pwz}, surface\cite{Pannell:2024hbu} and interface\cite{Harribey:2024gjn} defect theories. In part, studies of the gradient properties of these theories are limited by the fact that the beta functions have not previously been investigated to high loop order. However, it is still possible to set up a gradient flow problem using arbitrary $\nc{\mathfrak{b}}{n}{m}$ to parametrise the unknown beta function coefficients and solve the resulting equations as before. One benefit of these theories is that their interaction tensors contain fewer indices than $\lambda_{ijkl}$, so that the diagrammatic method of computing $R$ outlined in Appendix \ref{sec:diagcalc} becomes more straightforward. Indeed, for line defects the interaction tensor is simply a vector, $h_i$, so that the dimension of theory space will be identical to the number of fields. This simpler setting may serve as a sort of toy model for the information captured by the metric in the multiscalar models studied in this paper. Perhaps by computing $R$ more completely, and studying the geodesics of the metric more exactly it will be possible to fully understand how the geometry of theory space captures CFT and RG data.

\ack{We thank Petr Kravchuk, Slava Rychkov, and Balt van Rees for discussions. We are particularly grateful to Hugh Osborn for enlightening discussions and comments on the manuscript. AS is supported by the Royal Society under grant URF\textbackslash{}R1\textbackslash211417 and by STFC under grant ST/X000753/1.}

\begin{appendix}

\section{Solution to gradient flow through three loops}\label{app:gradsol}
In order to properly define the various coefficients which appear throughout the paper, we will set out explicitly the expansions for $A$, $G_{IJ}$ and $\beta^I$ which participate in the gradient flow equation though three loops. The $A$ function we will use is given through $\text{O}(\lambda^5)$ by
\begin{equation}
\begin{split}
A=&\nc{a}{2}{}\,
+\text{O}(\lambda^6)\,.
\end{split}
\end{equation}
The metric, $G_{IJ}$, through $\text{O}(\lambda^3)$, is given by
\begin{alignat}{2}
G_{IJ}=&\nc{g}{0\lnsp}{}\lsp%
+\text{O}(\lambda^4)\,.
\end{alignat}
Finally, the beta function through $\text{O}(\lambda^4)$ is given by
\begin{equation}
\begin{split}
    \beta=&(\nc{\mathfrak{b}}{1}{}+\varepsilon\nc{\mathfrak{e}}{1}{})\,%
+\text{O}(\lambda^5)\,,
\end{split}
\end{equation}
where for the beta function each diagram represents a sum over distinct permutations of the external indices and we have allowed $\veps$ contributions in all coefficients as would be the case in a general scheme.

Let us now try to solve the gradient flow equation (\ref{eq:gradflowtosolve}) through $\text{O}(\lambda^4)$. At $\text{O}(\lambda)$ there is only a single equation as there is only a single tensor structure which can appear, namely $\lambda_{ijkl}$ itself. The equation (\ref{eq:gradflowtosolve}) then becomes
\begin{equation}
    2\nc{a}{2}{}\,\begin{tikzpicture}[scale=0.5,baseline=(vert_cent.base)]
        \node (vert_cent) at (0,0) {$\phantom{\cdot}$};
        \filldraw (0,0) circle (2pt);
        \draw (-1,1)--(0,0);
        \draw (-1,-1)--(0,0);
        \draw (1,1)--(0,0);
        \draw (1,-1)--(0,0);
    \end{tikzpicture}=-\varepsilon\nc{g}{0\lnsp}{}\,\begin{tikzpicture}[scale=0.5,baseline=(vert_cent.base)]
        \node (vert_cent) at (0,0) {$\phantom{\cdot}$};
        \filldraw (0,0) circle (2pt);
        \draw (-1,1)--(0,0);
        \draw (-1,-1)--(0,0);
        \draw (1,1)--(0,0);
        \draw (1,-1)--(0,0);
    \end{tikzpicture}\,,
\end{equation}
where we have used the fact that $\nc{\mathfrak{b}
}{1}{}=0$ and $\nc{\mathfrak{e}}{1}{}=-1$ in all schemes. To make positive definiteness of the metric manifest, we will want to look for a solution of the form $\nc{g}{0\lnsp}{}=1+\text{O}(\varepsilon)$. As we will not permit factors of $1/\varepsilon$ in the solution, this forces us to also assume that $\nc{a}{2}{}=-\varepsilon/2+\text{O}(\varepsilon^2)$, which leads us to the generic zero-loop solution
\begin{equation}
    \nc{a}{2\llnsp}{}=-\frac{\varepsilon}{2}+\nc{{a}}{2\llnsp}{}'\varepsilon^2\,,\qquad \nc{g}{0\lnsp}{}=1-2\nc{{a}}{2\llnsp}{}'\varepsilon\,,
\end{equation}
for unfixed $\nc{{a}}{2\llnsp}{}'$. At $\text{O}(\lambda^2)$ there is again only a single tensor structure which may appear, leading to the equation
\begin{equation}
\nc{a}{3\lnsp}{}\lsp\begin{tikzpicture}[scale=0.5,baseline=(vert_cent.base)]
        \node (vert_cent) at (0,0) {$\phantom{\cdot}$};
        \draw (0,0) circle (1cm);
        \filldraw (-1,0) circle (2pt);
        \filldraw (1,0) circle (2pt);
        \draw (-1.7,0.7)--(-1,0);
        \draw (-1.7,-0.7)--(-1,0);
        \draw (1.7,0.7)--(1,0);
        \draw (1.7,-0.7)--(1,0);
    \end{tikzpicture}=\bigg(\nc{\mathfrak{b}}{2}{}+\varepsilon\nc{\mathfrak{e}}{2}{}-\frac{\varepsilon}{3}\nc{g}{1\lnsp}{}-2\varepsilon\nc{{a}}{2\llnsp}{}'\nc{\mathfrak{b}}{2}{}-2\varepsilon^2\nc{{a}}{2\llnsp}{}'\nc{\mathfrak{e}}{2}{}\bigg)\lsp\begin{tikzpicture}[scale=0.5,baseline=(vert_cent.base)]
        \node (vert_cent) at (0,0) {$\phantom{\cdot}$};
        \draw (0,0) circle (1cm);
        \filldraw (-1,0) circle (2pt);
        \filldraw (1,0) circle (2pt);
        \draw (-1.7,0.7)--(-1,0);
        \draw (-1.7,-0.7)--(-1,0);
        \draw (1.7,0.7)--(1,0);
        \draw (1.7,-0.7)--(1,0);
    \end{tikzpicture}\,,
\end{equation}
where we note that the factor of $3$ arising from the derivative in $A$ is cancelled by the factor of $1/3$ required to symmetrise the resulting free indices. This has the obvious solution
\begin{equation}
    \nc{a}{3\lnsp}{}=\nc{\mathfrak{b}}{2}{}+\varepsilon\nc{\mathfrak{e}}{2}{}-\frac{\varepsilon}{3}\nc{g}{1\lnsp}{}-2\lsp\varepsilon\nc{{a}}{2\llnsp}{}'\nc{\mathfrak{b}}{2}{}-2\lsp\varepsilon^2\nc{{a}}{2\llnsp}{}'\nc{\mathfrak{e}}{2}{}\,.
\end{equation}
Beginning at $\text{O}(\lambda^3)$, multiple tensor structures, and thus multiple equations, appear. However, at this order each vacuum bubble still yields a single tensor structure when acted upon by a derivative, as can be seen by the symmetry between the different vertices in (\ref{eq:A2loops}), so that it is still possible to solve the gradient flow equations purely in the $\nc{a}{4\lnsp}{m}$. One finds the solution
\begin{equation}
\begin{split}
    \nc{a}{4\lnsp}{1}&=-\tfrac{1}{4} \varepsilon  (8 \nc{{a}}{2\llnsp}{}' \varepsilon  \nc{\mathfrak{e}}{3}{1}+\nc{g}{2\lnsp}{1}+\nc{g}{2\lnsp}{3}+\nc{g}{2\lnsp}{7}-4 \nc{\mathfrak{e}}{3}{1})-2 \nc{{a}}{2\llnsp}{}' \varepsilon  \nc{\mathfrak{b}}{3}{1}+\nc{\mathfrak{b}}{3}{1}\,,\\\nc{a}{4\lnsp}{2}&=\tfrac{1}{4} (-\varepsilon  (6 \nc{{a}}{2\llnsp}{}' \varepsilon  \nc{\mathfrak{e}}{3}{2}+\nc{g}{2\lnsp}{2}+\nc{g}{2\lnsp}{5}-3 \nc{\mathfrak{e}}{3}{2})+\nc{\mathfrak{b}}{3}{2} (3-6 \nc{{a}}{2\llnsp}{}' \varepsilon )+\nc{g}{1\lnsp}{} (\nc{\mathfrak{b}}{2}{}+\varepsilon  \nc{\mathfrak{e}}{2}{}))\,,\\ \nc{a}{4\lnsp}{3}&=\tfrac{1}{4} (-\varepsilon  (12 \nc{{a}}{2\llnsp}{}' \varepsilon  \nc{\mathfrak{e}}{3}{3}+\nc{g}{2\lnsp}{4}+\nc{g}{2\lnsp}{6}-6 \nc{\mathfrak{e}}{3}{3})+\nc{\mathfrak{b}}{3}{3} (6-12 \nc{{a}}{2\llnsp}{}' \varepsilon )+2 \nc{g}{1\lnsp}{} (\nc{\mathfrak{b}}{2}{}+\varepsilon  \nc{\mathfrak{e}}{2}{}))\,.
\end{split}
\end{equation}
At $\text{O}(\lambda^4)$, as recognised by Wallace and Zia, this nice property for the vacuum diagrams fails, and one is no longer able to generically solve the gradient flow equations in terms of $A$ alone. Using the metric coefficients, one is still able to find the solution
\begin{alignat}{2}
\nc{g}{2\lnsp}{2}&=\frac{1}{4 (\nc{\mathfrak{b}}{2}{}+\varepsilon  \nc{\mathfrak{e}}{2}{})}\bigg(6 \nc{\mathfrak{b}}{4}{6} \big(2 \nc{{a\smash{\mathrlap{'}}}}{2\llnsp}{} \lsp\varepsilon ^3-\varepsilon ^2-1\big)+12 \nc{{a\smash{\mathrlap{'}}}}{2\llnsp}{}\lsp \varepsilon ^4 \nc{\mathfrak{e}}{4}{6}+2 \nc{a}{5\lnsp}{3}-2 \nc{g}{1\lnsp}{} \nc{\mathfrak{b}}{3}{3}-4 \varepsilon  \nc{g}{1\lnsp}{}\nc{\mathfrak{e}}{3}{3}\nonumber\\
&\quad+\varepsilon  \nc{g}{3\lnsp}{6}+\varepsilon  \nc{g}{3\lnsp}{16}-6 \varepsilon ^3 \nc{\mathfrak{e}}{4}{6}-6 \varepsilon  \nc{\mathfrak{e}}{4}{6}\bigg)\,,\nonumber \\
\nc{g}{2\lnsp}{3}&=\frac{1}{6 (\nc{\mathfrak{b}}{2}{}+\varepsilon  \nc{\mathfrak{e}}{2}{})}\bigg(-24 \nc{{a\smash{\mathrlap{'}}}}{2\llnsp}{}\lsp \varepsilon ^3 \nc{\mathfrak{b}}{4}{1}+16 \nc{{a\smash{\mathrlap{'}}}}{2\llnsp}{}\lsp \varepsilon ^3 \nc{\mathfrak{b}}{4}{3}-24 \nc{{a\smash{\mathrlap{'}}}}{2\llnsp}{}\lsp \varepsilon ^4 \nc{\mathfrak{e}}{4}{1}+16 \nc{{a\smash{\mathrlap{'}}}}{2\llnsp}{}\lsp \varepsilon ^4 \nc{\mathfrak{e}}{4}{3}+8 \nc{g}{1\lnsp}{}(\nc{\mathfrak{b}}{3}{1}+3 \varepsilon  \nc{\mathfrak{e}}{3}{1})\nonumber \\&\quad-\varepsilon  \nc{g}{3\lnsp}{4}+\varepsilon  \nc{g}{3\lnsp}{8}-6 \nc{g}{2\lnsp}{1} (\nc{\mathfrak{b}}{2}{}+\varepsilon  \nc{\mathfrak{e}}{2}{})-2 \varepsilon  \nc{g}{3\lnsp}{13}+2 \varepsilon \nc{g}{3\lnsp}{18}+\varepsilon  \nc{g}{3\lnsp}{1}+12 \varepsilon ^2 \nc{\mathfrak{b}}{4}{1}+12 \nc{\mathfrak{b}}{4}{1}-8 \varepsilon ^2 \nc{\mathfrak{b}}{4}{3}\nonumber \\&\quad-8 \nc{\mathfrak{b}}{4}{3}+12 \varepsilon ^3 \nc{\mathfrak{e}}{4}{1}+12 \varepsilon \nc{\mathfrak{e}}{4}{1}-8 \varepsilon ^3 \nc{\mathfrak{e}}{4}{3}-8 \varepsilon  \nc{\mathfrak{e}}{4}{3}\bigg)\,,\nonumber \\
\nc{g}{2\lnsp}{4}&=\frac{1}{2 (\nc{\mathfrak{b}}{2}{}+\varepsilon \nc{\mathfrak{e}}{2}{})}\bigg(12 \nc{\mathfrak{b}}{4}{7} \big(2 \nc{{a\smash{\mathrlap{'}}}}{2\llnsp}{}\lsp \varepsilon ^3-\varepsilon ^2-1\big)+24 \nc{{a\smash{\mathrlap{'}}}}{2\llnsp}{}\lsp \varepsilon ^4 \nc{\mathfrak{e}}{4}{7}+4 \nc{a}{5\lnsp}{3}-4 \nc{g}{1\lnsp}{} \nc{\mathfrak{b}}{3}{2}-16 \varepsilon  \nc{g}{1\lnsp}{}\nc{\mathfrak{e}}{3}{2}+\varepsilon  \nc{g}{3\lnsp}{6}\nonumber \\&\quad+\varepsilon  \nc{g}{3\lnsp}{9}+2 \varepsilon  \nc{g}{3\lnsp}{15}-12 \varepsilon ^3 \nc{\mathfrak{e}}{4}{7}-12 \varepsilon  \nc{\mathfrak{e}}{4}{7}\bigg)\,,\nonumber \\
\nc{g}{2\lnsp}{5}&=\frac{1}{32 (\nc{\mathfrak{b}}{2}{}+\varepsilon  \nc{\mathfrak{e}}{2}{})}\bigg(24 \nc{\mathfrak{b}}{4}{7} \big(-2 \nc{{a\smash{\mathrlap{'}}}}{2\llnsp}{}\lsp \varepsilon ^3+\varepsilon ^2+1\big)+48 \nc{{a\smash{\mathrlap{'}}}}{2\llnsp}{}\lsp \varepsilon ^3 \nc{\mathfrak{b}}{4}{8}-96 \nc{{a\smash{\mathrlap{'}}}}{2\llnsp}{}\lsp \varepsilon ^3 \nc{\mathfrak{b}}{4}{9}+192 \nc{{a\smash{\mathrlap{'}}}}{2\llnsp}{}\lsp \varepsilon ^3\nc{\mathfrak{b}}{4}{5}\nonumber \\&\quad-48 \nc{{a\smash{\mathrlap{'}}}}{2\llnsp}{}\lsp \varepsilon ^4 \nc{\mathfrak{e}}{4}{7}+48 \nc{{a\smash{\mathrlap{'}}}}{2\llnsp}{}\lsp \varepsilon ^4 \nc{\mathfrak{e}}{4}{8}-96 \nc{{a\smash{\mathrlap{'}}}}{2\llnsp}{}\lsp \varepsilon ^4 \nc{\mathfrak{e}}{4}{9}+192 \nc{{a\smash{\mathrlap{'}}}}{2\llnsp}{}\lsp \varepsilon ^4\nc{\mathfrak{e}}{4}{5}+24 \nc{a}{5\lnsp}{3}+8 \nc{g}{1\lnsp}{} \nc{\mathfrak{b}}{3}{2}-24 \nc{g}{1\lnsp}{} \nc{\mathfrak{b}}{3}{3}\nonumber \\&\quad+32 \varepsilon  \nc{g}{1\lnsp}{} \nc{\mathfrak{e}}{3}{2}-48 \varepsilon  \nc{g}{1\lnsp}{} \nc{\mathfrak{e}}{3}{3}+16 \varepsilon \nc{g}{3\lnsp}{3}+2 \varepsilon  \nc{g}{3\lnsp}{5}-2 \varepsilon  \nc{g}{3\lnsp}{6}+6 \varepsilon  \nc{g}{3\lnsp}{9}-3 \varepsilon  \nc{g}{3\lnsp}{10}+2 \varepsilon  \nc{g}{3\lnsp}{11}-4 \varepsilon  \nc{g}{3\lnsp}{15} \nonumber\\&\quad+8 \varepsilon  \nc{g}{3\lnsp}{16}+2 \varepsilon  \nc{g}{3\lnsp}{17}-24\varepsilon ^2 \nc{\mathfrak{b}}{4}{8}-24 \nc{\mathfrak{b}}{4}{8}+48 \varepsilon ^2 \nc{\mathfrak{b}}{4}{9}+48 \nc{\mathfrak{b}}{4}{9}-96 \varepsilon ^2 \nc{\mathfrak{b}}{4}{5}-96 \nc{\mathfrak{b}}{4}{5}+24 \varepsilon ^3\nc{\mathfrak{e}}{4}{7}\nonumber \\&\quad+24 \varepsilon  \nc{\mathfrak{e}}{4}{7}-24 \varepsilon ^3 \nc{\mathfrak{e}}{4}{8}-24 \varepsilon  \nc{\mathfrak{e}}{4}{8}+48 \varepsilon ^3 \nc{\mathfrak{e}}{4}{9}+48 \varepsilon \nc{\mathfrak{e}}{4}{9}-96 \varepsilon ^3 \nc{\mathfrak{e}}{4}{5}-96 \varepsilon  \nc{\mathfrak{e}}{4}{5}\bigg)\,,\nonumber \\
\nc{g}{2\lnsp}{6}&=\frac{1}{16(\nc{\mathfrak{b}}{2}{}+\varepsilon  \nc{\mathfrak{e}}{2}{})}\bigg(24 \nc{\mathfrak{b}}{4}{7} \big(2 \nc{{a\smash{\mathrlap{'}}}}{2\llnsp}{}\lsp \varepsilon ^3-\varepsilon ^2-1\big)-48 \nc{{a\smash{\mathrlap{'}}}}{2\llnsp}{}\lsp \varepsilon ^3 \nc{\mathfrak{b}}{4}{8}+96 \nc{{a\smash{\mathrlap{'}}}}{2\llnsp}{}\lsp \varepsilon ^3 \nc{\mathfrak{b}}{4}{9}+48 \nc{{a\smash{\mathrlap{'}}}}{2\llnsp}{}\lsp \varepsilon ^4\nc{\mathfrak{e}}{4}{7}\nonumber \\&\quad-48 \nc{{a\smash{\mathrlap{'}}}}{2\llnsp}{}\lsp \varepsilon ^4 \nc{\mathfrak{e}}{4}{8}+96 \nc{{a\smash{\mathrlap{'}}}}{2\llnsp}{}\lsp \varepsilon ^4 \nc{\mathfrak{e}}{4}{9}+8 \nc{a}{5\lnsp}{3}-8 \nc{g}{1\lnsp}{} \nc{\mathfrak{b}}{3}{2}+8 \nc{g}{1\lnsp}{} \nc{\mathfrak{b}}{3}{3}-32 \varepsilon \nc{g}{1\lnsp}{} \nc{\mathfrak{e}}{3}{2}+16 \varepsilon  \nc{g}{1\lnsp}{} \nc{\mathfrak{e}}{3}{3}-2 \varepsilon  \nc{g}{3\lnsp}{5}\nonumber \\&\quad+2 \varepsilon  \nc{g}{3\lnsp}{6}+2 \varepsilon  \nc{g}{3\lnsp}{9}+3 \varepsilon  \nc{g}{3\lnsp}{10}-2 \varepsilon  \nc{g}{3\lnsp}{11}+4 \varepsilon\nc{g}{3\lnsp}{15}-2 \varepsilon  \nc{g}{3\lnsp}{17}+24 \varepsilon ^2 \nc{\mathfrak{b}}{4}{8}+24 \nc{\mathfrak{b}}{4}{8}-48 \varepsilon ^2 \nc{\mathfrak{b}}{4}{9}\nonumber \\&\quad-48 \nc{\mathfrak{b}}{4}{9}-24 \varepsilon ^3 \nc{\mathfrak{e}}{4}{7}-24\varepsilon  \nc{\mathfrak{e}}{4}{7}+24 \varepsilon ^3 \nc{\mathfrak{e}}{4}{8}+24 \varepsilon  \nc{\mathfrak{e}}{4}{8}-48 \varepsilon ^3 \nc{\mathfrak{e}}{4}{9}-48 \varepsilon  \nc{\mathfrak{e}}{4}{9}\bigg)\,,\nonumber \\
\nc{g}{2\lnsp}{7}&=\frac{1}{6(\nc{\mathfrak{b}}{2}{}+\varepsilon  \nc{\mathfrak{e}}{2}{})}\bigg(12 \nc{\mathfrak{b}}{4}{1} \big(-2 \nc{{a\smash{\mathrlap{'}}}}{2\llnsp}{}\lsp \varepsilon ^3+\varepsilon ^2+1\big)+48 \nc{{a\smash{\mathrlap{'}}}}{2\llnsp}{}\lsp \varepsilon ^3 \nc{\mathfrak{b}}{4}{2}-24 \nc{{a\smash{\mathrlap{'}}}}{2\llnsp}{}\lsp \varepsilon ^4 \nc{\mathfrak{e}}{4}{1}+48 \nc{{a\smash{\mathrlap{'}}}}{2\llnsp}{}\lsp \varepsilon ^4\nc{\mathfrak{e}}{4}{2}\nonumber \\&\quad+4 \nc{g}{1\lnsp}{} (\nc{\mathfrak{b}}{3}{1}+3 \varepsilon  \nc{\mathfrak{e}}{3}{1})-2 \varepsilon  \nc{g}{3\lnsp}{4}+2 \varepsilon  \nc{g}{3\lnsp}{7}+\varepsilon  \nc{g}{3\lnsp}{8}-\varepsilon  \nc{g}{3\lnsp}{13}+\varepsilon  \nc{g}{3\lnsp}{1}-24\varepsilon ^2 \nc{\mathfrak{b}}{4}{2}-24 \nc{\mathfrak{b}}{4}{2}+12 \varepsilon ^3 \nc{\mathfrak{e}}{4}{1}\nonumber \\&\quad+12 \varepsilon  \nc{\mathfrak{e}}{4}{1}-24 \varepsilon ^3 \nc{\mathfrak{e}}{4}{2}-24 \varepsilon  \nc{\mathfrak{e}}{4}{2}\bigg)\,,\nonumber \\
\nc{a}{5\lnsp}{1}&=-\frac{1}{2} \varepsilon  \bigg(-6 \left(\varepsilon  \nc{\mathfrak{b}}{4}{1} (1-2 \nc{{a\smash{\mathrlap{'}}}}{2\llnsp}{}\lsp \varepsilon )+\nc{\mathfrak{e}}{4}{1} \big(-2 \nc{{a\smash{\mathrlap{'}}}}{2\llnsp}{}\lsp \varepsilon ^3+\varepsilon ^2+1\big)+2 \nc{g}{1\lnsp}{}\nc{\mathfrak{e}}{3}{1}\right)+\nc{g}{3\lnsp}{4}+\nc{g}{3\lnsp}{13}\bigg)\nonumber\\&\quad+2 \nc{g}{1\lnsp}{} \nc{\mathfrak{b}}{3}{1}+3 \nc{\mathfrak{b}}{4}{1}\,,\nonumber \\
\nc{a}{5\lnsp}{2}&=\frac{1}{160} \bigg(\varepsilon  \big(96 \nc{{a\smash{\mathrlap{'}}}}{2\llnsp}{}\lsp \varepsilon ^2 \nc{\mathfrak{b}}{4}{6}-48 \nc{{a\smash{\mathrlap{'}}}}{2\llnsp}{}\lsp \varepsilon ^2 \nc{\mathfrak{b}}{4}{7}+48 \nc{{a\smash{\mathrlap{'}}}}{2\llnsp}{}\lsp \varepsilon ^2 \nc{\mathfrak{b}}{4}{8}-96 \nc{{a\smash{\mathrlap{'}}}}{2\llnsp}{}\lsp \varepsilon ^2\nc{\mathfrak{b}}{4}{9}-192 \nc{{a\smash{\mathrlap{'}}}}{2\llnsp}{}\lsp \varepsilon ^2 \nc{\mathfrak{b}}{4}{4}+192 \nc{{a\smash{\mathrlap{'}}}}{2\llnsp}{}\lsp \varepsilon ^2 \nc{\mathfrak{b}}{4}{5}\nonumber \\&\quad+96 \nc{{a\smash{\mathrlap{'}}}}{2\llnsp}{}\lsp \varepsilon ^3 \nc{\mathfrak{e}}{4}{6}-48 \nc{{a\smash{\mathrlap{'}}}}{2\llnsp}{}\lsp \varepsilon ^3\nc{\mathfrak{e}}{4}{7}+48 \nc{{a\smash{\mathrlap{'}}}}{2\llnsp}{}\lsp \varepsilon ^3 \nc{\mathfrak{e}}{4}{8}-96 \nc{{a\smash{\mathrlap{'}}}}{2\llnsp}{}\lsp \varepsilon ^3 \nc{\mathfrak{e}}{4}{9}-192 \nc{{a\smash{\mathrlap{'}}}}{2\llnsp}{}\lsp \varepsilon ^3 \nc{\mathfrak{e}}{4}{4}+192 \nc{{a\smash{\mathrlap{'}}}}{2\llnsp}{}\lsp \varepsilon ^3\nc{\mathfrak{e}}{4}{5}\nonumber \\&\quad+160 \nc{g}{1\lnsp}{} \nc{\mathfrak{e}}{3}{2}-80 \nc{g}{1\lnsp}{} \nc{\mathfrak{e}}{3}{3}-32 \nc{g}{3\lnsp}{2}+16 \nc{g}{3\lnsp}{3}+2 \nc{g}{3\lnsp}{5}+6 \nc{g}{3\lnsp}{6}+6 \nc{g}{3\lnsp}{9}-3 \nc{g}{3\lnsp}{10}+2 \nc{g}{3\lnsp}{11}-32 \nc{g}{3\lnsp}{14}\nonumber \\&\quad-4 \nc{g}{3\lnsp}{15}+16\nc{g}{3\lnsp}{16}+2 \nc{g}{3\lnsp}{17}-48 \varepsilon  \nc{\mathfrak{b}}{4}{6}+24 \varepsilon  \nc{\mathfrak{b}}{4}{7}-24 \varepsilon  \nc{\mathfrak{b}}{4}{8}+48 \varepsilon  \nc{\mathfrak{b}}{4}{9}+96 \varepsilon  \nc{\mathfrak{b}}{4}{4}-96\varepsilon  \nc{\mathfrak{b}}{4}{5}\nonumber \\&\quad-48 \varepsilon ^2 \nc{\mathfrak{e}}{4}{6}-48 \nc{\mathfrak{e}}{4}{6}+24 \varepsilon ^2 \nc{\mathfrak{e}}{4}{7}+24 \nc{\mathfrak{e}}{4}{7}-24 \varepsilon ^2 \nc{\mathfrak{e}}{4}{8}-24\nc{\mathfrak{e}}{4}{8}+48 \varepsilon ^2 \nc{\mathfrak{e}}{4}{9}+48 \nc{\mathfrak{e}}{4}{9}+96 \varepsilon ^2 \nc{\mathfrak{e}}{4}{4}+96 \nc{\mathfrak{e}}{4}{4}\nonumber \\&\quad-96 \varepsilon ^2 \nc{\mathfrak{e}}{4}{5}-96\nc{\mathfrak{e}}{4}{5}\big)+40 \nc{a}{5\lnsp}{3}+40 \nc{g}{1\lnsp}{} (\nc{\mathfrak{b}}{3}{2}-\nc{\mathfrak{b}}{3}{3})\nonumber\\
&\quad-24 (2 \nc{\mathfrak{b}}{4}{6}-\nc{\mathfrak{b}}{4}{7}+\nc{\mathfrak{b}}{4}{8}-2 \nc{\mathfrak{b}}{4}{9}-4\nc{\mathfrak{b}}{4}{4}+4 \nc{\mathfrak{b}}{4}{5})\bigg),\nonumber \\
\nc{a}{5\lnsp}{4}&=\frac{1}{8} \varepsilon  \bigg(48 \nc{{a\smash{\mathrlap{'}}}}{2\llnsp}{}\lsp \varepsilon ^2 \nc{\mathfrak{b}}{4}{7}-48 \nc{{a\smash{\mathrlap{'}}}}{2\llnsp}{}\lsp \varepsilon ^2 \nc{\mathfrak{b}}{4}{8}+48 \nc{{a\smash{\mathrlap{'}}}}{2\llnsp}{}\lsp \varepsilon ^3 \nc{\mathfrak{e}}{4}{7}-48 \nc{{a\smash{\mathrlap{'}}}}{2\llnsp}{}\lsp \varepsilon ^3\nc{\mathfrak{e}}{4}{8}-32 \nc{g}{1\lnsp}{} \nc{\mathfrak{e}}{3}{2}+16 \nc{g}{1\lnsp}{} \nc{\mathfrak{e}}{3}{3}-2 \nc{g}{3\lnsp}{5}+2 \nc{g}{3\lnsp}{6}\nonumber \\&\quad+2 \nc{g}{3\lnsp}{9}-\nc{g}{3\lnsp}{10}-2 \nc{g}{3\lnsp}{11}+4 \nc{g}{3\lnsp}{15}-2 \nc{g}{3\lnsp}{17}-24 \varepsilon  \nc{\mathfrak{b}}{4}{7}+24\varepsilon  \nc{\mathfrak{b}}{4}{8}-24 \varepsilon ^2 \nc{\mathfrak{e}}{4}{7}-24 \nc{\mathfrak{e}}{4}{7}+24 \varepsilon ^2 \nc{\mathfrak{e}}{4}{8}\nonumber \\&\quad+24 \nc{\mathfrak{e}}{4}{8}\bigg)+\nc{a}{5\lnsp}{3}+\nc{g}{1\lnsp}{}(\nc{\mathfrak{b}}{3}{3}-\nc{\mathfrak{b}}{3}{2})-3 \nc{\mathfrak{b}}{4}{7}+3 \nc{\mathfrak{b}}{4}{8}\,,\nonumber \\
\nc{a}{5\lnsp}{5}&=\frac{1}{5} \left(\nc{\mathfrak{b}}{4}{10} \big(-2 \nc{{a\smash{\mathrlap{'}}}}{2\llnsp}{}\lsp \varepsilon ^3+\varepsilon ^2+1\big)+\varepsilon  \nc{\mathfrak{e}}{4}{10} \big(-2 \nc{{a\smash{\mathrlap{'}}}}{2\llnsp}{}\lsp \varepsilon ^3+\varepsilon ^2+1\big)+\varepsilon  (-\nc{g}{3\lnsp}{12})\right)\,.
\end{alignat}

\section{Explicit terms in the Ricci scalar expansion}\label{app:explicit}
One finds, either by using \href{https://www.nikhef.nl/~form/}{\texttt{FORM}} or from the diagrammatic calculation, that the leading order contribution to the Ricci scalar takes the generic form
\begin{alignat}{2}
\nc{R}{0\lnsp}{}&=N
   \left(\frac{31 (\nc{g}{1\lnsp}{})^2}{864(\nc{g}{0\lnsp}{})^3}-\frac{5 \nc{g}{2\lnsp}{1}}{8(\nc{g}{0\lnsp}{})^2}-\frac{13 \nc{g}{2\lnsp}{2}}{18(\nc{g}{0\lnsp}{})^2}+\frac{\nc{g}{2\lnsp}{3}}{8(\nc{g}{0\lnsp}{})^2}-\frac{\nc{g}{2\lnsp}{4}}{8(\nc{g}{0\lnsp}{})^2}+\frac{13 \nc{g}{2\lnsp}{5}}{36(\nc{g}{0\lnsp}{})^2}+\frac{\nc{g}{2\lnsp}{6}}{4(\nc{g}{0\lnsp}{})^2}+\frac{\nc{g}{2\lnsp}{7}}{2(\nc{g}{0\lnsp}{})^2}\right) \nonumber \\&+N^2 \bigg(\frac{41 (\nc{g}{1\lnsp}{})^2}{3456(\nc{g}{0\lnsp}{})^3}-\frac{7 \nc{g}{2\lnsp}{1}}{32(\nc{g}{0\lnsp}{})^2}-\frac{11 \nc{g}{2\lnsp}{2}}{72(\nc{g}{0\lnsp}{})^2}-\frac{5 \nc{g}{2\lnsp}{3}}{32(\nc{g}{0\lnsp}{})^2}-\frac{5 \nc{g}{2\lnsp}{4}}{96(\nc{g}{0\lnsp}{})^2}+\frac{11 \nc{g}{2\lnsp}{5}}{144(\nc{g}{0\lnsp}{})^2}+\frac{5 \nc{g}{2\lnsp}{6}}{48(\nc{g}{0\lnsp}{})^2} \nonumber \\&+\frac{3 \nc{g}{2\lnsp}{7}}{8(\nc{g}{0\lnsp}{})^2}\bigg)+N^3 \bigg(-\frac{167 (\nc{g}{1\lnsp}{})^2}{6912(\nc{g}{0\lnsp}{})^3}+\frac{223 \nc{g}{2\lnsp}{1}}{576(\nc{g}{0\lnsp}{})^2}+\frac{23 \nc{g}{2\lnsp}{2}}{48(\nc{g}{0\lnsp}{})^2}-\frac{83 \nc{g}{2\lnsp}{3}}{576(\nc{g}{0\lnsp}{})^2}+\frac{49 \nc{g}{2\lnsp}{4}}{576(\nc{g}{0\lnsp}{})^2}-\frac{23 \nc{g}{2\lnsp}{5}}{96(\nc{g}{0\lnsp}{})^2}\nonumber \\&-\frac{49\nc{g}{2\lnsp}{6}}{288(\nc{g}{0\lnsp}{})^2}-\frac{35 \nc{g}{2\lnsp}{7}}{144(\nc{g}{0\lnsp}{})^2}\bigg)+N^4 \bigg(-\frac{127 (\nc{g}{1\lnsp}{})^2}{6912(\nc{g}{0\lnsp}{})^3}+\frac{31 \nc{g}{2\lnsp}{1}}{96(\nc{g}{0\lnsp}{})^2}+\frac{5 \nc{g}{2\lnsp}{2}}{16(\nc{g}{0\lnsp}{})^2}+\frac{7 \nc{g}{2\lnsp}{3}}{96(\nc{g}{0\lnsp}{})^2}\nonumber \\&+\frac{41\nc{g}{2\lnsp}{4}}{576(\nc{g}{0\lnsp}{})^2}-\frac{5 \nc{g}{2\lnsp}{5}}{32(\nc{g}{0\lnsp}{})^2}-\frac{41 \nc{g}{2\lnsp}{6}}{288(\nc{g}{0\lnsp}{})^2}-\frac{19 \nc{g}{2\lnsp}{7}}{48(\nc{g}{0\lnsp}{})^2}\bigg)+N^5 \bigg(-\frac{11 (\nc{g}{1\lnsp}{})^2}{2304(\nc{g}{0\lnsp}{})^3}+\frac{\nc{g}{2\lnsp}{1}}{9(\nc{g}{0\lnsp}{})^2}\nonumber\\&+\frac{11\nc{g}{2\lnsp}{2}}{144(\nc{g}{0\lnsp}{})^2}+\frac{23 \nc{g}{2\lnsp}{3}}{288(\nc{g}{0\lnsp}{})^2}+\frac{11 \nc{g}{2\lnsp}{4}}{576(\nc{g}{0\lnsp}{})^2}-\frac{11 \nc{g}{2\lnsp}{5}}{288(\nc{g}{0\lnsp}{})^2}-\frac{11 \nc{g}{2\lnsp}{6}}{288(\nc{g}{0\lnsp}{})^2}-\frac{55 \nc{g}{2\lnsp}{7}}{288(\nc{g}{0\lnsp}{})^2}\bigg)\nonumber\\&+N^6 \bigg(-\frac{(\nc{g}{1\lnsp}{})^2}{2304(\nc{g}{0\lnsp}{})^3}+\frac{\nc{g}{2\lnsp}{1}}{48(\nc{g}{0\lnsp}{})^2}+\frac{\nc{g}{2\lnsp}{2}}{144(\nc{g}{0\lnsp}{})^2}+\frac{\nc{g}{2\lnsp}{3}}{48(\nc{g}{0\lnsp}{})^2}+\frac{\nc{g}{2\lnsp}{4}}{576(\nc{g}{0\lnsp}{})^2}-\frac{\nc{g}{2\lnsp}{5}}{288(\nc{g}{0\lnsp}{})^2}-\frac{\nc{g}{2\lnsp}{6}}{288(\nc{g}{0\lnsp}{})^2}\nonumber \\&-\frac{\nc{g}{2\lnsp}{7}}{24(\nc{g}{0\lnsp}{})^2}\bigg)+N^7 \left(\frac{\nc{g}{2\lnsp}{1}}{576(\nc{g}{0\lnsp}{})^2}+\frac{\nc{g}{2\lnsp}{3}}{576(\nc{g}{0\lnsp}{})^2}-\frac{\nc{g}{2\lnsp}{7}}{288(\nc{g}{0\lnsp}{})^2}\right)\,.
\label{eq:Ricci0}
\end{alignat}
Using the method outlined in the text, one then finds the generic expression for $\nc{R}{1\lnsp}{}$
\begin{alignat}{2}
    \nc{R}{1\lnsp}{}&=\lambda_{iijj}\bigg(-\frac{71 (\nc{g}{1\lnsp}{})^3 }{432 (\nc{g}{0\lnsp}{})^4}-\frac{3
   \nc{g}{3\lnsp}{2} }{2 (\nc{g}{0\lnsp}{})^2}-\frac{13 \nc{g}{3\lnsp}{3} }{9 (\nc{g}{0\lnsp}{})^2}-\frac{5 \nc{g}{3\lnsp}{4} }{4 (\nc{g}{0\lnsp}{})^2}-\frac{95 \nc{g}{3\lnsp}{5} }{72 (\nc{g}{0\lnsp}{})^2}-\frac{25 \nc{g}{3\lnsp}{6} }{18 (\nc{g}{0\lnsp}{})^2}-\frac{\nc{g}{3\lnsp}{7}
   }{6 (\nc{g}{0\lnsp}{})^2}+\frac{\nc{g}{3\lnsp}{8} }{3 (\nc{g}{0\lnsp}{})^2} \nonumber \\&\quad+\frac{\nc{g}{3\lnsp}{9} }{36 (\nc{g}{0\lnsp}{})^2}+\frac{7 \nc{g}{3\lnsp}{11} }{72 (\nc{g}{0\lnsp}{})^2}+\frac{49 \nc{g}{1\lnsp}{} \nc{g}{2\lnsp}{1} }{24 (\nc{g}{0\lnsp}{})^3}+\frac{5 \nc{g}{3\lnsp}{13}}{4 (\nc{g}{0\lnsp}{})^2}+\frac{3 \nc{g}{3\lnsp}{14} }{2 (\nc{g}{0\lnsp}{})^2}+\frac{4 \nc{g}{3\lnsp}{15} }{3 (\nc{g}{0\lnsp}{})^2}+\frac{25 \nc{g}{3\lnsp}{16} }{18 (\nc{g}{0\lnsp}{})^2}+\frac{11 \nc{g}{3\lnsp}{17} }{9 (\nc{g}{0\lnsp}{})^2}\nonumber \\&\quad+\frac{7 \nc{g}{3\lnsp}{18}}{3 (\nc{g}{0\lnsp}{})^2} +\frac{233 \nc{g}{1\lnsp}{} \nc{g}{2\lnsp}{2} }{108 (\nc{g}{0\lnsp}{})^3}+\frac{7 \nc{g}{1\lnsp}{} \nc{g}{2\lnsp}{3} }{24 (\nc{g}{0\lnsp}{})^3}+\frac{17 \nc{g}{1\lnsp}{} \nc{g}{2\lnsp}{4} }{27 (\nc{g}{0\lnsp}{})^3}-\frac{79 \nc{g}{1\lnsp}{} \nc{g}{2\lnsp}{5} }{108
   (\nc{g}{0\lnsp}{})^3}-\frac{67 \nc{g}{1\lnsp}{} \nc{g}{2\lnsp}{6} }{108 (\nc{g}{0\lnsp}{})^3}-\frac{41 \nc{g}{1\lnsp}{} \nc{g}{2\lnsp}{7} }{24 (\nc{g}{0\lnsp}{})^3}-\frac{17 \nc{g}{3\lnsp}{1} }{12 (\nc{g}{0\lnsp}{})^2}\bigg) \nonumber \\&\quad+ N\lambda_{iijj}\bigg(\frac{137  (\nc{g}{1\lnsp}{})^3}{5184 (\nc{g}{0\lnsp}{})^4}-\frac{9 \nc{g}{2\lnsp}{1}
    \nc{g}{1\lnsp}{}}{32 (\nc{g}{0\lnsp}{})^3}-\frac{281 \nc{g}{2\lnsp}{2}  \nc{g}{1\lnsp}{}}{432 (\nc{g}{0\lnsp}{})^3}+\frac{17 \nc{g}{2\lnsp}{3}  \nc{g}{1\lnsp}{}}{32 (\nc{g}{0\lnsp}{})^3}-\frac{13 \nc{g}{2\lnsp}{4}  \nc{g}{1\lnsp}{}}{216 (\nc{g}{0\lnsp}{})^3}+\frac{37 \nc{g}{2\lnsp}{5}
   \nc{g}{1\lnsp}{}}{144 (\nc{g}{0\lnsp}{})^3}+\frac{13 \nc{g}{2\lnsp}{6}  \nc{g}{1\lnsp}{}}{432 (\nc{g}{0\lnsp}{})^3} \nonumber \\&\quad-\frac{29 \nc{g}{2\lnsp}{7}  \nc{g}{1\lnsp}{}}{96 (\nc{g}{0\lnsp}{})^3}+\frac{13 \nc{g}{3\lnsp}{2} }{24 (\nc{g}{0\lnsp}{})^2}+\frac{\nc{g}{3\lnsp}{3} }{2 (\nc{g}{0\lnsp}{})^2}+\frac{5 \nc{g}{3\lnsp}{4}
   }{48 (\nc{g}{0\lnsp}{})^2}+\frac{91 \nc{g}{3\lnsp}{5} }{288 (\nc{g}{0\lnsp}{})^2}+\frac{23 \nc{g}{3\lnsp}{6} }{72 (\nc{g}{0\lnsp}{})^2}+\frac{11 \nc{g}{3\lnsp}{7} }{24 (\nc{g}{0\lnsp}{})^2}-\frac{7 \nc{g}{3\lnsp}{8} }{12 (\nc{g}{0\lnsp}{})^2} \nonumber \\&\quad-\frac{13 \nc{g}{3\lnsp}{9}}{144 (\nc{g}{0\lnsp}{})^2}-\frac{59 \nc{g}{3\lnsp}{11} }{288 (\nc{g}{0\lnsp}{})^2}-\frac{5 \nc{g}{3\lnsp}{13} }{48 (\nc{g}{0\lnsp}{})^2}-\frac{13 \nc{g}{3\lnsp}{14} }{24 (\nc{g}{0\lnsp}{})^2}-\frac{5 \nc{g}{3\lnsp}{15} }{36 (\nc{g}{0\lnsp}{})^2}-\frac{23 \nc{g}{3\lnsp}{16}
   }{72 (\nc{g}{0\lnsp}{})^2}-\frac{\nc{g}{3\lnsp}{17} }{9 (\nc{g}{0\lnsp}{})^2}+\frac{\nc{g}{3\lnsp}{18} }{4 (\nc{g}{0\lnsp}{})^2}\nonumber \\&\quad+\frac{11 \nc{g}{3\lnsp}{1} }{48 (\nc{g}{0\lnsp}{})^2}\bigg) +N^2\lambda_{iijj}\bigg(\frac{991  (\nc{g}{1\lnsp}{})^3}{10368 (\nc{g}{0\lnsp}{})^4}-\frac{107 \nc{g}{2\lnsp}{1}  \nc{g}{1\lnsp}{}}{96 (\nc{g}{0\lnsp}{})^3}-\frac{323
   \nc{g}{2\lnsp}{2}  \nc{g}{1\lnsp}{}}{288 (\nc{g}{0\lnsp}{})^3}-\frac{29 \nc{g}{2\lnsp}{3}  \nc{g}{1\lnsp}{}}{96 (\nc{g}{0\lnsp}{})^3}-\frac{331 \nc{g}{2\lnsp}{4}  \nc{g}{1\lnsp}{}}{864 (\nc{g}{0\lnsp}{})^3}\nonumber \\&\quad+\frac{311 \nc{g}{2\lnsp}{5}  \nc{g}{1\lnsp}{}}{864 (\nc{g}{0\lnsp}{})^3} +\frac{85 \nc{g}{2\lnsp}{6}  \nc{g}{1\lnsp}{}}{216 (\nc{g}{0\lnsp}{})^3}+\frac{305 \nc{g}{2\lnsp}{7}  \nc{g}{1\lnsp}{}}{288 (\nc{g}{0\lnsp}{})^3}+\frac{35 \nc{g}{3\lnsp}{2} }{48 (\nc{g}{0\lnsp}{})^2}+\frac{103 \nc{g}{3\lnsp}{3} }{144 (\nc{g}{0\lnsp}{})^2}+\frac{103 \nc{g}{3\lnsp}{4} }{144
   (\nc{g}{0\lnsp}{})^2}+\frac{139 \nc{g}{3\lnsp}{5} }{192 (\nc{g}{0\lnsp}{})^2}\nonumber \\&\quad+\frac{7 \nc{g}{3\lnsp}{6} }{9 (\nc{g}{0\lnsp}{})^2}+\frac{\nc{g}{3\lnsp}{7} }{72 (\nc{g}{0\lnsp}{})^2}-\frac{\nc{g}{3\lnsp}{8} }{18 (\nc{g}{0\lnsp}{})^2}+\frac{\nc{g}{3\lnsp}{9} }{32
   (\nc{g}{0\lnsp}{})^2}+\frac{\nc{g}{3\lnsp}{11} }{192 (\nc{g}{0\lnsp}{})^2}-\frac{103 \nc{g}{3\lnsp}{13} }{144 (\nc{g}{0\lnsp}{})^2}-\frac{35 \nc{g}{3\lnsp}{14} }{48 (\nc{g}{0\lnsp}{})^2}-\frac{121 \nc{g}{3\lnsp}{15} }{144 (\nc{g}{0\lnsp}{})^2}\nonumber \\&\quad-\frac{7 \nc{g}{3\lnsp}{16} }{9
   (\nc{g}{0\lnsp}{})^2} -\frac{35 \nc{g}{3\lnsp}{17} }{48 (\nc{g}{0\lnsp}{})^2}-\frac{17 \nc{g}{3\lnsp}{18} }{12 (\nc{g}{0\lnsp}{})^2}+\frac{109 \nc{g}{3\lnsp}{1} }{144 (\nc{g}{0\lnsp}{})^2}\bigg) +N^3\lambda_{iijj} \bigg(\frac{ (\nc{g}{1\lnsp}{})^3}{27 (\nc{g}{0\lnsp}{})^4}-\frac{25 \nc{g}{2\lnsp}{1}  \nc{g}{1\lnsp}{}}{48 (\nc{g}{0\lnsp}{})^3}-\frac{151 \nc{g}{2\lnsp}{2}  \nc{g}{1\lnsp}{}}{432
   (\nc{g}{0\lnsp}{})^3}\nonumber \\&\quad-\frac{19 \nc{g}{2\lnsp}{3}  \nc{g}{1\lnsp}{}}{48 (\nc{g}{0\lnsp}{})^3}-\frac{139 \nc{g}{2\lnsp}{4}  \nc{g}{1\lnsp}{}}{864 (\nc{g}{0\lnsp}{})^3}+\frac{5 \nc{g}{2\lnsp}{5}  \nc{g}{1\lnsp}{}}{48 (\nc{g}{0\lnsp}{})^3}+\frac{49 \nc{g}{2\lnsp}{6}  \nc{g}{1\lnsp}{}}{288 (\nc{g}{0\lnsp}{})^3} +\frac{107
   \nc{g}{2\lnsp}{7}  \nc{g}{1\lnsp}{}}{144 (\nc{g}{0\lnsp}{})^3}+\frac{5 \nc{g}{3\lnsp}{2} }{24 (\nc{g}{0\lnsp}{})^2}+\frac{5 \nc{g}{3\lnsp}{3} }{24 (\nc{g}{0\lnsp}{})^2}\nonumber \\&\quad+\frac{25 \nc{g}{3\lnsp}{4} }{72 (\nc{g}{0\lnsp}{})^2}+\frac{47 \nc{g}{3\lnsp}{5} }{192
   (\nc{g}{0\lnsp}{})^2}+\frac{19 \nc{g}{3\lnsp}{6} }{72 (\nc{g}{0\lnsp}{})^2}-\frac{2 \nc{g}{3\lnsp}{7} }{9 (\nc{g}{0\lnsp}{})^2}+\frac{2 \nc{g}{3\lnsp}{8} }{9 (\nc{g}{0\lnsp}{})^2}+\frac{\nc{g}{3\lnsp}{9} }{36 (\nc{g}{0\lnsp}{})^2}+\frac{5 \nc{g}{3\lnsp}{11} }{64
   (\nc{g}{0\lnsp}{})^2}-\frac{25 \nc{g}{3\lnsp}{13} }{72 (\nc{g}{0\lnsp}{})^2}\nonumber \\&\quad-\frac{5 \nc{g}{3\lnsp}{14} }{24 (\nc{g}{0\lnsp}{})^2}-\frac{23 \nc{g}{3\lnsp}{15} }{72 (\nc{g}{0\lnsp}{})^2}-\frac{19 \nc{g}{3\lnsp}{16} }{72 (\nc{g}{0\lnsp}{})^2}-\frac{31 \nc{g}{3\lnsp}{17} }{96
   (\nc{g}{0\lnsp}{})^2}-\frac{11 \nc{g}{3\lnsp}{18} }{12 (\nc{g}{0\lnsp}{})^2}+\frac{25 \nc{g}{3\lnsp}{1} }{72 (\nc{g}{0\lnsp}{})^2}\bigg) +N^4\lambda_{iijj}\bigg(\frac{53  (\nc{g}{1\lnsp}{})^3}{10368 (\nc{g}{0\lnsp}{})^4}\nonumber\\&\quad-\frac{11 \nc{g}{2\lnsp}{1}  \nc{g}{1\lnsp}{}}{96 (\nc{g}{0\lnsp}{})^3}-\frac{31 \nc{g}{2\lnsp}{2}  \nc{g}{1\lnsp}{}}{864 (\nc{g}{0\lnsp}{})^3}-\frac{11 \nc{g}{2\lnsp}{3}  \nc{g}{1\lnsp}{}}{96
   (\nc{g}{0\lnsp}{})^3}-\frac{7 \nc{g}{2\lnsp}{4}  \nc{g}{1\lnsp}{}}{288 (\nc{g}{0\lnsp}{})^3}+\frac{\nc{g}{2\lnsp}{5}  \nc{g}{1\lnsp}{}}{96 (\nc{g}{0\lnsp}{})^3}+\frac{11 \nc{g}{2\lnsp}{6}  \nc{g}{1\lnsp}{}}{432 (\nc{g}{0\lnsp}{})^3} +\frac{55 \nc{g}{2\lnsp}{7}  \nc{g}{1\lnsp}{}}{288
   (\nc{g}{0\lnsp}{})^3}\nonumber \\&\quad+\frac{\nc{g}{3\lnsp}{2} }{48 (\nc{g}{0\lnsp}{})^2}+\frac{\nc{g}{3\lnsp}{3} }{48 (\nc{g}{0\lnsp}{})^2}+\frac{11 \nc{g}{3\lnsp}{4} }{144 (\nc{g}{0\lnsp}{})^2}+\frac{19 \nc{g}{3\lnsp}{5} }{576 (\nc{g}{0\lnsp}{})^2}+\frac{\nc{g}{3\lnsp}{6} }{36
   (\nc{g}{0\lnsp}{})^2}-\frac{11 \nc{g}{3\lnsp}{7} }{144 (\nc{g}{0\lnsp}{})^2}+\frac{11 \nc{g}{3\lnsp}{8} }{144 (\nc{g}{0\lnsp}{})^2} +\frac{\nc{g}{3\lnsp}{9} }{288 (\nc{g}{0\lnsp}{})^2}\nonumber \\&\quad+\frac{13 \nc{g}{3\lnsp}{11} }{576 (\nc{g}{0\lnsp}{})^2}-\frac{11 \nc{g}{3\lnsp}{13} }{144
   (\nc{g}{0\lnsp}{})^2}-\frac{\nc{g}{3\lnsp}{14} }{48 (\nc{g}{0\lnsp}{})^2}-\frac{5 \nc{g}{3\lnsp}{15} }{144 (\nc{g}{0\lnsp}{})^2}-\frac{\nc{g}{3\lnsp}{16} }{36 (\nc{g}{0\lnsp}{})^2}-\frac{\nc{g}{3\lnsp}{17} }{18 (\nc{g}{0\lnsp}{})^2}-\frac{11 \nc{g}{3\lnsp}{18} }{48
   (\nc{g}{0\lnsp}{})^2}\nonumber \\&\quad+\frac{11 \nc{g}{3\lnsp}{1} }{144 (\nc{g}{0\lnsp}{})^2}\bigg)+N^5\lambda_{iijj}\bigg(\frac{ (\nc{g}{1\lnsp}{})^3}{5184 (\nc{g}{0\lnsp}{})^4}-\frac{\nc{g}{2\lnsp}{1}  \nc{g}{1\lnsp}{}}{96 (\nc{g}{0\lnsp}{})^3}-\frac{\nc{g}{2\lnsp}{3}  \nc{g}{1\lnsp}{}}{96 (\nc{g}{0\lnsp}{})^3} -\frac{\nc{g}{2\lnsp}{4}  \nc{g}{1\lnsp}{}}{864 (\nc{g}{0\lnsp}{})^3}+\frac{\nc{g}{2\lnsp}{6}
   \nc{g}{1\lnsp}{}}{864 (\nc{g}{0\lnsp}{})^3}\nonumber \\&\quad+\frac{5 \nc{g}{2\lnsp}{7}  \nc{g}{1\lnsp}{}}{288 (\nc{g}{0\lnsp}{})^3}+\frac{\nc{g}{3\lnsp}{4} }{144 (\nc{g}{0\lnsp}{})^2} +\frac{\nc{g}{3\lnsp}{5} }{576 (\nc{g}{0\lnsp}{})^2}-\frac{\nc{g}{3\lnsp}{7} }{144 (\nc{g}{0\lnsp}{})^2}+\frac{\nc{g}{3\lnsp}{8}}{144 (\nc{g}{0\lnsp}{})^2}+\frac{\nc{g}{3\lnsp}{11} }{576 (\nc{g}{0\lnsp}{})^2}\nonumber \\&\quad-\frac{\nc{g}{3\lnsp}{13} }{144 (\nc{g}{0\lnsp}{})^2}-\frac{\nc{g}{3\lnsp}{17} }{288 (\nc{g}{0\lnsp}{})^2}-\frac{\nc{g}{3\lnsp}{18} }{48 (\nc{g}{0\lnsp}{})^2}+\frac{\nc{g}{3\lnsp}{1} }{144
   (\nc{g}{0\lnsp}{})^2}\bigg)\,.
    \label{eq:Ricci1}
\end{alignat}

\section{Ricci scalar at \texorpdfstring{$\text{O}(\boldsymbol{\lambda^0)}$}{O(lambda0)}}\label{sec:diagcalc}
To demonstrate the power the diagrammatic notation offers for simplifying tensor expressions, we will use it to compute explicitly the terms appearing in (\ref{eq:Ricci0}). This notation efficiently encodes the relevant information in a given tensor structure, so that contracting indices, either between two tensor structures or within a single tensor structure, becomes a simple exercise in combinatorics. In this notation uncontracted indices are represented by free lines, often bundled into 'sides' representing generalised free indices $I=(ijkl)$. We are to always understand these external sides to be symmetrised over possible permutations, e.g.\ we have that
\begin{equation}
    \begin{tikzpicture}[scale=0.65,
    vertex/.style={draw,circle,fill=black,minimum size=2pt,inner sep=0pt},
    arc/.style={thick},baseline=(vert_cent.base)]
    \node (vert_cent) at (0,-0.75) {$\phantom{\cdot}$};
    \foreach [count=\i] \coord in {
(1.00,0), (-1,0),(-1,-0.5),(1,-0.5)}{
        \node[] (p\i) at \coord {};
    }
    \foreach [count=\i] \coord in {
(-1,-1), (1,-1),(-1,-1.5),(1,-1.5)}{
        \node[] (d\i) at \coord {};
    }
    \draw (d1) edge (d2);
    \draw (d3) edge (d4);
    \draw (p1) edge (p2);
    \draw (p3) edge (p4);
    \node[xshift=-7pt] at (-1,-0.75) {$I$};
    \node[xshift=7pt] at (1,-0.75) {$J$};
\end{tikzpicture}=\frac{1}{24}\left(\begin{tikzpicture}[scale=0.65,
    vertex/.style={draw,circle,fill=black,minimum size=2pt,inner sep=0pt},
    arc/.style={thick},baseline=(vert_cent.base)]
    \node (vert_cent) at (0,-0.75) {$\phantom{\cdot}$};
    \foreach [count=\i] \coord in {
(1.00,0), (-1,0),(-1,-0.5),(1,-0.5)}{
        \node[] (p\i) at \coord {};
    }
    \foreach [count=\i] \coord in {
(-1,-1), (1,-1),(-1,-1.5),(1,-1.5)}{
        \node[] (d\i) at \coord {};
    }
    \draw (d1) edge (d2);
    \draw (d3) edge (d4);
    \draw (p1) edge (p2);
    \draw (p3) edge (p4);
    \node[xshift=7pt] at (1,0) {$m$};
    \node[xshift=7pt] at (1,-0.5) {$n$};
    \node[xshift=7pt] at (1,-1) {$o$};
    \node[xshift=7pt] at (1,-1.5) {$p$};
    \node[xshift=-7pt] at (-1,0) {$i$};
    \node[xshift=-7pt] at (-1,-0.5) {$j$};
    \node[xshift=-7pt] at (-1,-1) {$k$};
    \node[xshift=-7pt] at (-1,-1.5) {$l$};
\end{tikzpicture}+\begin{tikzpicture}[scale=0.65,
    vertex/.style={draw,circle,fill=black,minimum size=2pt,inner sep=0pt},
    arc/.style={thick},baseline=(vert_cent.base)]
    \node (vert_cent) at (0,-0.75) {$\phantom{\cdot}$};
    \foreach [count=\i] \coord in {
(1.00,0), (-1,0),(-1,-0.5),(1,-0.5)}{
        \node[] (p\i) at \coord {};
    }
    \foreach [count=\i] \coord in {
(-1,-1), (1,-1),(-1,-1.5),(1,-1.5)}{
        \node[] (d\i) at \coord {};
    }
    \node (c) at (0,-1.25) {$\phantom{\cdot}$};
    \draw (d1) edge (d4);
    \draw (d3) edge (c);
    \draw (c) edge (d2);
    \draw (p1) edge (p2);
    \draw (p3) edge (p4);
    \node[xshift=7pt] at (1,0) {$m$};
    \node[xshift=7pt] at (1,-0.5) {$n$};
    \node[xshift=7pt] at (1,-1) {$o$};
    \node[xshift=7pt] at (1,-1.5) {$p$};
    \node[xshift=-7pt] at (-1,0) {$i$};
    \node[xshift=-7pt] at (-1,-0.5) {$j$};
    \node[xshift=-7pt] at (-1,-1) {$k$};
    \node[xshift=-7pt] at (-1,-1.5) {$l$};
\end{tikzpicture}+\lsp\cdots\right)\,,
\end{equation}
where the dots indicate the remaining $22$ distinct permutations. When it is obvious or otherwise unimportant which side is associated with which external generalised vertex, we will not include the index explicitly. As we are dealing with generalised indices, the contraction of two free sides with one another involves the connecting of lines in all possible ways, and then dividing by the total number of permutations. We represent this index contraction diagrammatically by leaving a small empty space between the sets of indices being contracted, for instance in the diagram
\begin{equation}
    \begin{tikzpicture}[scale=0.65,
    vertex/.style={draw,circle,fill=black,minimum size=2pt,inner sep=0pt},
    arc/.style={thick},baseline=(vert_cent.base)]
    \node (vert_cent) at (0,-0.75) {$\phantom{\cdot}$};
    \node[vertex] (c) at (0,-0.25) {};
    \foreach [count=\i] \coord in {
(1.00,0), (-1,0),(-1,-0.5),(1,-0.5)}{
        \node[] (p\i) at \coord {};
    }
    \foreach [count=\i] \coord in {
(-1,-1), (1,-1),(-1,-1.5),(1,-1.5)}{
        \node[] (d\i) at \coord {};
    }
    \draw (c) edge (p1)
                   edge (p2)
                   edge (p3)
                   edge (p4);
    \draw (d1) edge (d2);
    \draw (d3) edge (d4);
    \node[vertex] (c2) at (2.25,-0.25) {};
    \foreach [count=\i] \coord in {
(3.25,0), (1.25,0),(1.25,-0.5),(3.25,-0.5)}{
        \node[] (r\i) at \coord {};
    }
    \foreach [count=\i] \coord in {
(1.25,-1), (3.25,-1),(1.25,-1.5),(3.25,-1.5)}{
        \node[] (l\i) at \coord {};
    }
    \draw (c2) edge (r1)
                   edge (r2)
                   edge (r3)
                   edge (r4);
    \draw (l1) edge (l2);
    \draw (l3) edge (l4);
\end{tikzpicture}\,.
\end{equation}
A quick combinatorial argument allows one to conclude that the contraction in this diagram can be evaluated to give
\begin{equation}
    \begin{tikzpicture}[scale=0.65,
    vertex/.style={draw,circle,fill=black,minimum size=2pt,inner sep=0pt},
    arc/.style={thick},baseline=(vert_cent.base)]
    \node (vert_cent) at (0,-0.75) {$\phantom{\cdot}$};
    \node[vertex] (c) at (0,-0.25) {};
    \foreach [count=\i] \coord in {
(1.00,0), (-1,0),(-1,-0.5),(1,-0.5)}{
        \node[] (p\i) at \coord {};
    }
    \foreach [count=\i] \coord in {
(-1,-1), (1,-1),(-1,-1.5),(1,-1.5)}{
        \node[] (d\i) at \coord {};
    }
    \draw (c) edge (p1)
                   edge (p2)
                   edge (p3)
                   edge (p4);
    \draw (d1) edge (d2);
    \draw (d3) edge (d4);
    \node[vertex] (c2) at (2.25,-0.25) {};
    \foreach [count=\i] \coord in {
(3.25,0), (1.25,0),(1.25,-0.5),(3.25,-0.5)}{
        \node[] (r\i) at \coord {};
    }
    \foreach [count=\i] \coord in {
(1.25,-1), (3.25,-1),(1.25,-1.5),(3.25,-1.5)}{
        \node[] (l\i) at \coord {};
    }
    \draw (c2) edge (r1)
                   edge (r2)
                   edge (r3)
                   edge (r4);
    \draw (l1) edge (l2);
    \draw (l3) edge (l4);
\end{tikzpicture}=\frac{4}{24}\lsp\begin{tikzpicture}[scale=0.65,
    vertex/.style={draw,circle,fill=black,minimum size=2pt,inner sep=0pt},
    arc/.style={thick},baseline=(vert_cent.base)]
    \node (vert_cent) at (0,-0.75) {$\phantom{\cdot}$};
    \node[vertex] (c1) at (-0.25,-0.25) {};
    \node[vertex] (c2) at (0.25,-0.25) {};
    \foreach [count=\i] \coord in {(-1,0),(-1,-0.5),(-1,-1)}{
        \node[] (p\i) at \coord {};
    }
    \foreach [count=\i] \coord in {(1,0),(1,-0.5),(1,-1)}{
        \node[] (d\i) at \coord {};
    }
    \node[] (d4) at (1,-1.5) {};
    \node[] (p4) at (-1,-1.5) {};
    \draw (c1) edge (p1)
                   edge (p2)
                   edge[bend right=60] (c2)
                   edge[bend left=60] (c2);
    \draw (c2) edge (d1)
                   edge (d2);
    \draw (p3) edge (d3);
    \draw (p4) edge (d4);
\end{tikzpicture}+\frac{4}{24}\lsp\begin{tikzpicture}[scale=0.65,
    vertex/.style={draw,circle,fill=black,minimum size=2pt,inner sep=0pt},
    arc/.style={thick},baseline=(vert_cent.base)]
    \node (vert_cent) at (0,-0.75) {$\phantom{\cdot}$};
    \node[vertex] (c1) at (0,-0.25) {};
    \node[vertex] (c2) at (0,-1.25) {};
    \foreach [count=\i] \coord in {
(1.00,0), (-1,0),(-1,-0.5),(1,-0.5)}{
        \node[] (p\i) at \coord {};
    }
    \foreach [count=\i] \coord in {
(-1,-1), (1,-1),(-1,-1.5),(1,-1.5)}{
        \node[] (d\i) at \coord {};
    }
    \draw (c1) edge (p1)
                   edge (p2)
                   edge (p3)
                   edge (p4);
    \draw (c2) edge (d1)
                   edge (d2)
                   edge (d3)
                   edge (d4);
\end{tikzpicture}+\frac{16}{24}\begin{tikzpicture}[scale=0.65,
    vertex/.style={draw,circle,fill=black,minimum size=2pt,inner sep=0pt},
    arc/.style={thick},baseline=(vert_cent.base)]
    \node (vert_cent) at (0,-0.75) {$\phantom{\cdot}$};
    \node[vertex] (c1) at (0,0) {};
    \node[vertex] (c2) at (0,-1) {};
    \foreach [count=\i] \coord in {(-1,0),(-1,-0.5),(-1,-1)}{
        \node[] (p\i) at \coord {};
    }
    \foreach [count=\i] \coord in {(1,0),(1,-0.5),(1,-1)}{
        \node[] (d\i) at \coord {};
    }
    \node[] (d4) at (1,-1.5) {};
    \node[] (p4) at (-1,-1.5) {};
    \draw (c1) edge (p1)
                   edge (p2)
                   edge (d1)
                   edge (c2);
    \draw (c2) edge (p3)
                   edge (d2)
                   edge (d3);
    \draw (p4) edge (d4);
\end{tikzpicture}\lsp\,,
\end{equation}
where the free generalised indices on either side of the individual diagrams are carried through from the left-hand side to the diagrams right-hand side.

As the Ricci scalar involves at most two derivatives acting on the metric, to compute $R^{0}$ we will need an expression for $G_{IJ}$ through $\text{O}(\lambda^2)$. Demanding that the expansion for $G_{IJ}$ begins like the trivial metric $\delta_{IJ}$, the most generic form that the metric can take is
\begin{equation}
\begin{split}
G_{IJ}=&\nc{g}{0\lnsp}{}\lsp\begin{tikzpicture}[scale=0.65,
    vertex/.style={draw,circle,fill=black,minimum size=2pt,inner sep=0pt},
    arc/.style={thick},baseline=(vert_cent.base)]
    \node (vert_cent) at (0,-0.75) {$\phantom{\cdot}$};
    \foreach [count=\i] \coord in {
(1.00,0), (-1,0),(-1,-0.5),(1,-0.5)}{
        \node[] (p\i) at \coord {};
    }
    \foreach [count=\i] \coord in {
(-1,-1), (1,-1),(-1,-1.5),(1,-1.5)}{
        \node[] (d\i) at \coord {};
    }
    \draw (d1) edge (d2);
    \draw (d3) edge (d4);
    \draw (p1) edge (p2);
    \draw (p3) edge (p4);
\end{tikzpicture}+\nc{g}{1\lnsp}{}\lsp\begin{tikzpicture}[scale=0.65,
    vertex/.style={draw,circle,fill=black,minimum size=2pt,inner sep=0pt},
    arc/.style={thick},baseline=(vert_cent.base)]
    \node (vert_cent) at (0,-0.75) {$\phantom{\cdot}$};
    \node[vertex] (c) at (0,-0.25) {};
    \foreach [count=\i] \coord in {
(1.00,0), (-1,0),(-1,-0.5),(1,-0.5)}{
        \node[] (p\i) at \coord {};
    }
    \foreach [count=\i] \coord in {
(-1,-1), (1,-1),(-1,-1.5),(1,-1.5)}{
        \node[] (d\i) at \coord {};
    }
    \draw (c) edge (p1)
                   edge (p2)
                   edge (p3)
                   edge (p4);
    \draw (d1) edge (d2);
    \draw (d3) edge (d4);
\end{tikzpicture}+\nc{g}{2\lnsp}{1}\lsp
    \begin{tikzpicture}[scale=0.65,
    vertex/.style={draw,circle,fill=black,minimum size=2pt,inner sep=0pt},
    arc/.style={thick},baseline=(vert_cent.base)]
    \node (vert_cent) at (0,-0.75) {$\phantom{\cdot}$};
    \node[vertex] (c1) at (0,-0.25) {};
    \node[vertex] (c2) at (0,-1.25) {};
    \foreach [count=\i] \coord in {
(1.00,0), (-1,0),(-1,-0.5),(1,-0.5)}{
        \node[] (p\i) at \coord {};
    }
    \foreach [count=\i] \coord in {
(-1,-1), (1,-1),(-1,-1.5),(1,-1.5)}{
        \node[] (d\i) at \coord {};
    }
    \draw (c1) edge (p1)
                   edge (p2)
                   edge (p3)
                   edge (d1);
    \draw (c2) edge (p4)
                   edge (d2)
                   edge (d3)
                   edge (d4);
\end{tikzpicture}+\nc{g}{2\lnsp}{2}\lsp
    \begin{tikzpicture}[scale=0.65,
    vertex/.style={draw,circle,fill=black,minimum size=2pt,inner sep=0pt},
    arc/.style={thick},baseline=(vert_cent.base)]
    \node (vert_cent) at (0,-0.75) {$\phantom{\cdot}$};
    \node[vertex] (c1) at (0,-0.25) {};
    \node[vertex] (c2) at (0,-1.25) {};
    \foreach [count=\i] \coord in {
(1.00,0), (-1,0),(-1,-0.5),(1,-0.5)}{
        \node[] (p\i) at \coord {};
    }
    \foreach [count=\i] \coord in {
(-1,-1), (1,-1),(-1,-1.5),(1,-1.5)}{
        \node[] (d\i) at \coord {};
    }
    \draw (c1) edge (p1)
                   edge (p2)
                   edge (p3)
                   edge (p4);
    \draw (c2) edge (d1)
                   edge (d2)
                   edge (d3)
                   edge (d4);
\end{tikzpicture}+\nc{g}{2\lnsp}{3}\lsp
    \begin{tikzpicture}[scale=0.65,
    vertex/.style={draw,circle,fill=black,minimum size=2pt,inner sep=0pt},
    arc/.style={thick},baseline=(vert_cent.base)]
    \node (vert_cent) at (0,-0.75) {$\phantom{\cdot}$};
    \node[vertex] (c1) at (-0.25,-0.5) {};
    \node[vertex] (c2) at (0.25,-0.5) {};
    \foreach [count=\i] \coord in {(-1,0),(-1,-0.5),(-1,-1)}{
        \node[] (p\i) at \coord {};
    }
    \foreach [count=\i] \coord in {(1,0),(1,-0.5),(1,-1)}{
        \node[] (d\i) at \coord {};
    }
        \node[] (d4) at (1,-1.5) {};
    \node[] (p4) at (-1,-1.5) {};
    \draw (c1) edge (p1)
                   edge (p2)
                   edge (p3)
                   edge (c2);
    \draw (c2) edge (d1)
                   edge (d2)
                   edge (d3);
    \draw (p4) edge (d4);
\end{tikzpicture}\\
    &+\nc{g}{2\lnsp}{4}\lsp\begin{tikzpicture}[scale=0.65,
    vertex/.style={draw,circle,fill=black,minimum size=2pt,inner sep=0pt},
    arc/.style={thick},baseline=(vert_cent.base)]
    \node (vert_cent) at (0,-0.75) {$\phantom{\cdot}$};
    \node[vertex] (c1) at (0,0) {};
    \node[vertex] (c2) at (0,-1) {};
    \foreach [count=\i] \coord in {(-1,0),(-1,-0.5),(-1,-1)}{
        \node[] (p\i) at \coord {};
    }
    \foreach [count=\i] \coord in {(1,0),(1,-0.5),(1,-1)}{
        \node[] (d\i) at \coord {};
    }
    \node[] (d4) at (1,-1.5) {};
    \node[] (p4) at (-1,-1.5) {};
    \draw (c1) edge (p1)
                   edge (p2)
                   edge (d1)
                   edge (c2);
    \draw (c2) edge (p3)
                   edge (d2)
                   edge (d3);
    \draw (p4) edge (d4);
\end{tikzpicture}+\nc{g}{2\lnsp}{5}\lsp
    \begin{tikzpicture}[scale=0.65,
    vertex/.style={draw,circle,fill=black,minimum size=2pt,inner sep=0pt},
    arc/.style={thick},baseline=(vert_cent.base)]
    \node (vert_cent) at (0,-0.75) {$\phantom{\cdot}$};
    \node[vertex] (c1) at (-0.25,-0.25) {};
    \node[vertex] (c2) at (0.25,-0.25) {};
    \foreach [count=\i] \coord in {(-1,0),(-1,-0.5),(-1,-1)}{
        \node[] (p\i) at \coord {};
    }
    \foreach [count=\i] \coord in {(1,0),(1,-0.5),(1,-1)}{
        \node[] (d\i) at \coord {};
    }
    \node[] (d4) at (1,-1.5) {};
    \node[] (p4) at (-1,-1.5) {};
    \draw (c1) edge (p1)
                   edge (p2)
                   edge[bend right=60] (c2)
                   edge[bend left=60] (c2);
    \draw (c2) edge (d1)
                   edge (d2);
    \draw (p3) edge (d3);
    \draw (p4) edge (d4);
\end{tikzpicture}+\nc{g}{2\lnsp}{6}\lsp
    \begin{tikzpicture}[scale=0.65,
    vertex/.style={draw,circle,fill=black,minimum size=2pt,inner sep=0pt},
    arc/.style={thick},baseline=(vert_cent.base)]
    \node (vert_cent) at (0,-0.75) {$\phantom{\cdot}$};
    \node[vertex] (c1) at (0,0) {};
    \node[vertex] (c2) at (0,-0.5) {};
    \foreach [count=\i] \coord in {(-1,0),(-1,-0.5),(-1,-1)}{
        \node[] (p\i) at \coord {};
    }
    \foreach [count=\i] \coord in {(1,0),(1,-0.5),(1,-1)}{
        \node[] (d\i) at \coord {};
    }
    \node[] (d4) at (1,-1.5) {};
    \node[] (p4) at (-1,-1.5) {};
    \draw (c1) edge (p1)
                   edge (d1)
                   edge[bend right=60] (c2)
                   edge[bend left=60] (c2);
    \draw (c2) edge (p2)
                   edge (d2);
    \draw (p3) edge (d3);
    \draw (p4) edge (d4);
\end{tikzpicture}+\nc{g}{2\lnsp}{7}\lsp
    \begin{tikzpicture}[scale=0.65,
    vertex/.style={draw,circle,fill=black,minimum size=2pt,inner sep=0pt},
    arc/.style={thick},baseline=(vert_cent.base)]
    \node (vert_cent) at (0,-0.75) {$\phantom{\cdot}$};
    \node[vertex] (c1) at (-0.5,0) {};
    \node[vertex] (c2) at (0.5,0) {};
    \foreach [count=\i] \coord in {(-1,0),(-1,-0.5),(-1,-1)}{
        \node[] (p\i) at \coord {};
    }
    \foreach [count=\i] \coord in {(1,0),(1,-0.5),(1,-1)}{
        \node[] (d\i) at \coord {};
    }
    \node[] (d4) at (1,-1.5) {};
    \node[] (p4) at (-1,-1.5) {};
    \draw (c1) edge (p1)
                   edge (c2)
                   edge[bend right=60] (c2)
                   edge[bend left=60] (c2);
    \draw (c2) edge (d1);
    \draw (p2) edge (d2);
    \draw (p3) edge (d3);
    \draw (p4) edge (d4);
\end{tikzpicture}+\text{O}(\lambda^3)\,,
\end{split}
\label{eq:metricl2}
\end{equation}
where in this analysis we will not demand that $G_{IJ}$ is related to a beta function and keep the various coefficients unfixed. There will exist an analogous expansion for the inverse metric,
\begin{equation}
\begin{split}
G^{IJ}=&\nc{\mathfrak{g}}{0}{}\lsp\begin{tikzpicture}[scale=0.65,
    vertex/.style={draw,circle,fill=black,minimum size=2pt,inner sep=0pt},
    arc/.style={thick},baseline=(vert_cent.base)]
    \node (vert_cent) at (0,-0.75) {$\phantom{\cdot}$};
    \foreach [count=\i] \coord in {
(1.00,0), (-1,0),(-1,-0.5),(1,-0.5)}{
        \node[] (p\i) at \coord {};
    }
    \foreach [count=\i] \coord in {
(-1,-1), (1,-1),(-1,-1.5),(1,-1.5)}{
        \node[] (d\i) at \coord {};
    }
    \draw (d1) edge (d2);
    \draw (d3) edge (d4);
    \draw (p1) edge (p2);
    \draw (p3) edge (p4);
\end{tikzpicture}+\nc{\mathfrak{g}}{1}{}\lsp\begin{tikzpicture}[scale=0.65,
    vertex/.style={draw,circle,fill=black,minimum size=2pt,inner sep=0pt},
    arc/.style={thick},baseline=(vert_cent.base)]
    \node (vert_cent) at (0,-0.75) {$\phantom{\cdot}$};
    \node[vertex] (c) at (0,-0.25) {};
    \foreach [count=\i] \coord in {
(1.00,0), (-1,0),(-1,-0.5),(1,-0.5)}{
        \node[] (p\i) at \coord {};
    }
    \foreach [count=\i] \coord in {
(-1,-1), (1,-1),(-1,-1.5),(1,-1.5)}{
        \node[] (d\i) at \coord {};
    }
    \draw (c) edge (p1)
                   edge (p2)
                   edge (p3)
                   edge (p4);
    \draw (d1) edge (d2);
    \draw (d3) edge (d4);
\end{tikzpicture}+\nc{\mathfrak{g}}{2}{1}\lsp
    \begin{tikzpicture}[scale=0.65,
    vertex/.style={draw,circle,fill=black,minimum size=2pt,inner sep=0pt},
    arc/.style={thick},baseline=(vert_cent.base)]
    \node (vert_cent) at (0,-0.75) {$\phantom{\cdot}$};
    \node[vertex] (c1) at (0,-0.25) {};
    \node[vertex] (c2) at (0,-1.25) {};
    \foreach [count=\i] \coord in {
(1.00,0), (-1,0),(-1,-0.5),(1,-0.5)}{
        \node[] (p\i) at \coord {};
    }
    \foreach [count=\i] \coord in {
(-1,-1), (1,-1),(-1,-1.5),(1,-1.5)}{
        \node[] (d\i) at \coord {};
    }
    \draw (c1) edge (p1)
                   edge (p2)
                   edge (p3)
                   edge (d1);
    \draw (c2) edge (p4)
                   edge (d2)
                   edge (d3)
                   edge (d4);
\end{tikzpicture}+\nc{\mathfrak{g}}{2}{2}\lsp
    \begin{tikzpicture}[scale=0.65,
    vertex/.style={draw,circle,fill=black,minimum size=2pt,inner sep=0pt},
    arc/.style={thick},baseline=(vert_cent.base)]
    \node (vert_cent) at (0,-0.75) {$\phantom{\cdot}$};
    \node[vertex] (c1) at (0,-0.25) {};
    \node[vertex] (c2) at (0,-1.25) {};
    \foreach [count=\i] \coord in {
(1.00,0), (-1,0),(-1,-0.5),(1,-0.5)}{
        \node[] (p\i) at \coord {};
    }
    \foreach [count=\i] \coord in {
(-1,-1), (1,-1),(-1,-1.5),(1,-1.5)}{
        \node[] (d\i) at \coord {};
    }
    \draw (c1) edge (p1)
                   edge (p2)
                   edge (p3)
                   edge (p4);
    \draw (c2) edge (d1)
                   edge (d2)
                   edge (d3)
                   edge (d4);
\end{tikzpicture}+\nc{\mathfrak{g}}{2}{3}\lsp
    \begin{tikzpicture}[scale=0.65,
    vertex/.style={draw,circle,fill=black,minimum size=2pt,inner sep=0pt},
    arc/.style={thick},baseline=(vert_cent.base)]
    \node (vert_cent) at (0,-0.75) {$\phantom{\cdot}$};
    \node[vertex] (c1) at (-0.25,-0.5) {};
    \node[vertex] (c2) at (0.25,-0.5) {};
    \foreach [count=\i] \coord in {(-1,0),(-1,-0.5),(-1,-1)}{
        \node[] (p\i) at \coord {};
    }
    \foreach [count=\i] \coord in {(1,0),(1,-0.5),(1,-1)}{
        \node[] (d\i) at \coord {};
    }
        \node[] (d4) at (1,-1.5) {};
    \node[] (p4) at (-1,-1.5) {};
    \draw (c1) edge (p1)
                   edge (p2)
                   edge (p3)
                   edge (c2);
    \draw (c2) edge (d1)
                   edge (d2)
                   edge (d3);
    \draw (p4) edge (d4);
\end{tikzpicture}\\
    &+\nc{\mathfrak{g}}{2}{4}\lsp\begin{tikzpicture}[scale=0.65,
    vertex/.style={draw,circle,fill=black,minimum size=2pt,inner sep=0pt},
    arc/.style={thick},baseline=(vert_cent.base)]
    \node (vert_cent) at (0,-0.75) {$\phantom{\cdot}$};
    \node[vertex] (c1) at (0,0) {};
    \node[vertex] (c2) at (0,-1) {};
    \foreach [count=\i] \coord in {(-1,0),(-1,-0.5),(-1,-1)}{
        \node[] (p\i) at \coord {};
    }
    \foreach [count=\i] \coord in {(1,0),(1,-0.5),(1,-1)}{
        \node[] (d\i) at \coord {};
    }
    \node[] (d4) at (1,-1.5) {};
    \node[] (p4) at (-1,-1.5) {};
    \draw (c1) edge (p1)
                   edge (p2)
                   edge (d1)
                   edge (c2);
    \draw (c2) edge (p3)
                   edge (d2)
                   edge (d3);
    \draw (p4) edge (d4);
\end{tikzpicture}+\nc{\mathfrak{g}}{2}{5}\lsp
    \begin{tikzpicture}[scale=0.65,
    vertex/.style={draw,circle,fill=black,minimum size=2pt,inner sep=0pt},
    arc/.style={thick},baseline=(vert_cent.base)]
    \node (vert_cent) at (0,-0.75) {$\phantom{\cdot}$};
    \node[vertex] (c1) at (-0.25,-0.25) {};
    \node[vertex] (c2) at (0.25,-0.25) {};
    \foreach [count=\i] \coord in {(-1,0),(-1,-0.5),(-1,-1)}{
        \node[] (p\i) at \coord {};
    }
    \foreach [count=\i] \coord in {(1,0),(1,-0.5),(1,-1)}{
        \node[] (d\i) at \coord {};
    }
    \node[] (d4) at (1,-1.5) {};
    \node[] (p4) at (-1,-1.5) {};
    \draw (c1) edge (p1)
                   edge (p2)
                   edge[bend right=60] (c2)
                   edge[bend left=60] (c2);
    \draw (c2) edge (d1)
                   edge (d2);
    \draw (p3) edge (d3);
    \draw (p4) edge (d4);
\end{tikzpicture}+\nc{\mathfrak{g}}{2}{6}\lsp
    \begin{tikzpicture}[scale=0.65,
    vertex/.style={draw,circle,fill=black,minimum size=2pt,inner sep=0pt},
    arc/.style={thick},baseline=(vert_cent.base)]
    \node (vert_cent) at (0,-0.75) {$\phantom{\cdot}$};
    \node[vertex] (c1) at (0,0) {};
    \node[vertex] (c2) at (0,-0.5) {};
    \foreach [count=\i] \coord in {(-1,0),(-1,-0.5),(-1,-1)}{
        \node[] (p\i) at \coord {};
    }
    \foreach [count=\i] \coord in {(1,0),(1,-0.5),(1,-1)}{
        \node[] (d\i) at \coord {};
    }
    \node[] (d4) at (1,-1.5) {};
    \node[] (p4) at (-1,-1.5) {};
    \draw (c1) edge (p1)
                   edge (d1)
                   edge[bend right=60] (c2)
                   edge[bend left=60] (c2);
    \draw (c2) edge (p2)
                   edge (d2);
    \draw (p3) edge (d3);
    \draw (p4) edge (d4);
\end{tikzpicture}+\nc{\mathfrak{g}}{2}{7}\lsp
    \begin{tikzpicture}[scale=0.65,
    vertex/.style={draw,circle,fill=black,minimum size=2pt,inner sep=0pt},
    arc/.style={thick},baseline=(vert_cent.base)]
    \node (vert_cent) at (0,-0.75) {$\phantom{\cdot}$};
    \node[vertex] (c1) at (-0.5,0) {};
    \node[vertex] (c2) at (0.5,0) {};
    \foreach [count=\i] \coord in {(-1,0),(-1,-0.5),(-1,-1)}{
        \node[] (p\i) at \coord {};
    }
    \foreach [count=\i] \coord in {(1,0),(1,-0.5),(1,-1)}{
        \node[] (d\i) at \coord {};
    }
    \node[] (d4) at (1,-1.5) {};
    \node[] (p4) at (-1,-1.5) {};
    \draw (c1) edge (p1)
                   edge (c2)
                   edge[bend right=60] (c2)
                   edge[bend left=60] (c2);
    \draw (c2) edge (d1);
    \draw (p2) edge (d2);
    \draw (p3) edge (d3);
    \draw (p4) edge (d4);
\end{tikzpicture}+\text{O}(\lambda^3)\,,
\end{split}
\end{equation}
whose coefficients can be determined by demanding that $G_{IK}G^{KJ}=\delta_I{\!}^J$,
\begin{equation}
\begin{split}
    \nc{\mathfrak{g}}{0}{}&=\frac{1}{\nc{g}{0\lnsp}{}}\,,\quad\nc{\mathfrak{g}}{1}{}=-\frac{\nc{g}{1\lnsp}{}}{(\nc{g}{0\lnsp}{})^2}\,,\quad\nc{\mathfrak{g}}{2}{1}=-\frac{\nc{g}{2\lnsp}{1}}{(\nc{g}{0\lnsp}{})^2}\,,\\\nc{\mathfrak{g}}{2}{2}&=\frac{\left((\nc{g}{1\lnsp}{})^2-6 \nc{g}{2\lnsp}{2}\right)}{6(\nc{g}{0\lnsp}{})^2} \,,\quad\nc{\mathfrak{g}}{2}{3}=-\frac{\nc{g}{2\lnsp}{3}}{(\nc{g}{0\lnsp}{})^2}\,, \quad\nc{\mathfrak{g}}{2}{4}=\frac{\left(2 (\nc{g}{1\lnsp}{})^2-3 \nc{g}{2\lnsp}{4}\right)}{3(\nc{g}{0\lnsp}{})^2} \,, \\ \nc{\mathfrak{g}}{2}{5}&=\frac{\left((\nc{g}{1\lnsp}{})^2-6
   \nc{g}{2\lnsp}{5}\right)}{6(\nc{g}{0\lnsp}{})^2} \,,\quad\nc{\mathfrak{g}}{2}{6}=-\frac{\nc{g}{2\lnsp}{6}}{(\nc{g}{0\lnsp}{})^2}\,,\quad\nc{\mathfrak{g}}{2}{7}=-\frac{\nc{g}{2\lnsp}{7}}{(\nc{g}{0\lnsp}{})^2}\,.
\end{split}
\label{eq:MettoInvMet}
\end{equation}
Written in terms of the Christoffel symbols, the Ricci scalar is given by
\begin{equation}
    R=G^{IJ}\left(\Gamma^{K}_{IK}\Gamma^{L}_{JL}-\Gamma^{K}_{LI}\Gamma^{L}_{KJ}+\partial_K\Gamma^{K}_{IJ}-\partial_I\Gamma^{K}_{JK}\right)\,,
\end{equation}
which can be written as an expansion in the interaction tensor
\begin{equation}
    R=\sum_{n=0}^\infty\sum_m \nc{R}{n\lnsp}{m}\lsp\nc{T}{n}{m}
\end{equation}
where $m$ indexes a sum over the distinct vacuum bubbles with $n$ vertices. To extract the $\text{O}(\lambda^0)$ term $\nc{R}{0\lnsp}{}$ we can set the overall factor of the inverse metric to be $\nc{\mathfrak{g}}{0}{}\delta^{IJ}$, and then include the Christoffel symbols only through $\text{O}(\lambda)$:
\begin{equation}
\nc{R}{0\lnsp}{}=\nc{\mathfrak{g}}{0}{}\lsp\delta^{IJ}\left({\nc{\Gamma}{0}{}}^{K}_{IK}{\nc{\Gamma}{0}{}}^{L}_{JL}-{\nc{\Gamma}{0}{}}^{K}_{LI}{\nc{\Gamma}{0}{}}^{L}_{KJ}+\partial_K{\nc{\Gamma}{1}{}}^{K}_{IJ}-\partial_I{\nc{\Gamma}{1}{}}^{K}_{JK}\right)\,.
\label{R0genexp}
\end{equation}
We will consider these terms individually, as the different methods of contracting indices will give rise to distinct diagrams. Using the expression for the Christoffel symbols in terms of the metric and inverse metric, we see that the lowest order symbols must be given by
\begin{equation}
    {\nc{\Gamma}{0}{}}^{I}_{JK}=\frac{\nc{\mathfrak{g}}{0}{}}{2}\delta^{IL}\left(\partial_J (\nc{G}{1}{})_{KL}+\partial_{K}(\nc{G}{1}{})_{JL}-\partial_L (\nc{G}{1}{})_{JK}\right)\,.
\end{equation}
This can be represented most straightforwardly diagrammatically, where one appreciates that the symmetry of $G_{IJ}$ with respect to exchange of $I$ and $J$ and the underlying symmetrisation of the generalised indices $I$, $J$ and $K$ forces the three terms in the parentheses to be equal
\begin{equation}
    {\nc{\Gamma}{0}{}}^{I}_{JK}=\frac{\nc{\mathfrak{g}}{0}{}\nc{g}{1\lnsp}{}}{2}\,
    \begin{tikzpicture}[scale=0.65,
    vertex/.style={draw,circle,fill=black,minimum size=2pt,inner sep=0pt},
    arc/.style={thick},baseline=(vert_cent.base)]
    \node (vert_cent) at (0,0.26) {$\phantom{\cdot}$};
    \foreach [count=\i] \coord in {
(-0.75,1), (-0.25,1),(0.25,1),(0.75,1)}{
        \node[inner sep=0pt] (t\i) at \coord {};
    }
    \foreach [count=\i] \coord in {
(1.375,0.375), (1.125,0.125),(0.875,-0.125),(0.521,-0.479)}{
        \node[inner sep=0pt] (r\i) at \coord {};
    }
    \foreach [count=\i] \coord in {
(-1.375,0.375), (-1.125,0.125),(-0.875,-0.125),(-0.521,-0.479)}{
        \node[inner sep=0pt] (l\i) at \coord {};
    }
    \draw (l1) edge (t1);
    \draw (l2) edge (t2);
    \draw (r1) edge (t4);
    \draw (r2) edge (t3);
    \draw (r3) edge (l3);
    \draw (r4) edge (l4);
    \node[xshift=-7pt,yshift=-7pt] at (-1,0) {$I$};
    \node[xshift=7pt,yshift=-7pt] at (1,0) {$K$};
    \node[yshift=7pt] at (0,1) {$J$};
\end{tikzpicture}\,.
\end{equation}
Tracing over an upper and a lower index then yields
\begin{equation}
{\nc{\Gamma}{0}{}}^{I}_{JI}=\frac{\nc{\mathfrak{g}}{0}{}\nc{g}{1\lnsp}{}}{2}\,
    \begin{tikzpicture}[scale=0.65,
    vertex/.style={draw,circle,fill=black,minimum size=2pt,inner sep=0pt},
    arc/.style={thick},baseline=(vert_cent.base)]
    \node (vert_cent) at (0,0) {$\phantom{\cdot}$};
    \foreach [count=\i] \coord in {
(-0.75,1), (-0.25,1),(0.25,1),(0.75,1)}{
        \node[fill=white] (t\i) at \coord {$\phantom{\cdot}$};
    }
    \foreach [count=\i] \coord in {
(0,0)}{
        \node[fill=white] (b\i) at \coord {$\phantom{\cdot}$};
    }
    \draw (t4) to[out=270, in=0] (b1);
    \draw (t3) to[out=270, in=30] (b1);
    \draw (t1) to[out=270, in=180] (b1);
    \draw (t2) to[out=270, in=150] (b1);
    \draw (0,-0.87) circle (0.8cm);
    \draw (0,-0.7) circle (0.5cm);
    \node[yshift=7pt] at (0,1) {$J$};
    \node[fill=white] at (0,0) {$\phantom{\cdot}$};
\end{tikzpicture}\,,
\end{equation}
an expression that may be simplified with the help of a contraction rule
\begin{equation}
    \begin{tikzpicture}[scale=0.5,
    vertex/.style={draw,circle,fill=black,minimum size=2pt,inner sep=0pt},
    arc/.style={thick},baseline=(vert_cent.base)]
    \node (vert_cent) at (0,0) {$\phantom{\cdot}$};
    \foreach [count=\i] \coord in {
(-0.75,0.5), (-0.25,0.5),(0.25,0.5),(0.75,0.5)}{
        \node[inner sep=0pt] (t\i) at \coord {};
    }
    \foreach [count=\i] \coord in {
(-0.75,1.5), (-0.25,1.5),(0.25,1.5),(0.75,1.5)}{
        \node[inner sep=0pt] (tt\i) at \coord {};
    }
    \foreach [count=\i] \coord in {
(-0.75,-0.5), (-0.25,-0.5),(0.25,-0.5),(0.75,-0.5)}{
        \node[inner sep=0pt] (b\i) at \coord {};
    }
    \foreach [count=\i] \coord in {
(-0.75,-1.5), (-0.25,-1.5),(0.25,-1.5),(0.75,-1.5)}{
        \node[inner sep=0pt] (bb\i) at \coord {};
    }
    \draw (tt1) edge (t1);
    \draw (tt2) edge (t2);
    \draw (bb1) edge (b1);
    \draw (bb2) edge (b2);
    \draw (1.25,0) circle (1.25cm);
    \draw (1.25,0) circle (0.75cm);
    \node[fill=white] at (0.25,0) {$\phantom{\cdot}$};
    \node[fill=white] at (0.75,0) {$\phantom{\cdot}$};
\end{tikzpicture}=\frac{(N+3)(N+2)}{24}\,\begin{tikzpicture}[scale=0.5,
    vertex/.style={draw,circle,fill=black,minimum size=2pt,inner sep=0pt},
    arc/.style={thick},baseline=(vert_cent.base)]
    \node (vert_cent) at (0,0) {$\phantom{\cdot}$};
    \foreach [count=\i] \coord in {
(-0.75,1.5), (-0.25,1.5)}{
        \node[inner sep=0pt] (tt\i) at \coord {};
    }
    \foreach [count=\i] \coord in {
(-0.75,-1.5), (-0.25,-1.5)}{
        \node[inner sep=0pt] (bb\i) at \coord {};
    }
    \draw (tt1) edge (bb1);
    \draw (tt2) edge (bb2);
\end{tikzpicture}+\frac{(N+3)(N+2)}{24}\,\begin{tikzpicture}[scale=0.5,
    vertex/.style={draw,circle,fill=black,minimum size=2pt,inner sep=0pt},
    arc/.style={thick},baseline=(vert_cent.base)]
    \node (vert_cent) at (0,0) {$\phantom{\cdot}$};
    \foreach [count=\i] \coord in {
(-0.75,1.5), (-0.25,1.5)}{
        \node[inner sep=0pt] (tt\i) at \coord {};
    }
    \foreach [count=\i] \coord in {
(-0.75,-1.5), (-0.25,-1.5)}{
        \node[inner sep=0pt] (bb\i) at \coord {};
    }
    \node[] (m) at (-0.5,0) {$\phantom{\cdot}$};
    \draw (tt1) edge (bb2);
    \draw (tt2) edge (m);
    \draw (m) edge (bb1);
\end{tikzpicture}
\label{eq:contrule2bubbles}
\end{equation}
to yield
\begin{equation}
{\nc{\Gamma}{0}{}}^{I}_{JI}=\frac{\nc{\mathfrak{g}}{0}{}\nc{g}{1\lnsp}{}(N+3)(N+2)}{24}\,\,
    \begin{tikzpicture}[scale=0.65,
    vertex/.style={draw,circle,fill=black,minimum size=2pt,inner sep=0pt},
    arc/.style={thick},baseline=(vert_cent.base)]
    \node (vert_cent) at (0,0.8) {$\phantom{\cdot}$};
    \foreach [count=\i] \coord in {
(-0.75,1), (-0.25,1),(0.25,1),(0.75,1)}{
        \node[inner sep=0pt] (t\i) at \coord {};
    }
    \draw (t4) to[out=270, in=270,looseness=4] (t3);
    \draw (t1) to[out=270, in=270,looseness=4] (t2);
    \node[yshift=7pt] at (0,1) {$J$};
\end{tikzpicture}\,.
\end{equation}
One may then straightforwardly evaluate the first term in (\ref{R0genexp}) to be
\begin{equation}
    \nc{\mathfrak{g}}{0}{}\lsp\delta^{IJ}{\nc{\Gamma}{0}{}}^{K}_{IK}{\nc{\Gamma}{0}{}}^{L}_{JL}=\frac{(\nc{\mathfrak{g}}{0}{})^3(\nc{g}{1\lnsp}{})^2(N+3)^2(N+2)^2(8N^2+16N)}{(24)^3}\,,
\label{eq:RicciM1cont1}
\end{equation}
where we have used the contraction rule
\begin{equation}
    \begin{tikzpicture}[scale=0.65,
    vertex/.style={draw,circle,fill=black,minimum size=2pt,inner sep=0pt},
    arc/.style={thick},baseline=(vert_cent.base)]
    \node (vert_cent) at (0,0) {$\phantom{\cdot}$};
    \foreach [count=\i] \coord in {
(-0.75,0.5), (-0.25,0.5),(0.25,0.5),(0.75,0.5)}{
        \node[inner sep=0pt] (t\i) at \coord {};
    }
    \foreach [count=\i] \coord in {
(-0.75,-0.5), (-0.25,-0.5),(0.25,-0.5),(0.75,-0.5)}{
        \node[inner sep=0pt] (b\i) at \coord {};
    }
    \draw (b4) to[out=270, in=270,looseness=4] (b3);
    \draw (b1) to[out=270, in=270,looseness=4] (b2);
    \draw (t4) to[out=90, in=90,looseness=4] (t3);
    \draw (t1) to[out=90, in=90,looseness=4] (t2);
\end{tikzpicture}=\frac{8N^2+16N}{24}\,.
\label{eq:contrule4cups}
\end{equation}
The second term may then be evaluated in a very similar way. Ignoring for now the final contraction with $\delta^{IJ}$ this term is
\begin{equation}
\begin{split}
    \left(\frac{\nc{\mathfrak{g}}{0}{}\nc{g}{1\lnsp}{}}{2}\right)^{-2}{\nc{\Gamma}{0}{}}^{K}_{LI}{\nc{\Gamma}{0}{}}^{L}_{KJ}=& \begin{tikzpicture}[scale=0.65,
    vertex/.style={draw,circle,fill=black,minimum size=2pt,inner sep=0pt},
    arc/.style={thick},baseline=(vert_cent.base)]
    \node (vert_cent) at (0,0) {$\phantom{\cdot}$};
    \foreach [count=\i] \coord in {
(0,1), (0,-1)}{
        \node[] (c\i) at \coord {$\phantom{\cdot}$};
    }
    \foreach [count=\i] \coord in {
(1.5,-0.75), (1.5,-0.25),(1.5,0.25),(1.5,0.75)}{
        \node[inner sep=0pt] (r\i) at \coord {};
    }
    \foreach [count=\i] \coord in {
(-1.5,-0.75), (-1.5,-0.25),(-1.5,0.25),(-1.5,0.75)}{
        \node[inner sep=0pt] (l\i) at \coord {};
    }
    \draw (l1) edge (c2);
    \draw (l2) edge (c2);
    \draw (l3) edge (c1);
    \draw (l4) edge (c1);
    \draw (r1) edge (c2);
    \draw (r2) edge (c2);
    \draw (r3) edge (c1);
    \draw (r4) edge (c1);
    \draw (c1) to[out=230, in=130] (c2);
    \draw (c1) to[out=240, in=120] (c2);
    \draw (c1) to[out=300, in=60] (c2);
    \draw (c1) to[out=310, in=50] (c2);
    \node[xshift=-7pt] at (-1.5,0) {$I$};
    \node[xshift=7pt] at (1.5,0) {$J$};
\end{tikzpicture}\\=&\frac{4}{24}\,\begin{tikzpicture}[scale=0.65,
    vertex/.style={draw,circle,fill=black,minimum size=2pt,inner sep=0pt},
    arc/.style={thick},baseline=(vert_cent.base)]
    \node (vert_cent) at (0,0) {$\phantom{\cdot}$};
    \foreach [count=\i] \coord in {
(0,1), (0,-1)}{
        \node[] (c\i) at \coord {$\phantom{\cdot}$};
    }
    \foreach [count=\i] \coord in {
(1.5,-0.75), (1.5,-0.25),(1.5,0.25),(1.5,0.75)}{
        \node[inner sep=0pt] (r\i) at \coord {};
    }
    \foreach [count=\i] \coord in {
(-1.5,-0.75), (-1.5,-0.25),(-1.5,0.25),(-1.5,0.75)}{
        \node[inner sep=0pt] (l\i) at \coord {};
    }
    \draw (l1) edge (c2);
    \draw (l2) edge (c2);
    \draw (l3) edge (r3);
    \draw (l4) edge (r4);
    \draw (r1) edge (c2);
    \draw (r2) edge (c2);
    \draw (c2) to[out=120, in=60,looseness=4] (c2);
    \draw (c2) to[out=140, in=40,looseness=6] (c2);
    \node[xshift=-7pt] at (-1.5,0) {$I$};
    \node[xshift=7pt] at (1.5,0) {$J$};
\end{tikzpicture}+\frac{4}{24}\,\begin{tikzpicture}[scale=0.65,
    vertex/.style={draw,circle,fill=black,minimum size=2pt,inner sep=0pt},
    arc/.style={thick},baseline=(vert_cent.base)]
    \node (vert_cent) at (0,0) {$\phantom{\cdot}$};
    \foreach [count=\i] \coord in {
(0,1), (0,-1)}{
        \node[] (c\i) at \coord {$\phantom{\cdot}$};
    }
    \foreach [count=\i] \coord in {
(1.5,-0.75), (1.5,-0.25),(1.5,0.25),(1.5,0.75)}{
        \node[inner sep=0pt] (r\i) at \coord {};
    }
    \foreach [count=\i] \coord in {
(-1.5,-0.75), (-1.5,-0.25),(-1.5,0.25),(-1.5,0.75)}{
        \node[inner sep=0pt] (l\i) at \coord {};
    }
    \draw (l1) edge (c2);
    \draw (l2) edge (c2);
    \draw (r1) edge (c2);
    \draw (r2) edge (c2);
    \draw (l3) to[out=0, in=50] (c2);
    \draw (l4) to[out=0, in=40] (c2);
    \node[fill=white] at (0,0) {$\phantom{\cdot}$};
    \draw (r3) to[out=180, in=130] (c2);
    \draw (r4) to[out=180, in=140] (c2);
    \node[xshift=-7pt] at (-1.5,0) {$I$};
    \node[xshift=7pt] at (1.5,0) {$J$};
\end{tikzpicture}+\frac{16}{24}\,\begin{tikzpicture}[scale=0.65,
    vertex/.style={draw,circle,fill=black,minimum size=2pt,inner sep=0pt},
    arc/.style={thick},baseline=(vert_cent.base)]
    \node (vert_cent) at (0,0) {$\phantom{\cdot}$};
    \foreach [count=\i] \coord in {
(0,1), (0,-1)}{
        \node[] (c\i) at \coord {$\phantom{\cdot}$};
    }
    \foreach [count=\i] \coord in {
(1.5,-0.75), (1.5,-0.25),(1.5,0.25),(1.5,0.75)}{
        \node[inner sep=0pt] (r\i) at \coord {};
    }
    \foreach [count=\i] \coord in {
(-1.5,-0.75), (-1.5,-0.25),(-1.5,0.25),(-1.5,0.75)}{
        \node[inner sep=0pt] (l\i) at \coord {};
    }
    \draw (l1) edge (c2);
    \draw (l2) edge (c2);
    \draw (r1) edge (c2);
    \draw (r2) edge (c2);
    \draw (r4) edge (l4);
    \draw (l3) to[out=0, in=40] (c2);
    \node[fill=white] at (0,0) {$\phantom{\cdot}$};
    \draw (r3) to[out=180, in=130] (c2);
    \node[xshift=-7pt] at (-1.5,0) {$I$};
    \node[xshift=7pt] at (1.5,0) {$J$};
\end{tikzpicture}\,.
\end{split}
\end{equation}
These diagrams we can then evaluate individually as
\begin{equation}
    \begin{tikzpicture}[scale=0.65,
    vertex/.style={draw,circle,fill=black,minimum size=2pt,inner sep=0pt},
    arc/.style={thick},baseline=(vert_cent.base)]
    \node (vert_cent) at (0,0) {$\phantom{\cdot}$};
    \foreach [count=\i] \coord in {
(0,1), (0,-1)}{
        \node[] (c\i) at \coord {$\phantom{\cdot}$};
    }
    \foreach [count=\i] \coord in {
(1.5,-0.75), (1.5,-0.25),(1.5,0.25),(1.5,0.75)}{
        \node[inner sep=0pt] (r\i) at \coord {};
    }
    \foreach [count=\i] \coord in {
(-1.5,-0.75), (-1.5,-0.25),(-1.5,0.25),(-1.5,0.75)}{
        \node[inner sep=0pt] (l\i) at \coord {};
    }
    \draw (l1) edge (c2);
    \draw (l2) edge (c2);
    \draw (l3) edge (r3);
    \draw (l4) edge (r4);
    \draw (r1) edge (c2);
    \draw (r2) edge (c2);
    \draw (c2) to[out=120, in=60,looseness=4] (c2);
    \draw (c2) to[out=140, in=40,looseness=6] (c2);
    \node[xshift=-7pt] at (-1.5,0) {$I$};
    \node[xshift=7pt] at (1.5,0) {$J$};
\end{tikzpicture}=\frac{2(N+3)(N+2)}{24}\,\begin{tikzpicture}[scale=0.65,
    vertex/.style={draw,circle,fill=black,minimum size=2pt,inner sep=0pt},
    arc/.style={thick},baseline=(vert_cent.base)]
    \node (vert_cent) at (0,0) {$\phantom{\cdot}$};
    \foreach [count=\i] \coord in {
(1.5,-0.75), (1.5,-0.25),(1.5,0.25),(1.5,0.75)}{
        \node[inner sep=0pt] (r\i) at \coord {};
    }
    \foreach [count=\i] \coord in {
(-1.5,-0.75), (-1.5,-0.25),(-1.5,0.25),(-1.5,0.75)}{
        \node[inner sep=0pt] (l\i) at \coord {};
    }
    \draw (l1) edge (r1);
    \draw (l2) edge (r2);
    \draw (l3) edge (r3);
    \draw (l4) edge (r4);
    \node[xshift=-7pt] at (-1.5,0) {$I$};
    \node[xshift=7pt] at (1.5,0) {$J$};
\end{tikzpicture}\,,
\end{equation}
\begin{equation}
    \begin{tikzpicture}[scale=0.65,
    vertex/.style={draw,circle,fill=black,minimum size=2pt,inner sep=0pt},
    arc/.style={thick},baseline=(vert_cent.base)]
    \node (vert_cent) at (0,0) {$\phantom{\cdot}$};
    \foreach [count=\i] \coord in {
(0,1), (0,-1)}{
        \node[] (c\i) at \coord {$\phantom{\cdot}$};
    }
    \foreach [count=\i] \coord in {
(1.5,-0.75), (1.5,-0.25),(1.5,0.25),(1.5,0.75)}{
        \node[inner sep=0pt] (r\i) at \coord {};
    }
    \foreach [count=\i] \coord in {
(-1.5,-0.75), (-1.5,-0.25),(-1.5,0.25),(-1.5,0.75)}{
        \node[inner sep=0pt] (l\i) at \coord {};
    }
    \draw (l1) edge (c2);
    \draw (l2) edge (c2);
    \draw (r1) edge (c2);
    \draw (r2) edge (c2);
    \draw (l3) to[out=0, in=50] (c2);
    \draw (l4) to[out=0, in=40] (c2);
    \node[fill=white] at (0,0) {$\phantom{\cdot}$};
    \draw (r3) to[out=180, in=130] (c2);
    \draw (r4) to[out=180, in=140] (c2);
    \node[xshift=-7pt] at (-1.5,0) {$I$};
    \node[xshift=7pt] at (1.5,0) {$J$};
\end{tikzpicture}=\frac{16}{24}\,\begin{tikzpicture}[scale=0.65,
    vertex/.style={draw,circle,fill=black,minimum size=2pt,inner sep=0pt},
    arc/.style={thick},baseline=(vert_cent.base)]
    \node (vert_cent) at (0,0) {$\phantom{\cdot}$};
    \foreach [count=\i] \coord in {
(1.5,-0.75), (1.5,-0.25),(1.5,0.25),(1.5,0.75)}{
        \node[inner sep=0pt] (r\i) at \coord {};
    }
    \foreach [count=\i] \coord in {
(-1.5,-0.75), (-1.5,-0.25),(-1.5,0.25),(-1.5,0.75)}{
        \node[inner sep=0pt] (l\i) at \coord {};
    }
    \draw (l1) edge (r1);
    \draw (l2) to[out=0, in=0,looseness=4] (l3);
    \draw (r2) to[out=180, in=180,looseness=4] (r3);
    \draw (l4) edge (r4);
    \node[xshift=-7pt] at (-1.5,0) {$I$};
    \node[xshift=7pt] at (1.5,0) {$J$};
\end{tikzpicture}+\frac{4}{24}\,\begin{tikzpicture}[scale=0.65,
    vertex/.style={draw,circle,fill=black,minimum size=2pt,inner sep=0pt},
    arc/.style={thick},baseline=(vert_cent.base)]
    \node (vert_cent) at (0,0) {$\phantom{\cdot}$};
    \foreach [count=\i] \coord in {
(1.5,-0.75), (1.5,-0.25),(1.5,0.25),(1.5,0.75)}{
        \node[inner sep=0pt] (r\i) at \coord {};
    }
    \foreach [count=\i] \coord in {
(-1.5,-0.75), (-1.5,-0.25),(-1.5,0.25),(-1.5,0.75)}{
        \node[inner sep=0pt] (l\i) at \coord {};
    }
    \draw (l1) edge (r1);
    \draw (l2) edge (r2);
    \draw (l3) edge (r3);
    \draw (l4) edge (r4);
    \node[xshift=-7pt] at (-1.5,0) {$I$};
    \node[xshift=7pt] at (1.5,0) {$J$};
\end{tikzpicture}+\frac{4}{24}\,\begin{tikzpicture}[scale=0.65,
    vertex/.style={draw,circle,fill=black,minimum size=2pt,inner sep=0pt},
    arc/.style={thick},baseline=(vert_cent.base)]
    \node (vert_cent) at (0,0) {$\phantom{\cdot}$};
    \foreach [count=\i] \coord in {
(1.5,-0.75), (1.5,-0.25),(1.5,0.25),(1.5,0.75)}{
        \node[inner sep=0pt] (r\i) at \coord {};
    }
    \foreach [count=\i] \coord in {
(-1.5,-0.75), (-1.5,-0.25),(-1.5,0.25),(-1.5,0.75)}{
        \node[inner sep=0pt] (l\i) at \coord {};
    }
    \draw (l1) to[out=0, in=0,looseness=4] (l2);
    \draw (l4) to[out=0, in=0,looseness=4] (l3);
    \draw (r2) to[out=180, in=180,looseness=4] (r1);
    \draw (r4) to[out=180, in=180,looseness=4] (r3);
    \node[xshift=-7pt] at (-1.5,0) {$I$};
    \node[xshift=7pt] at (1.5,0) {$J$};
\end{tikzpicture}\,,
\end{equation}
\begin{equation}
    \begin{tikzpicture}[scale=0.65,
    vertex/.style={draw,circle,fill=black,minimum size=2pt,inner sep=0pt},
    arc/.style={thick},baseline=(vert_cent.base)]
    \node (vert_cent) at (0,0) {$\phantom{\cdot}$};
    \foreach [count=\i] \coord in {
(0,1), (0,-1)}{
        \node[] (c\i) at \coord {$\phantom{\cdot}$};
    }
    \foreach [count=\i] \coord in {
(1.5,-0.75), (1.5,-0.25),(1.5,0.25),(1.5,0.75)}{
        \node[inner sep=0pt] (r\i) at \coord {};
    }
    \foreach [count=\i] \coord in {
(-1.5,-0.75), (-1.5,-0.25),(-1.5,0.25),(-1.5,0.75)}{
        \node[inner sep=0pt] (l\i) at \coord {};
    }
    \draw (l1) edge (c2);
    \draw (l2) edge (c2);
    \draw (r1) edge (c2);
    \draw (r2) edge (c2);
    \draw (r4) edge (l4);
    \draw (l3) to[out=0, in=40] (c2);
    \node[fill=white] at (0,0) {$\phantom{\cdot}$};
    \draw (r3) to[out=180, in=130] (c2);
    \node[xshift=-7pt] at (-1.5,0) {$I$};
    \node[xshift=7pt] at (1.5,0) {$J$};
\end{tikzpicture}=\frac{4N+12}{24}\,\begin{tikzpicture}[scale=0.65,
    vertex/.style={draw,circle,fill=black,minimum size=2pt,inner sep=0pt},
    arc/.style={thick},baseline=(vert_cent.base)]
    \node (vert_cent) at (0,0) {$\phantom{\cdot}$};
    \foreach [count=\i] \coord in {
(1.5,-0.75), (1.5,-0.25),(1.5,0.25),(1.5,0.75)}{
        \node[inner sep=0pt] (r\i) at \coord {};
    }
    \foreach [count=\i] \coord in {
(-1.5,-0.75), (-1.5,-0.25),(-1.5,0.25),(-1.5,0.75)}{
        \node[inner sep=0pt] (l\i) at \coord {};
    }
    \draw (l1) edge (r1);
    \draw (l2) to[out=0, in=0,looseness=4] (l3);
    \draw (r2) to[out=180, in=180,looseness=4] (r3);
    \draw (l4) edge (r4);
    \node[xshift=-7pt] at (-1.5,0) {$I$};
    \node[xshift=7pt] at (1.5,0) {$J$};
\end{tikzpicture}+\frac{2N+6}{24}\,\begin{tikzpicture}[scale=0.65,
    vertex/.style={draw,circle,fill=black,minimum size=2pt,inner sep=0pt},
    arc/.style={thick},baseline=(vert_cent.base)]
    \node (vert_cent) at (0,0) {$\phantom{\cdot}$};
    \foreach [count=\i] \coord in {
(1.5,-0.75), (1.5,-0.25),(1.5,0.25),(1.5,0.75)}{
        \node[inner sep=0pt] (r\i) at \coord {};
    }
    \foreach [count=\i] \coord in {
(-1.5,-0.75), (-1.5,-0.25),(-1.5,0.25),(-1.5,0.75)}{
        \node[inner sep=0pt] (l\i) at \coord {};
    }
    \draw (l1) edge (r1);
    \draw (l2) edge (r2);
    \draw (l3) edge (r3);
    \draw (l4) edge (r4);
    \node[xshift=-7pt] at (-1.5,0) {$I$};
    \node[xshift=7pt] at (1.5,0) {$J$};
\end{tikzpicture}\,.
\end{equation}
Contracting the two remaining generalised indices and using (\ref{eq:contrule2bubbles}) and (\ref{eq:contrule4cups}) one finds
\begin{equation}
    -\nc{\mathfrak{g}}{0}{}\lsp\delta^{IJ}{\nc{\Gamma}{0}{}}^{K}_{LI}{\nc{\Gamma}{0}{}}^{L}_{KJ}=-\nc{\mathfrak{g}}{0}{}\left(\frac{\nc{\mathfrak{g}}{0}{}\nc{g}{1\lnsp}{}}{2}\right)^2\frac{8 N (536 + 706 N + 369 N^2 + 101 N^3 + 15 N^4 + N^5)}{(24)^3}\,.
\label{eq:RicciM1cont2}
\end{equation}
Turning now to the final two terms in (\ref{R0genexp}), we note that there will be two distinct types of diagrams contributing to ${\nc{\Gamma}{1}{}}^{I}_{JK}$: those looking schematically like $\delta\lsp \partial\llsp \nc{G}{2}{}$, and those of the form $\nc{G}{1}{}\partial \llsp\nc{G}{1}{}$. To finish the computation of the $(\nc{g}{1\lnsp}{})^3$ terms in (\ref{eq:Ricci0}), let us first consider the latter. As noted previously in the evaluation of ${\nc{\Gamma}{0}{}}^{I}_{JK}$, the simple form of ${\nc{G}{1}{}}_{JK}$ guarantees that $\partial_{I}{\nc{G}{1}{}}_{JK}$ is symmetric on its three indices, so that the $\nc{G}{1}{}\partial \llsp\nc{G}{1}{}$ terms are given by
\begin{equation}
    \begin{aligned}
    {\nc{\Gamma}{1}{}}^{I}_{JK}\supset\frac{\nc{\mathfrak{g}}{1}{}\nc{g}{1\lnsp}{}}{2}\,
    \begin{tikzpicture}[scale=0.65,
    vertex/.style={draw,circle,fill=black,minimum size=2pt,inner sep=0pt},
    arc/.style={thick},baseline=(vert_cent.base)]
    \node (vert_cent) at (0,0.75) {$\phantom{\cdot}$};
    \foreach [count=\i] \coord in {
(-0.75,1), (-0.25,1),(0.25,1),(0.75,1)}{
        \node[inner sep=0pt] (t\i) at \coord {};
    }
    \foreach [count=\i] \coord in {
(1.375,0.375), (1.125,0.125),(0.875,-0.125),(0.521,-0.479)}{
        \node[inner sep=0pt] (r\i) at \coord {};
    }
    \foreach [count=\i] \coord in {
(-1.375,0.375), (-1.125,0.125),(-0.875,-0.125),(-0.521,-0.479)}{
        \node[inner sep=0pt] (l\i) at \coord {};
    }
    \draw (l1) edge (t1);
    \draw (l2) edge (t2);
    \draw (r1) edge (t4);
    \draw (r2) edge (t3);
    \draw (r3) edge (l3);
    \draw (r4) edge (l4);
    \node[xshift=-7pt,yshift=-7pt] at (-1,0) {$J$};
    \node[xshift=7pt,yshift=-7pt] at (1,0) {$K$};
    \node[vertex] (c) at (0.5,1.75) {};
    \foreach [count=\i] \coord in {
(0.25,1.25), (0.25,2.25),(0.75,2.25),(0.75,1.25)}{
        \node[] (p\i) at \coord {};
    }
    \foreach [count=\i] \coord in {
(-0.75,2.25), (-0.75,1.25),(-0.25,2.25),(-0.25,1.25)}{
        \node[] (d\i) at \coord {};
    }
    \draw (c) edge (p1)
                   edge (p2)
                   edge (p3)
                   edge (p4);
    \draw (d1) edge (d2);
    \draw (d3) edge (d4);
    \node[yshift=7pt] at (0,2.5) {$I$};
\end{tikzpicture}&=\frac{4\nc{\mathfrak{g}}{1}{}\nc{g}{1\lnsp}{}}{48}\,\begin{tikzpicture}[scale=0.65,
    vertex/.style={draw,circle,fill=black,minimum size=2pt,inner sep=0pt},
    arc/.style={thick},baseline=(vert_cent.base)]
    \node (vert_cent) at (0,0.26) {$\phantom{\cdot}$};
    \foreach [count=\i] \coord in {
(-0.75,1), (-0.25,1),(0.25,1),(0.75,1)}{
        \node[inner sep=0pt] (t\i) at \coord {};
    }
    \foreach [count=\i] \coord in {
(1.375,0.375), (1.125,0.125),(0.875,-0.125),(0.521,-0.479)}{
        \node[inner sep=0pt] (r\i) at \coord {};
    }
    \foreach [count=\i] \coord in {
(-1.375,0.375), (-1.125,0.125),(-0.875,-0.125),(-0.521,-0.479)}{
        \node[inner sep=0pt] (l\i) at \coord {};
    }
    \draw (l1) edge (t1);
    \draw (l2) edge (t2);
    \draw (r3) edge (l3);
    \draw (r4) edge (l4);
    \node[vertex] (c) at (0.906,0.635) {};
    \draw (c) edge (r1)
                   edge (r2)
                   edge (t3)
                   edge (t4);
    \node[xshift=-7pt,yshift=-7pt] at (-1,0) {$J$};
    \node[xshift=7pt,yshift=-7pt] at (1,0) {$K$};
    \node[yshift=7pt] at (0,1) {$I$};
\end{tikzpicture}+\frac{4\nc{\mathfrak{g}}{1}{}\nc{g}{1\lnsp}{}}{48}\,\begin{tikzpicture}[scale=0.65,
    vertex/.style={draw,circle,fill=black,minimum size=2pt,inner sep=0pt},
    arc/.style={thick},baseline=(vert_cent.base)]
    \node (vert_cent) at (0,0.26) {$\phantom{\cdot}$};
    \foreach [count=\i] \coord in {
(-0.75,1), (-0.25,1),(0.25,1),(0.75,1)}{
        \node[inner sep=0pt] (t\i) at \coord {};
    }
    \foreach [count=\i] \coord in {
(1.375,0.375), (1.125,0.125),(0.875,-0.125),(0.521,-0.479)}{
        \node[inner sep=0pt] (r\i) at \coord {};
    }
    \foreach [count=\i] \coord in {
(-1.375,0.375), (-1.125,0.125),(-0.875,-0.125),(-0.521,-0.479)}{
        \node[inner sep=0pt] (l\i) at \coord {};
    }
    \draw (r1) edge (t4);
    \draw (r2) edge (t3);
    \draw (r3) edge (l3);
    \draw (r4) edge (l4);
    \node[vertex] (c) at (-0.906,0.635) {};
    \draw (c) edge (l1)
                   edge (l2)
                   edge (t1)
                   edge (t2);
    \node[xshift=-7pt,yshift=-7pt] at (-1,0) {$J$};
    \node[xshift=7pt,yshift=-7pt] at (1,0) {$K$};
    \node[yshift=7pt] at (0,1) {$I$};
\end{tikzpicture}\\&\quad+\frac{16\nc{\mathfrak{g}}{1}{}\nc{g}{1\lnsp}{}}{48}\,\begin{tikzpicture}[scale=0.65,
    vertex/.style={draw,circle,fill=black,minimum size=2pt,inner sep=0pt},
    arc/.style={thick},baseline=(vert_cent.base)]
    \node (vert_cent) at (0,0.26) {$\phantom{\cdot}$};
    \foreach [count=\i] \coord in {
(-0.75,1), (-0.25,1),(0.25,1),(0.75,1)}{
        \node[inner sep=0pt] (t\i) at \coord {};
    }
    \foreach [count=\i] \coord in {
(1.375,0.375), (1.125,0.125),(0.875,-0.125),(0.521,-0.479)}{
        \node[inner sep=0pt] (r\i) at \coord {};
    }
    \foreach [count=\i] \coord in {
(-1.375,0.375), (-1.125,0.125),(-0.875,-0.125),(-0.521,-0.479)}{
        \node[inner sep=0pt] (l\i) at \coord {};
    }
    \draw (r1) edge (t4);
    \draw (l1) edge (t1);
    \draw (r3) edge (l3);
    \draw (r4) edge (l4);
    \node[vertex] (c) at (0,0.5) {};
    \draw (c) edge (r2)
                   edge (l2)
                   edge (t3)
                   edge (t2);
    \node[xshift=-7pt,yshift=-7pt] at (-1,0) {$J$};
    \node[xshift=7pt,yshift=-7pt] at (1,0) {$K$};
    \node[yshift=7pt] at (0,1) {$I$};
\end{tikzpicture}\,.
\end{aligned}
\end{equation}
To determine the contribution to $\nc{R}{0\lnsp}{}$ we must find the difference between tracing over the lower indices and tracing over one upper and one lower index, and then remove the final index with a derivative. First, tracing over the lower indices yields
\begin{equation}
    {\nc{\Gamma}{1}{}}^{I}_{JJ}\supset\frac{4\nc{\mathfrak{g}}{1}{}\nc{g}{1\lnsp}{}}{48}\begin{tikzpicture}[scale=0.65,
    vertex/.style={draw,circle,fill=black,minimum size=2pt,inner sep=0pt},
    arc/.style={thick},baseline=(vert_cent.base)]
    \node (vert_cent) at (0,0) {$\phantom{\cdot}$};
    \foreach [count=\i] \coord in {
(-0.75,1), (-0.25,1),(0.25,1),(0.75,1)}{
        \node[inner sep=0pt] (t\i) at \coord {};
    }
    \foreach [count=\i] \coord in {
(0,0),(0,0),(0,0),(0,0)}{
        \node[fill=white] (r\i) at \coord {$\phantom{\cdot}$};
    }
    \foreach [count=\i] \coord in {
(0,0),(0,0),(0,0),(0,0)}{
        \node[fill=white] (l\i) at \coord {$\phantom{\cdot}$};
    }
    \draw (l1) to[out=165, in=270] (t1);
    \draw (l2) to[out=135, in=270] (t2);
    \node[vertex] (c) at (0.5,0.65) {};
    \draw (c) to[out=225, in=45] (r1);
    \draw (c) to[out=-45, in=15] (r2);
    \draw (c) edge (t3)
                   edge (t4);
    \node[yshift=7pt] at (0,1) {$I$};
    \draw (0,-0.6) circle (0.6cm);
    \draw (0,-0.52) circle (0.4cm);
    \node[fill=white] at (0,0) {$\phantom{\cdot}$};
\end{tikzpicture}+\frac{4\nc{\mathfrak{g}}{1}{}\nc{g}{1\lnsp}{}}{48}\begin{tikzpicture}[scale=0.65,
    vertex/.style={draw,circle,fill=black,minimum size=2pt,inner sep=0pt},
    arc/.style={thick},baseline=(vert_cent.base)]
    \node (vert_cent) at (0,0) {$\phantom{\cdot}$};
    \foreach [count=\i] \coord in {
(-0.75,1), (-0.25,1),(0.25,1),(0.75,1)}{
        \node[inner sep=0pt] (t\i) at \coord {};
    }
    \foreach [count=\i] \coord in {
(0,0),(0,0),(0,0),(0,0)}{
        \node[fill=white] (r\i) at \coord {$\phantom{\cdot}$};
    }
    \foreach [count=\i] \coord in {
(0,0),(0,0),(0,0),(0,0)}{
        \node[fill=white] (l\i) at \coord {$\phantom{\cdot}$};
    }
    \draw (r1) to[out=15, in=270] (t4);
    \draw (r2) to[out=45, in=270] (t3);
    \node[vertex] (c) at (-0.5,0.65) {};
    \draw (c) to[out=-45, in=135] (l1);
    \draw (c) to[out=225, in=165] (l2);
    \draw (c) edge (t1)
                   edge (t2);
    \node[yshift=7pt] at (0,1) {$I$};
    \draw (0,-0.6) circle (0.6cm);
    \draw (0,-0.52) circle (0.4cm);
    \node[fill=white] at (0,0) {$\phantom{\cdot}$};
\end{tikzpicture}+\frac{16\nc{\mathfrak{g}}{1}{}\nc{g}{1\lnsp}{}}{48}\begin{tikzpicture}[scale=0.65,
    vertex/.style={draw,circle,fill=black,minimum size=2pt,inner sep=0pt},
    arc/.style={thick},baseline=(vert_cent.base)]
    \node (vert_cent) at (0,0) {$\phantom{\cdot}$};
    \foreach [count=\i] \coord in {
(-0.75,1), (-0.25,1),(0.25,1),(0.75,1)}{
        \node[inner sep=0pt] (t\i) at \coord {};
    }
    \foreach [count=\i] \coord in {
(0,0),(0,0),(0,0),(0,0)}{
        \node[fill=white] (r\i) at \coord {$\phantom{\cdot}$};
    }
    \foreach [count=\i] \coord in {
(0,0),(0,0),(0,0),(0,0)}{
        \node[fill=white] (l\i) at \coord {$\phantom{\cdot}$};
    }
    \draw (r1) to[out=15, in=270] (t4);
    \draw (l1) to[out=165, in=270] (t1);
    \node[vertex] (c) at (0,0.75) {};
    \draw (c) edge (t3)
                   edge (t2);
    \draw (c) to[out=225, in=135] (l2);
    \draw (c) to[out=-45, in=45] (r2);
    \node[yshift=7pt] at (0,1) {$I$};
    \draw (0,-0.6) circle (0.6cm);
    \draw (0,-0.52) circle (0.4cm);
    \node[fill=white] at (0,0) {$\phantom{\cdot}$};
\end{tikzpicture}\,,
\end{equation}
which one can easily simplify using (\ref{eq:contrule2bubbles}) to find
\begin{equation}
    {\nc{\Gamma}{1}{}}^{I}_{JJ}\supset\frac{16\nc{\mathfrak{g}}{1}{}\nc{g}{1\lnsp}{}(N+3)(N+2)}{(24)^2}\,\begin{tikzpicture}[scale=0.65,
    vertex/.style={draw,circle,fill=black,minimum size=2pt,inner sep=0pt},
    arc/.style={thick},baseline=(vert_cent.base)]
    \node (vert_cent) at (0,0.5) {$\phantom{\cdot}$};
    \foreach [count=\i] \coord in {
(-0.75,1), (-0.25,1),(0.25,1),(0.75,1)}{
        \node[inner sep=0pt] (t\i) at \coord {};
    }
    \node[vertex] (c) at (0,0) {};
    \draw (c) edge (t1)
                   edge (t2)
                   edge (t3)
                   edge (t4);
    \node[yshift=7pt] at (0,1) {$I$};
\end{tikzpicture}+\frac{8\nc{\mathfrak{g}}{1}{}\nc{g}{1\lnsp}{}(N+3)(N+2)}{(24)^2}\,\begin{tikzpicture}[scale=0.65,
    vertex/.style={draw,circle,fill=black,minimum size=2pt,inner sep=0pt},
    arc/.style={thick},baseline=(vert_cent.base)]
    \node (vert_cent) at (0,0.25) {$\phantom{\cdot}$};
    \foreach [count=\i] \coord in {
(-0.75,1), (-0.25,1),(0.25,1),(0.75,1)}{
        \node[inner sep=0pt] (t\i) at \coord {};
    }
    \node[vertex] (c) at (-0.5,0) {};
    \draw (c) edge (t1)
                   edge (t2);
    \draw (t3) to[out=270, in=270,looseness=4] (t4);
    \draw (-0.5,-0.25) circle (0.25cm);
    \node[yshift=7pt] at (0,1) {$I$};
\end{tikzpicture}\,.
\end{equation}
Then, tracing over an upper and a lower index yields
\begin{equation}
    {\nc{\Gamma}{1}{}}^{I}_{JI}\supset\frac{4\nc{\mathfrak{g}}{1}{}\nc{g}{1\lnsp}{}}{48}\,\begin{tikzpicture}[scale=0.65,
    vertex/.style={draw,circle,fill=black,minimum size=2pt,inner sep=0pt},
    arc/.style={thick},baseline=(vert_cent.base)]
    \node (vert_cent) at (0,1) {$\phantom{\cdot}$};
    \foreach [count=\i] \coord in {
(0,1), (0,1),(0,1),(0,1)}{
        \node[fill=white] (t\i) at \coord {$\phantom{\cdot}$};
    }
    \foreach [count=\i] \coord in {
(0,1), (0,1),(0,1),(0,1)}{
        \node[fill=white] (r\i) at \coord {$\phantom{\cdot}$};
    }
    \foreach [count=\i] \coord in {
(-1.375,0.375), (-1.125,0.125),(-0.875,-0.125),(-0.625,-0.375)}{
        \node[inner sep=0pt] (l\i) at \coord {};
    }
    \draw (t1) to[out=195, in=60] (l1);
    \draw (t2) to[out=225, in=60] (l2);
    \draw (r3) to[out=-45, in=0] (l3);
    \draw (r4) to[out=-15, in=0,looseness=1.5] (l4);
    \node[vertex] (c) at (0,2) {};
    \draw (c)  to[out=0, in=15,looseness=2] (r1);
    \draw (c)  to[out=330, in=45,looseness=1] (r2);
    \draw (c)  to[out=180, in=165,looseness=2] (t3);
    \draw (c)  to[out=210, in=135,looseness=1] (t4);
    \node[xshift=-7pt,yshift=-7pt] at (-1,0) {$J$};
\end{tikzpicture}+\frac{4\nc{\mathfrak{g}}{1}{}\nc{g}{1\lnsp}{}}{48}\,\begin{tikzpicture}[scale=0.65,
    vertex/.style={draw,circle,fill=black,minimum size=2pt,inner sep=0pt},
    arc/.style={thick},baseline=(vert_cent.base)]
    \node (vert_cent) at (0,1) {$\phantom{\cdot}$};
    \foreach [count=\i] \coord in {
(0,1), (0,1),(0,1),(0,1)}{
        \node[fill=white] (t\i) at \coord {$\phantom{\cdot}$};
    }
    \foreach [count=\i] \coord in {
(0,1), (0,1),(0,1),(0,1)}{
        \node[fill=white] (r\i) at \coord {$\phantom{\cdot}$};
    }
    \foreach [count=\i] \coord in {
(-1.375,0.375), (-1.125,0.125),(-0.875,-0.125),(-0.521,-0.479)}{
        \node[inner sep=0pt] (l\i) at \coord {};
    }
    \draw (0,1.6) circle (0.6cm);
    \draw (0,1.52) circle (0.4cm);
    \node[fill=white] at (0,1) {$\phantom{\cdot}$};
    \draw (r3) to[out=-45,in=30] (l3);
    \draw (r4) to[out=-15,in=30,looseness=2] (l4);
    \node[vertex] (c) at (-0.906,0.635) {};
    \draw (c) edge (l1)
                   edge (l2);
    \draw (c) to[out=30,in=195] (t1);
    \draw (c) to[out=0,in=225] (t2);
    \node[xshift=-7pt,yshift=-7pt] at (-1,0) {$J$};
\end{tikzpicture}+\frac{16\nc{\mathfrak{g}}{1}{}\nc{g}{1\lnsp}{}}{48}\,\begin{tikzpicture}[scale=0.65,
    vertex/.style={draw,circle,fill=black,minimum size=2pt,inner sep=0pt},
    arc/.style={thick},baseline=(vert_cent.base)]
    \node (vert_cent) at (0,1) {$\phantom{\cdot}$};
    \foreach [count=\i] \coord in {
(0,1), (0,1),(0,1),(0,1)}{
        \node[fill=white] (t\i) at \coord {$\phantom{\cdot}$};
    }
    \foreach [count=\i] \coord in {
(0,1), (0,1),(0,1),(0,1)}{
        \node[fill=white] (r\i) at \coord {$\phantom{\cdot}$};
    }
    \foreach [count=\i] \coord in {
(-1.375,0.375), (-1.125,0.125),(-0.875,-0.125),(-0.521,-0.479)}{
        \node[inner sep=0pt] (l\i) at \coord {};
    }
    \draw (0,1.52) circle (0.4cm);
    \node[fill=white] at (0,1) {$\phantom{\cdot}$};
    \draw (l1) to[in=165] (t1);
    \draw (r3) to[out=-15,in=0,looseness=2] (l3);
    \draw (r4) to[out=15,in=0,looseness=2.5] (l4);
    \node[vertex] (c) at (0,0.3) {};
    \draw (c) edge (l2);
    \draw (c) to[out=30,in=-45] (r2);
    \draw (c) to[out=180,in=195,looseness=2] (t3);
    \draw (c) to[out=150,in=225] (t2);
    \node[xshift=-7pt,yshift=-7pt] at (-1,0) {$J$};
\end{tikzpicture}\,,
\end{equation}
which can be simplified to
\begin{equation}
    {\nc{\Gamma}{1}{}}^{I}_{JI}\supset\frac{4\nc{\mathfrak{g}}{1}{}\nc{g}{1\lnsp}{}(N+5)(N+4)}{(24)^2}\,\begin{tikzpicture}[scale=0.65,
    vertex/.style={draw,circle,fill=black,minimum size=2pt,inner sep=0pt},
    arc/.style={thick},baseline=(vert_cent.base)]
    \node (vert_cent) at (0,0.5) {$\phantom{\cdot}$};
    \foreach [count=\i] \coord in {
(-0.75,1), (-0.25,1),(0.25,1),(0.75,1)}{
        \node[inner sep=0pt] (t\i) at \coord {};
    }
    \node[vertex] (c) at (0,0) {};
    \draw (c) edge (t1)
                   edge (t2)
                   edge (t3)
                   edge (t4);
    \node[yshift=7pt] at (0,1) {$J$};
\end{tikzpicture}+\frac{32\nc{\mathfrak{g}}{1}{}\nc{g}{1\lnsp}{}(N+4)}{(24)^2}\,\begin{tikzpicture}[scale=0.65,
    vertex/.style={draw,circle,fill=black,minimum size=2pt,inner sep=0pt},
    arc/.style={thick},baseline=(vert_cent.base)]
    \node (vert_cent) at (0,0.25) {$\phantom{\cdot}$};
    \foreach [count=\i] \coord in {
(-0.75,1), (-0.25,1),(0.25,1),(0.75,1)}{
        \node[inner sep=0pt] (t\i) at \coord {};
    }
    \node[vertex] (c) at (-0.5,0) {};
    \draw (c) edge (t1)
                   edge (t2);
    \draw (t3) to[out=270, in=270,looseness=4] (t4);
    \draw (-0.5,-0.25) circle (0.25cm);
    \node[yshift=7pt] at (0,1) {$J$};
\end{tikzpicture}+\frac{8\nc{\mathfrak{g}}{1}{}\nc{g}{1\lnsp}{}}{(24)^2}\begin{tikzpicture}[scale=0.65,
    vertex/.style={draw,circle,fill=black,minimum size=2pt,inner sep=0pt},
    arc/.style={thick},baseline=(vert_cent.base)]
    \node (vert_cent) at (0,0.25) {$\phantom{\cdot}$};
    \foreach [count=\i] \coord in {
(-0.75,1), (-0.25,1),(0.25,1),(0.75,1)}{
        \node[inner sep=0pt] (t\i) at \coord {};
    }
    \node[vertex] (c) at (0,0) {};
    \draw (t1) to[out=270, in=270,looseness=4] (t2);
    \draw (t3) to[out=270, in=270,looseness=4] (t4);
    \draw (-0.25,0) circle (0.25cm);
    \draw (0.25,0) circle (0.25cm);
    \node[yshift=7pt] at (0,1) {$J$};
\end{tikzpicture}\,.
\end{equation}
As we have reduced both expressions to the sum of three distinct diagrams, to evaluate the remaining derivative and trace, we simply must consider the derivative acting on these diagrams. Using (\ref{eq:contrule2bubbles}) and (\ref{eq:contrule4cups}), after some combinatorial algebra one finds that
\begin{equation}
\begin{split}
    \partial_I\,\begin{tikzpicture}[scale=0.65,
    vertex/.style={draw,circle,fill=black,minimum size=2pt,inner sep=0pt},
    arc/.style={thick},baseline=(vert_cent.base)]
    \node (vert_cent) at (0,0.5) {$\phantom{\cdot}$};
    \foreach [count=\i] \coord in {
(-0.75,1), (-0.25,1),(0.25,1),(0.75,1)}{
        \node[inner sep=0pt] (t\i) at \coord {};
    }
    \node[vertex] (c) at (0,0) {};
    \draw (c) edge (t1)
                   edge (t2)
                   edge (t3)
                   edge (t4);
    \node[yshift=7pt] at (0,1) {$I$};
\end{tikzpicture}&=\frac{N(N+1)(N+2)(N+3)}{24}\,,\\
\partial_I\,\begin{tikzpicture}[scale=0.65,
    vertex/.style={draw,circle,fill=black,minimum size=2pt,inner sep=0pt},
    arc/.style={thick},baseline=(vert_cent.base)]
    \node (vert_cent) at (0,0.25) {$\phantom{\cdot}$};
    \foreach [count=\i] \coord in {
(-0.75,1), (-0.25,1),(0.25,1),(0.75,1)}{
        \node[inner sep=0pt] (t\i) at \coord {};
    }
    \node[vertex] (c) at (-0.5,0) {};
    \draw (c) edge (t1)
                   edge (t2);
    \draw (t3) to[out=270, in=270,looseness=4] (t4);
    \draw (-0.5,-0.25) circle (0.25cm);
    \node[yshift=7pt] at (0,1) {$I$};
\end{tikzpicture}&=\frac{2N(N+2)(N+3)}{24}\,,\\
\partial_I\,\begin{tikzpicture}[scale=0.65,
    vertex/.style={draw,circle,fill=black,minimum size=2pt,inner sep=0pt},
    arc/.style={thick},baseline=(vert_cent.base)]
    \node (vert_cent) at (0,0.25) {$\phantom{\cdot}$};
    \foreach [count=\i] \coord in {
(-0.75,1), (-0.25,1),(0.25,1),(0.75,1)}{
        \node[inner sep=0pt] (t\i) at \coord {};
    }
    \node[vertex] (c) at (0,0) {};
    \draw (t1) to[out=270, in=270,looseness=4] (t2);
    \draw (t3) to[out=270, in=270,looseness=4] (t4);
    \draw (-0.25,0) circle (0.25cm);
    \draw (0.25,0) circle (0.25cm);
    \node[yshift=7pt] at (0,1) {$I$};
\end{tikzpicture}&=\frac{8N(N+2)}{24}\,.
\end{split}
\label{eq:derivcontraction}
\end{equation}
The contribution to the Ricci scalar will then be
\begin{equation}
\begin{split}
    \nc{\mathfrak{g}}{0}{}\lsp\delta^{IJ}\left(\partial_K{\nc{\Gamma}{1}{}}^{K}_{IJ}-\partial_I{\nc{\Gamma}{1}{}}^{K}_{JK}\right)\supset\nc{\mathfrak{g}}{0}{}\nc{g}{1\lnsp}{} \nc{\mathfrak{g}}{1}{}\bigg(&-\frac{31}{432}N-\frac{41}{1728}N^2+\frac{167}{3456}N^3\\&+\frac{127}{3456}N^4+\frac{11}{1152}N^5+\frac{1}{1152}N^6\bigg)\,.
\end{split}
\label{eq:RicciM1cont3}
\end{equation}

Finally, we may turn our eye towards the contribution to the Ricci scalar from $\text{O}(\lambda^2)$ terms in the metric. We note that their presence in (\ref{R0genexp}) simplifies to
\begin{equation}
    \nc{\mathfrak{g}}{0}{}\lsp\delta^{IJ}\left(\partial_K{\nc{\Gamma}{1}{}}^{K}_{IJ}-\partial_I{\nc{\Gamma}{1}{}}^{K}_{JK}\right)\supset \big(\nc{\mathfrak{g}}{0}{}\big)^2\left(\partial^I\partial^J\lsp (\nc{G}{2}{})_{IJ}-\partial_I\partial^I\big(\delta^{JK}(\nc{G}{2}{})_{JK}\big)\right)\,,
\end{equation}
so that there are only two terms per $\text{O}(\lambda^2)$ diagram in (\ref{eq:metricl2}) that we must compute. For the sake of brevity we will refrain here from listing all of the steps explicitly, and provide only the results, but one may straightforwardly re-derive them by performing the same manipulations as before. For the traces, $\delta^{JK}(\nc{G}{2}{})_{JK}$, one finds
\begin{equation}
    \delta^{JK}(\nc{G}{2}{1})_{JK}=\nc{g}{2\lnsp}{1}\,
\,.
\end{equation}
The derivatives acting on these traces can then be evaluated using the three master formulas
\begin{equation}
\begin{split}
    \partial_I\partial^I%
&=\frac{N(N+1)(N+2)(N+3)}{12}\,.
\end{split}
\end{equation}
For the terms involving contractions with a derivative, $\partial^I\lsp (\nc{G}{2}{})_{IJ}$, one finds
\begin{equation}
    \partial^I(\nc{G}{2}{1})_{IJ}=\nc{g}{2\lnsp}{1}\lsp\frac{(N+3)(8+N(3+N))}{24}\,%
\right)\,.
\end{equation}
The remaining derivative can then be evaluated using (\ref{eq:derivcontraction}). In the end one finds that the $\text{O}(\lambda^2)$ terms in the metric provide the contributions shown in \eqref{eq:Ricci0} to the lowest-order term in the Ricci scalar.

\end{appendix}

\bibliography{main}

\end{document}